\documentclass[fleqn,usenatbib]{mnras}

\usepackage{newtxtext,newtxmath}

\usepackage[T1]{fontenc}
\usepackage{ae,aecompl}

\usepackage{graphicx}	
\usepackage{amsmath}	
\usepackage{amssymb}	

\def\surfs{{\sc surfs}}
\def\shark{{\sc Shark}}
\def\viperfish{{\sc Viperfish}}
\def\prospect{{\sc ProSpect}}

\title[FUV-to-FIR emission of \shark\ galaxies]{From the far-ultraviolet to the far-infrared - galaxy emission at $0\le z\le 10$ in the \shark\ semi-analytic model}

\author[Claudia del P. Lagos et al.]{Claudia del P. Lagos$^{1,2,3}$\thanks{E-mail: claudia.lagos@icrar.org},
Aaron S.~G. Robotham$^{1,2}$, 
James W. Trayford$^4$,
\newauthor{Rodrigo Tobar$^1$,
Mat\'ias Bravo$^1$,
Sabine Bellstedt$^1$,
Luke J. M. Davies$^1$,}
\newauthor{Simon P. Driver$^1$,
Pascal J. Elahi$^{1,2}$,
Danail Obreschkow$^{1,2}$,
Chris Power$^{1,2}$}
\\
$^{1}$International Centre for Radio Astronomy Research (ICRAR), M468, University of Western Australia, 35 Stirling Hwy, Crawley, \\WA 6009, Australia.\\
$^{2}$ARC Centre of Excellence for All Sky Astrophysics in 3 Dimensions (ASTRO 3D).\\
$^{3}$Cosmic Dawn Center (DAWN). \\
$^{4}$Leiden Observatory, Leiden University, PO Box 9513, NL-2300 RA Leiden, the Netherlands.
}

\date{Accepted XXX. Received YYY; in original form ZZZ}

\pubyear{2019}

\begin{document}
\label{firstpage}
\pagerange{\pageref{firstpage}--\pageref{lastpage}}
\maketitle

\begin{abstract}
We combine the \shark\ semi-analytic model of galaxy formation with the \prospect\ software tool for spectral energy distribution (SED) generation to study the multi-wavelength emission of galaxies from the far-ultraviolet (FUV) to the far-infrared (FIR) at $0\le z\le 10$. We produce a physical model for the attenuation of galaxies across cosmic time by combining a local Universe empirical relation to compute the dust mass of galaxies from their gas metallicity and mass,  attenuation curves derived from radiative transfer calculations of galaxies in the {\sc EAGLE} hydrodynamic simulation suite, and the properties of \shark\ galaxies. We are able to produce a wide range of galaxies, from the $z=8$ star-forming galaxies with almost no extinction, $z=2$ submillimeter galaxies, down to the normal star-forming and red sequence galaxies at $z=0$. Quantitatively, we find that \shark\ reproduces the observed (i) the $z=0$ FUV-to-FIR, (ii) $0\le z\le 3$ rest-frame $K$-band, and (iii) $0\le z\le 10$ rest-frame FUV luminosity functions, (iv) $z\le 8$ UV slopes, (v) the FUV-to-FIR number counts (including the widely disputed 850$\mu$m), (vi) redshift distribution of bright $850\mu$m galaxies and (vii) the integrated cosmic SED from $z=0$ to $z=1$ to an unprecedented level. This is achieved without the need to invoke changes in the stellar initial mass function, dust-to-metal mass ratio, or metal enrichment timescales. Our model predicts star formation in galaxy disks to dominate in the FUV-to-optical, while bulges dominate at the NIR at all redshifts. The FIR sees a strong evolution in which disks dominate at $z\le 1$ and starbursts (triggered by both galaxy mergers and disk instabilities, in an even mix) dominate at higher redshifts, even out to $z=10$. 
 \end{abstract}




\section{Introduction}


Galaxy formation and evolution is one of the most outstanding questions in astrophysics. Galaxies are thought to form in the centre of the gravitational potential of dark matter (DM)-dominated halos, and hence are significantly affected by the growth of structures in the Universe. They are also subject to highly non-linear, complex astrophysical processes, such as gas accretion, star formation, energetic events that change the thermodynamics of the gas, just to mention a few (see \citealt{Somerville15} for a review on the topic). The clues we get about how galaxies form and evolve come mostly from the electromagnetic spectrum produced by the integrated contribution of gas, dust and stars in galaxies.

This integrated electromagnetic spectrum, also called spectral energy distribution (SED), encodes information of a galaxy's stellar populations, via the light emitted by stars, as well as its interstellar medium (ISM) (both in terms of content and composition) through the absorption of the far-ultraviolet (FUV)-to-optical light, the re-emission in the infrared (IR) and via emission lines in the optical, IR and radio. In addition to this, bright events, such as active galactic nuclei (AGN) can significantly affect the observed SEDs of galaxies (see \citealt{Conroy13} for a review on galaxy SEDs).

Truly multi-wavelength surveys, such as GAMA \citep{Driver09} in the local Universe and COSMOS \citep{Scoville07}, CANDELS \citep{Koekemoer11} and DEVILS \citep{Davies18} in the high-redshift Universe, are becoming more common, and attempt to get a full picture of galaxy properties across the electromagnetic spectrum and cosmic time. This has allowed a full reconstruction of how the stellar mass, star formation rate (SFR), ISM and dust masses evolve with time for the overall population of galaxies \citep{Santini13,Scoville16,Driver17}, the integrated SEDs (referred to as cosmic SED, CSEDs) of galaxies as a function of time \citep{Andrews17}, the size-luminosity correlation as a function of wavelength in the local Universe \citep{Lange15}, the IR-UV correlation as a function of redshift \citep{Capak15}, among many others. The multi-wavelength nature of these surveys can also unveil the contribution from different galaxy populations to the cosmic SFR density of the Universe: at $z=0$ most star formation takes place in galaxies that are bright in the UV-to-optical, while at $z\gtrsim 1$ IR-bright galaxies tend to dominate (e.g. \citealt{Casey12,Magnelli13,Madau14}). These observations require cosmological galaxy formation simulations to be able to reliably predict SEDs of galaxies in as much of the electromagnetic spectrum as possible in order to offer a physical framework in which to interpret these observations, and to truly exploit their constraining power. 

Multi-wavelength predictions covering from the FUV to the FIR have been challenging to produce because of the associated computational cost and uncertainties in the modelling process. In semi-analytic models (SAMs) of galaxy formation, a tool used to follow the formation and evolution of galaxies in DM halo merger trees from cosmological $N$-body simulations, this has been notoriously difficult. Early on \citet{Baugh05}, using GALFORM, noticed that there was significant tension arising when attempting to reproduce the FUV-to-near IR (NIR) and the FIR emission of galaxies simultaneously, and suggested that allowing for deviations from a universal initial stellar mass function (IMF) of stars in the case of starbursts helped solve the tension. This was done using a full radiative transfer (RT) approach in SAM galaxies, assuming a two-phase dust model in idealised geometries and employing the code GRASIL \citep{Granato00}. \citet{Lacey15} confirmed this conclusion in an updated version of GALFORM by adopting a more simplified method to predicting the FIR emission of galaxies. \citet{Cowley19} showed that this tension also impacted the CSED and extra-galactic background light predictions. 

Other SAMs, such as that of \citet{Somerville12} have also attempted to predict the full FUV-to-FIR SEDs of galaxies. They used a different approach to Baugh et al., in that they used 
an attenuation model similar to that of \citet{Charlot00}, with attenuation parameters varying with galaxy properties, and used observed dust templates to inform their model on how to re-emit the light in the IR. \citet{Somerville12} scaled the optical depth with sensible galaxy properties, such as gas metallicity, gas mass and galaxy size, but without a theoretical motivation for their exact scaling. Despite this uncertainty, they found their model to provide a good match to the FUV-to-NIR emission of galaxies, but systematically underpredicted the emission at the FIR, finding a similar tension to that reported by \citet{Baugh05}.

In cosmological hydrodynamical simulations of galaxy formation the situation is not less different. \citet{Trayford17} presented a full RT treatment of galaxies in the {\sc EAGLE} simulations, which allowed the authors to produce FUV-to-FIR SEDs for all their galaxies. \citet{Camps16,Baes19,Cowley19} showed that {\sc EAGLE} was capable of reproducing the FUV-to-NIR emission of galaxies, but under-predicted the FIR emission, possibly suggesting the need for changes in their physical model by e.g. invoking a varying IMF. 

A clear difficulty in providing predictions over the full FUV-to-FIR SED is how to simultaneously model the attenuation of stellar light and re-emission in the mid-to-far IR. To avoid this difficulty, many other SAMs and cosmological hydrodynamical simulations of galaxy formation limit themselves to modelling only the optical-to-NIR emission by using a slab or \citet{Charlot00}-like attenuation curves (see e.g. \citealt{DeLucia07,Croton16,Henriques15,Yung19} for examples from SAMs and \citealt{Trayford15,Nelson18,Vogelsberger19} for hydrodynamical simulations). Although the latter may be a pragmatic approach to tackle traditional galaxy surveys (e.g. SDSS, HST-based), future surveys are likely to move towards a more panchromatic view of galaxies, not only at $z\lesssim 2$ (e.g. GAMA and DEVILS, COSMOS, CANDELS, WAVES), but also at high redshift using the unprecedented combination of HST, JWST and ALMA.

Here, we use the recently introduced SAM of galaxy formation \shark\ \citep{Lagos18b} in combination with RT results from the {\sc EAGLE} simulations of \citet{Trayford17} to produce a physically-motivated model for the attenuation of light in galaxies from the FUV to the NIR, and adopt an energy-conserving approach combined with observational IR templates \citep{Dale14} to re-emit the light in the mid-to-far IR. Our aim is to understand to what extent our state-of-the-art model can reproduce the observed FUV-to-FIR emission of galaxies and whether fine tuning and/or changes in the physical model (such as invoking a varying IMF) are required. Our approach is similar to  \citet{Somerville12} in that we start by adopting the \citet{Charlot00} parametric attenuation form, but we instead use the RT-predicted attenuation curves of {\sc EAGLE} to inform \shark\ on how to scale the attenuation parameters with galaxy properties. 

The advantage of using {\sc EAGLE} to inform \shark, is that 
in {\sc EAGLE} there is no need for assumptions about the geometry of the gas in galaxies and hence the derived attenuation parameters should not be biased by those assumptions {(e.g. axi-symmetry, exponential radial profiles)}, which is a major risk in the case of RT applied to SAMs. {Although using {\sc EAGLE} allows us to relax typical assumptions made in SAMs, there are still important limitations. Most notably is the ISM model, which is sub-grid in simulations of coarse resolution such as {\sc EAGLE}, directly impacting how ``clumpy'' the ISM of galaxies can be. Other sub-grid physical processes, such as stellar and active galactic
nuclei (AGN) feedback, also impact the distribution of gas in galaxies, affecting the predicted attenuation. Hence, we ought to continue testing the validity of the attenuation model adopted here as simulations of higher resolution and improved ISM physics become available.}
Note that we do not attempt to tune to observations and instead combine the {\sc EAGLE} RT results with \shark\ and, when necessary, adopt standard attenuation parameters widely adopted in the literature. The \shark\ model and SEDs presented here will be used to create panchromatic lightcones for the upcoming surveys DEVILS, WAVES, among others.

This paper is organized as follows. $\S$~\ref{sharksec} introduces \shark, describing the main physical processes included in the model, highlighting some key features and successes. We also describe how dust masses are computed. $\S$~\ref{sec:viperfish} describes how we generate SEDs and the models we use for extinction and re-emission in the FIR. $\S$~\ref{sec:LFs} presents a comprehensive study of the galaxy LF from the FUV to the FIR, and from $z=0$ to $z=10$. We compare with available observations and analyse the physical drivers behind the predicted LF evolution. $\S$~\ref{cseds} presents an analysis of galaxy number counts from the NUV-to-FIR, and the cosmic SED, how it is affected by extinction, compare with observations when available, and break down the total light budget into the contribution from different galaxy components. Finally, in  $\S$~\ref{conclusions} we discuss the implications of our main findings, and the main successes and limitations of our work.

\section{The \shark\ semi-analytic model}\label{sharksec}

\shark, introduced by \citet{Lagos18b}, is an open source, flexible and highly modular SAM\footnote{\href{https://github.com/ICRAR/shark}{\url{https://github.com/ICRAR/shark}}}. 
The model includes all the physical processes that we think shape the formation and evolution of galaxies. These are (i) the collapse and merging of DM halos; (ii) the accretion of gas onto halos, which is modulated by the DM accretion rate; (iii) the shock heating and radiative cooling of gas inside DM halos, leading to the
formation of galactic disks via conservation of specific angular momentum of the cooling gas; (iv) star formation in galaxy disks; (v) stellar feedback from the evolving stellar populations; (vi) chemical enrichment of stars and gas; (vii) the growth via gas accretion and merging of supermassive black holes; (viii) heating by AGN; (ix) photoionization of the intergalactic medium; (x) galaxy mergers
driven by dynamical friction within common DM halos which can
trigger starbursts and the formation and/or growth of spheroids; (xi) collapse of globally unstable disks that also lead to starbursts and the formation and/or growth of bulges. \shark\ adopts a universal \citet{Chabrier03} IMF. \citet{Lagos18b} include several different models for gas cooling, AGN,
stellar and photo-ionisation feedback, and star formation. Here, we adopt the default \shark\ model (see models and parameters adopted in \citealt{Lagos18b}; their Table~$2$). 

An important assumption in \shark\ and any SAM is that galaxies can be described as a disk plus bulge at any time. The main distinction between these two components is their origin, while disks form stars from gas that is accreted onto the galaxy from the halo, bulges are built by stars that are accreted from satellite galaxies and starbursts that are driven by galaxy mergers or disk instabilities. 
Both disks and bulges in \shark\ form stars based on the surface density of molecular hydrogen, with the only difference being that in the latter the efficiency of conversion into stars is $10$ higher than for star formation in disks. In our default \shark\ model, we use the pressure relation of \citet{Blitz06} to estimate the radial breakdown between atomic and molecular gas. The higher H$_2$-stars conversion efficiency in starbursts is found to be key to reproduce the cosmic star formation rate density (CSFRD) at $z\gtrsim 1.5$ in \shark\ \citep{Lagos18b}. As mentioned above, bulges can grow via disk instabilities, which happen when self-gravity dominates over centrifugal forces. This is evaluated by a global Toomre's instability parameter \citep{Ostriker73, Efstathiou82},

\begin{equation}
\epsilon=\frac{V_{\rm circ}}{\sqrt{1.68\,G\, M_{\rm disk}/r_{\rm disk}}},
\label{DisKins}
\end{equation}

\noindent where $V_{\rm circ}$ is the maximum circular velocity,
$r_{\rm disk}$ is the half-baryon mass disk radius and
$M_{\rm disk}$ is the total baryon disk mass. Here baryon corresponds to gas plus stars. The numerical factor $1.68$ converts the disk half-baryon mass radius into a scalelength, assuming an exponential profile.
If $\epsilon<\epsilon_{\rm disk}$ the disk is considered to be unstable. In the default \shark\ model used here, $\epsilon_{\rm disk} = 0.8$. Simple theoretical arguments suggest $\epsilon_{\rm disk}\approx 1$ \citep{Efstathiou82}. However, because the process of bar creation and thickening of the disk
can be a very complex phenomenon \citep{Bournaud11} that can easily lead to the gas and stars not having the same $\epsilon$ parameter \citep{Romeo11,Romeo18}, in \shark\ we treat $\epsilon_{\rm disk}$ as a free parameter. Note that many other SAMs do not include the effect of disk instabilities (e.g. \citealt{Henriques15,Xie17}), though \citet{Fanidakis10b} and \citet{Griffin18}, using the GALFORM SAM \citep{Cole00,Lacey15}, argue that disk instabilities are a key physical processes required to obtain a realistic population of QSOs throughout cosmic time.

{In \shark, we numerically solve the differential equations (DEs) of mass, metals and angular momentum exchange between the different baryon reservoirs (see Eqs. $49$-$64$ in \citealt{Lagos18b}), only setting an accuracy to which these equations are solved. The baryon reservoirs in the model are: gas outside halos, hot and cold gas inside halos but outside galaxies, ionised/atomic/molecular gas and stars in disks and bulges in galaxies, and super-massive black holes. This approach makes our model less sensitive to the time-stepping of the $N$-body simulation compared to other models, and also means that the star-formation histories (SFH) of galaxies can have as complex shape as required to solve the DEs.}

The model parameters of our default \shark\ model were tuned to the $z=0,\,1,\,2$ stellar mass functions (SMFs), the $z=0$ the black hole-bulge mass relation and the mass-size relations. The model also reproduces very well observational results that are independent from those used for the tuning, such as the total neutral, atomic and molecular hydrogen-stellar mass scaling relations at z=0, the cosmic star formation rate (SFR) density evolution at $z\approx 0-4$, the cosmic density evolution of the atomic and molecular hydrogen at $z\lesssim 2$ or higher in the case of the latter, the mass-metallicity relations for the gas and stars, the contribution to the stellar mass by bulges and the SFR-stellar mass relation in the local Universe (see \citealt{Lagos18b} for more details). In addition, \citet{Davies19} show that \shark\ also reproduces the scatter around the main sequence of star formation in the SFR-stellar mass plane, \citet{Chauhan19} show that \shark\ reproduces very well the HI mass and velocity width of galaxies observed in the ALFALFA survey and \citet{Amarantidis19} show that the AGN LFs agree well with observations in the X-rays and radio wavelengths. These represent true successes of the model as none of these observations were used in the processes of tuning the free parameters.

With the aim of building the SEDs of galaxies, \shark\ produces an output file {\tt star$_{-}$formation$_{-}$histories}, which contain the amount of stars that formed and the metallicity with which they formed throughout all the epochs sampled by the snapshots of the simulation until the point in which the output is being written. This is done separately for stars that end up in the disk and the bulge by the time of the output. Bulges are separated into stars built up by galaxy mergers and by disk instabilities. If a galaxy has a bulge that was built up by these two processes, then both arrays will have non-zero inputs. This information is then used by \viperfish\ (described in $\S$~\ref{sec:viperfish}) to create the SEDs and consequently calculate the galaxies' emission in a large range of bands going from the far-UV (FUV) to the far-IR (FIR). {Because we solve the DEs numerically, the arrays in {\tt star$_{-}$formation$_{-}$histories} show the average SFR and metallicity from which stars formed in the $200$ snapshots of the $N$-body simulation (see details below).}

\subsection{The \surfs\ simulations}

The results presented in \citet{Lagos18b} were produced using the \surfs\ suite of N-body, DM-only simulations \citep{Elahi18}, most of which have cubic volumes of $210\,\rm cMpc/h$ on a side, and span a range in particle number,
currently up to $8.5$ billion particles using a $\Lambda$CDM \citet{Planck15} cosmology. These correspond to a total matter, baryon and $\Lambda$ densities of $\Omega_{\rm m}=0.3121$, $\Omega_{\rm b}=0.0491$ and $\Omega_{\ L}=0.6751$, respectively, with a Hubble parameter of $h=100\,\rm Mpc\, km\,s^{-1}$ with $h=0.6751$, scalar spectral index of $n_{\rm s}=0.9653$ and a power spectrum normalization of $\sigma_{\rm 8}=0.8150$. 
All simulations were run with a memory lean version of the {\sc gadget2} code on the Magnus supercomputer at the Pawsey Supercomputing Centre. In this paper, we use the L210N1536 simulation, which has a cosmological volume of $(210\,\rm cMpc/h)^3$, $1536^3$ DM particles with a mass of $2.21\times10^8\,\rm h^{-1}\,M_{\odot}$ and a softening length of 
$4.5\,\rm h^{-1}\,ckpc$. Here, cMpc and ckpc denote comoving Mpc and kpc, respectively. 
\surfs\ produces $200$ snapshots for each simulation, typically having a time span between snapshots in the range of $\approx 6-80$~Myr. 

Merger trees and halo catalogs, which are the basis for \shark\ (and generally any SAM), were constructed using the phase-space finder {\sc VELOCIraptor}\footnote{{\url{https://github.com/icrar/VELOCIraptor-STF/}}} \citep{Elahi19a,Canas18} and the halo merger tree code
{\sc TreeFrog}\footnote{\href{https://github.com/pelahi/TreeFrog}{\url{https://github.com/pelahi/TreeFrog}}}, developed to work on {\sc VELOCIraptor} \citep{Elahi19b}. \citet{Poulton18} show that {\sc TreeFrog}+{\sc VELOCIraptor} lead to very well behaved merger trees, with orbits that
are well reconstructed. \citet{Elahi18b} also show that these orbits reproduce the velocity dispersion vs. halo mass inferred in observations. 
\citet{Canas18} show that the same code can be applied to hydrodynamical simulations to identify galaxies and that the performance of {\sc VELOCIraptor} is superior to space-finders, even in complex merger cases. We refer to \citet{Lagos18b} for more details on how the merger trees and halo catalogs are constructed for \shark, and to \citet{Elahi19a,Elahi19b,Canas18,Poulton18} for more details on the {\sc VELOCIraptor} and {\sc TreeFrog} software. 

\begin{figure}
\begin{center}
\includegraphics[trim=8mm 8mm 11mm 15mm, clip,width=0.5\textwidth]{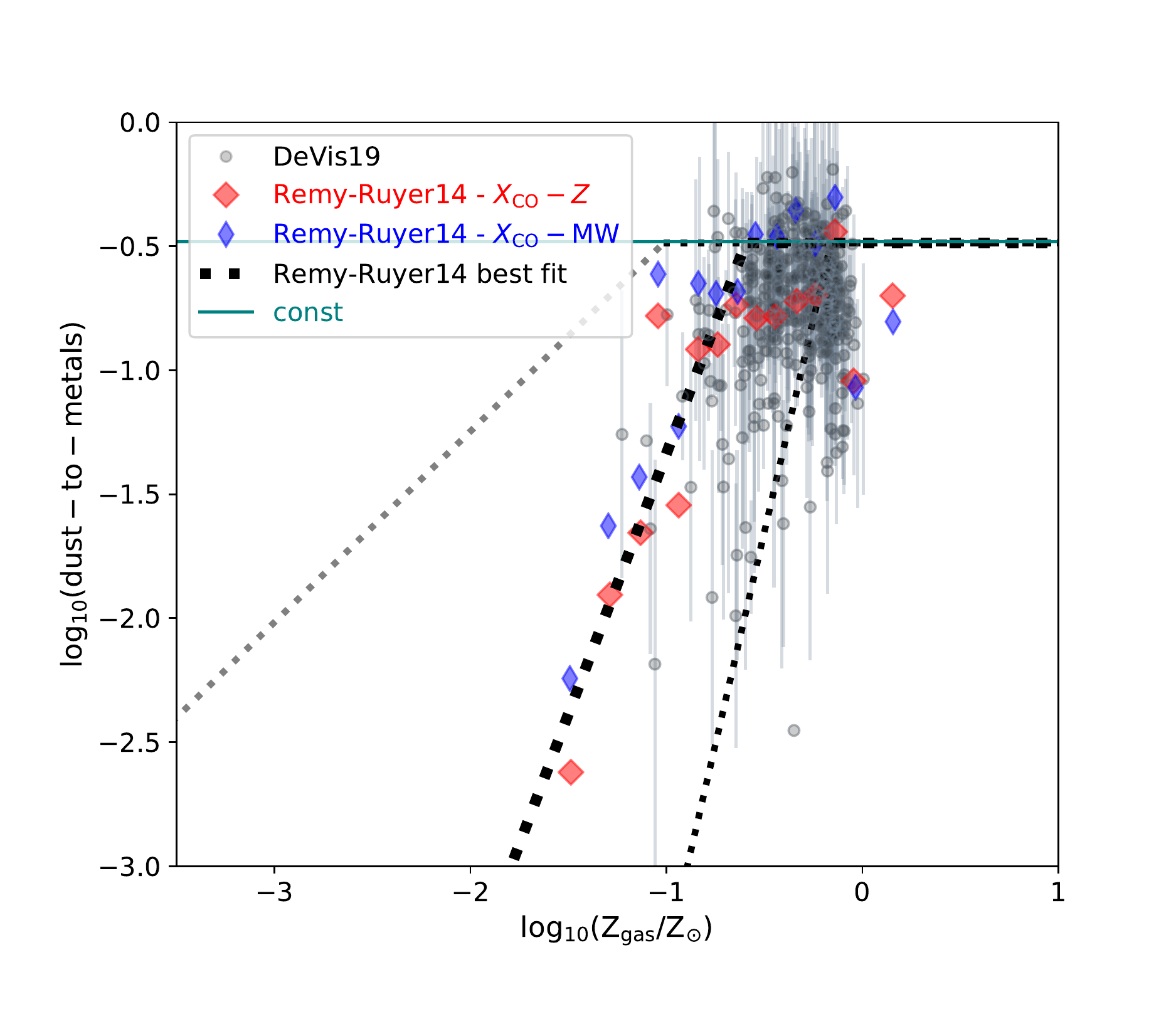}
\caption{Fraction of metals in dust as a function of gas metallicity. Local Universe observations of \citet{Remy-Ruyer14} are shown as diamonds, while their best fit relation is shown as thick dashed line. We also show the observations of \citet{DeVis19} as circles from the DustPedia of a large sample of local galaxies. The thin dotted lines show the $1\sigma$ uncertainty in the slope of the relation at low metallicities. The horizontal line shows the case of a constant dust-to-metal mass ratio at the Milky-Way value. For the observations of \citet{Remy-Ruyer14} we show two variants, one adopting a carbon monoxide-molecular hydrogen conversion adopting a Milky-Way conversion factor, and another one assuming a metallicity-dependent conversion factor. }
\label{DustDep}
\end{center}
\end{figure}

\begin{figure}
\begin{center}
\includegraphics[trim=9mm 0mm 15mm 11mm, clip,width=0.5\textwidth]{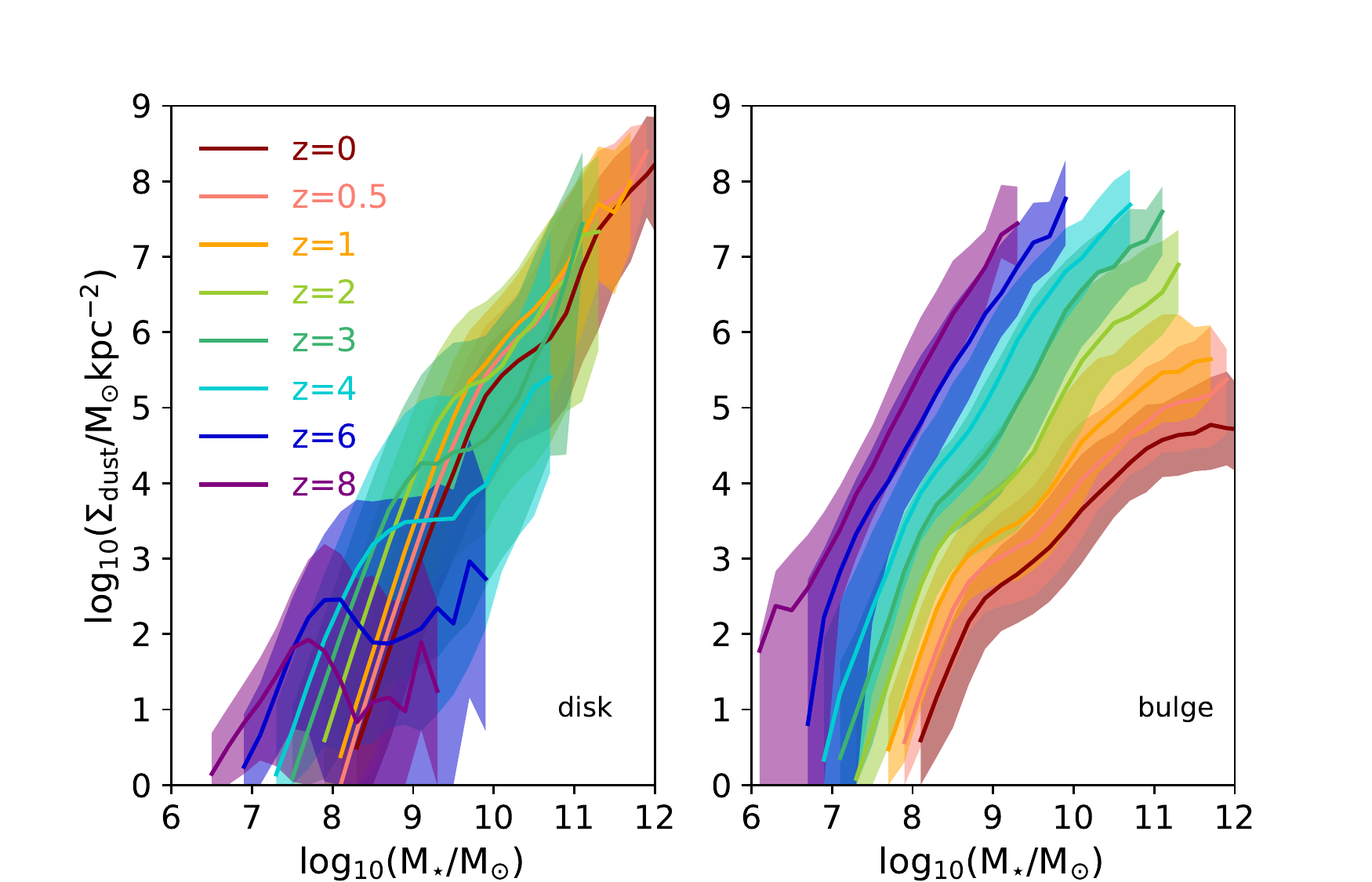}
\caption{Dust surface density (Eqs.~\ref{eqsigmadustdisk}~and~\ref{eqsigmadustbulge}) as a function of stellar mass from $z=8$ to $z=0$ for disks and bulges in \shark\, combined with the model RR14-steep to derive dust masses from the gas metallicity and surface density information. Lines show the medians, while shaded regions show the $16^{\rm th}-84^{\rm th}$ percentile ranges.}
\label{sigmadust}
\end{center}
\end{figure}

\subsection{Calculation of dust masses}

In this paper we consider three models to compute the dust mass from the mass in metals and the gas metallicity:

\begin{itemize}
    \item A constant dust-to-metals mass ratio, set to the Milky-Way value $M_{\rm dust}=0.33\,M_{\rm Z}$ \citep{Remy-Ruyer14} (referred to as $f_{\rm dust}$-const).
    \item The best fit $M_{\rm dust}/M_{\rm Z}-Z_{\rm gas}$ relation of \citet{Remy-Ruyer14} (see thick dotted line in Fig~\ref{DustDep}; see Table~$1$ in \citealt{Remy-Ruyer14}; referred to as RR14).
    \item The case in which a steeper relation is assumed with a break at higher gas metallicities, following the thin, black dotted line of Fig~\ref{DustDep}. This is motivated by the slope of the best fit $M_{\rm dust}/M_{\rm Z}-Z_{\rm gas}$ relation in \citet{Remy-Ruyer14} being quite significant and the recent observations of \citet{DeVis19} seemingly favouring a break in the dust-to-metal ratio at higher gas metallicities (referred to as RR14-steep).
\end{itemize}

The three different options above are shown in Fig~\ref{DustDep}, and are expected to make a difference only in galaxies with $Z_{\rm gas}/Z_{\odot} < 0.25$. This means that in the local Universe, only dwarf galaxies are expected to deviate from the constant dust-to-metal mass ratio significantly, and high redshift galaxies, as most of them have lower metallicities, deviating from $M_{\rm dust}=0.33\,M_{\rm Z}$.

Below, we describe how we compute $\Sigma_{\rm dust}$ for disks and bulges in \shark. 
\begin{itemize}
    \item {\bf Disks}. We compute an average $\Sigma_{\rm dust}$ for the disk as:
    \begin{equation}
        \Sigma_{\rm dust,disk} = \frac{0.5\,M_{\rm dust,disk}}{\pi\, r_{\rm 50,d}\,l_{\rm 50}},
        \label{eqsigmadustdisk}
    \end{equation}
    \noindent where $M_{\rm dust, disk}$ is the dust mass in the disk, $r_{\rm 50,d}$ is the half-gas mass radius of the disk and in a projected image represents the major axis, and $l_{\rm 50}$ is the projected minor axis, which is calculated as $l_{\rm 50}={\rm sin}(i)*(r_{\rm 50,d} - r_{\rm 50,d}/7.3)  +  r_{\rm 50,d}/7.3$, where $i$ is the inclination. The latter is $=r_{\rm 50,d}$ if the galaxy is perfectly face-on, and $=r_{\rm 50,d}/7.3$ if the galaxy is perfectly edge-on. The value $7.3$ comes from the scaleheight to scalelength observed relation in local galaxy disks \citet{Kregel02}. The inclination of a galaxy comes from the host subhalo angular momentum vector, or in the case of orphan galaxies, it is randomly chosen (see \citealt{Chauhan19} for details).
    \item {\bf Bulges}. We assume bulges to be spherically symmetric and hence the inclination is unimportant. We then compute the bulge dust surface density:
        \begin{equation}
        \Sigma_{\rm dust,bulge} = \frac{0.5\,M_{\rm dust,bulge}}{\pi\, r^2_{\rm 50,b}},
        \label{eqsigmadustbulge}
    \end{equation}
    \noindent where $M_{\rm dust,bulge}$ is the dust mass in the bulge, and $r_{\rm 50,b}$ is the half-gas mass radius of the bulge.
\end{itemize}

Fig.~\ref{sigmadust} shows the resulting dust surface density evolution for the disks and bulges, computed as in Eqs.~\ref{eqsigmadustdisk}~and~\ref{eqsigmadustbulge}, respectively, of \shark\ galaxies at $z=0$ to $z=8$, for the RR14-steep scaling.
Bulges display a monotonic evolution, with $\Sigma_{\rm dust}$ increasing with increasing redshift at fixed mass over the whole redshift range analysed here. This is due to a combination of the gas surface density evolution, in which high-z galaxies have higher $\Sigma_{\rm gas}$, and the fact that for bulges there is little evolution of the stellar mass-gas-metallicity relation. 

Galaxy disks on the other hand, display a more complex behavior. At $M_{\rm star}\gtrsim 10^{9.5}\,\rm M_{\odot}$, galaxies show a $\Sigma_{\rm dust}$ that increases from $z=0$ to $z\approx 2$, followed by a decrease towards higher redshift, at fixed stellar mass. At $10^8\,\rm M_{\odot} \lesssim M_{\star}\lesssim 10^{9.5}\,\rm M_{\odot}$, this reversal happens at higher redshift, $z\approx 4$. At lower stellar masses we see that the reversal moves to even higher redshift. However, those masses are below what we would consider as ``resolved'' in our simulation box. \citet{Lagos18b} showed that the box used here is reliable down to $M_{\star}\approx 10^{8}\,\rm M_{\odot}$, but below that the number density of galaxies artificially drops, deviating from the values obtained from a higher resolution box of the same cosmology and initial conditions. 
The evolution of $\Sigma_{\rm dust}$ for disks is driven by the competing effects of the gas metallicity and $\Sigma_{\rm gas}$ evolution. At fixed stellar mass, \shark\ galaxies exhibit a strong $Z_{\rm gas}$ evolution, with galaxies at $z=3$ being $0.6$~dex metal poorer than galaxies at $z=0$ at fixed stellar mass. However, in the same redshift range, $z=3$ galaxies have a $\Sigma_{\rm gas}$ that is $\approx 1.2$~dex larger than the $z=0$ counterparts of the same stellar mass. As a result, the evolution seen in $\Sigma_{\rm dust}$ is more modest than that obtained for $\Sigma_{\rm gas}$ and the reversal displayed is due to the metallicity evolution overcoming the increase in $\Sigma_{\rm gas}$.

\section{Lighting \shark\ galaxies through \viperfish}\label{sec:viperfish}

To generate SEDs for \shark, two packages have been developed: \prospect\footnote{\href{https://github.com/asgr/ProSpect}{\url{https://github.com/asgr/ProSpect}} and for an interactive \prospect\ web tool see {\url{http://prospect.icrar.org/}}, which is recommended as an education tool.} and \viperfish\footnote{\href{https://github.com/asgr/Viperfish}{\url{https://github.com/asgr/Viperfish}}}. \prospect\ (Robotham et al. in prep) is a low-level package that combines the popular GALEXev stellar synthesis libraries \citet{Bruzual03} (BC03 from hereafter) and/or EMILES \citep{Vazdekis16} with a multi-component dust attenuation model \citep{Charlot00} and dust re-emission \citep{Dale14}. On top of this sits \viperfish, which allows for simple extraction of \shark\ SFHs, metallicity histories (ZFH), and generation of the desired SED through target filters.

\prospect\ is designed in a pragmatic manner that allows for user-side flexibility in controlling the key components that affect the galaxy SED produced. 
Many of the design decisions were influenced by successful spectral fitting codes (e.g. MAGPHYS, \citealt{daCunha08}, and Cigale; \citealt{Noll09}) with the emphasis here on a code that works in a fully generative mode with the types of outputs available from SAMs. Other differences lie in the specific choice of dust modelling (in particular the re-emission templates) and the manner in which SFHs and ZFHs are incorporated (highly flexibly).

For the production of galaxy SEDs, the decision was made early on to focus efforts on the BC03 stellar population (SP) libraries using a \citet{Chabrier03} IMF since these are well understood in the community, have a broad spectral range that makes them useful for current and next generation multi-band surveys and are the default in \shark. \prospect\ can accept almost any functional form for the SFH or ZFH, which includes non-parametric, parametric or discontinuous specifications (the latter being most like the type produced in a modern SAM). The functional SFH or ZFH can in practice be arbitrarily complex, with internal interpolation schemes used to map the provided form onto the discrete library of temporal evolution available. For the ZFH, the metallicities are interpolated in log-space, producing a few tenths of a mag uncertainty at worst within the range available ($0.0001 \le Z \le 0.05$). {If the time-steps in which the SFH and ZFH are stored are too coarse, this interpolation may lead to large uncertainties in the predicted emission, particularly in the UV. Fortunately, the time-steps of our \surfs\ simulations are sufficiently fine so that the UV emission is accurately predicted. In the worst case scenario of an extreme recent starburst, the UV would still be converged to better than $30$\%, but in more common cases we expect an accuracy of $5$\% or better.}

The generative nature of \prospect\ means it can be used in a number of ways: either to fit real data using Bayesian modelling via optimisation of Markov-Chain Monte-Carlo (MCMC; see Bellstedt et al. in prep. and Davies et al. in prep.), or in a purely generative mode given a SFH and ZFH evolution of, e.g., a simulated galaxy. For producing lightcones with SEDs from SAMs, this generative mode is obviously of most interest. However, some sensible assumptions must be made regarding light attenuation due to dust, and its re-emission at longer wavelengths. How to do this in a fully physical sense, given the limited range of knowledge we have about any single SAM galaxy, is a matter of ongoing research, but for the current purposes of \shark\ SED generation we settle on a deliberately simplified fiducial model of dust processing.

Firstly, the dust is attenuated by the dust model of \citet{Charlot00}, in which the dust is assumed to be in a two-phase medium (birth clouds, BC, and diffuse ISM) in both the disk and the bulge (in which starbursts take place). Two different optical depths at $5500$\AA\ are assumed for these phases, $\hat{\tau}_{\rm BC}$ and $\hat{\tau}_{\rm ISM}$, respectively. The  absorption curves for the BCs and diffuse ISM are then defined as:

\begin{eqnarray}
\tau_{\rm ISM} &=& \hat{\tau}_{\rm ISM}\, (\lambda/5500\text{\AA})^{\eta_{\rm ISM}}
\label{taus_screen},\\
\tau_{\rm BC} &=& \tau_{\rm ISM} + \hat{\tau}_{\rm BC}\, (\lambda/5500\text{\AA})^{\eta_{\rm BC}}.\label{tau_bc}
\end{eqnarray}

\noindent The values we adopt as \citet{Charlot00} default are $\hat{\tau}_{\rm BC}=1$, $\hat{\tau}_{\rm ISM}=0.3$, $\eta_{\rm BS}={\eta_{\rm ISM}}=-0.7$ (suggested to be within a ``reasonable'' range in that paper). Stellar populations younger than $10$~Myr are in birth clouds, and hence their light is affected by both the optical depth of Eq.~\ref{tau_bc}, while older stars which are in the diffuse medium are attenuated by Eq.~\ref{taus_screen}.

With this model, light generated at different ages is attenuated differently, giving a natural means to simulate the effect of BC attenuation for younger stars. This absorbed light must then be re-emitted in a sensible fashion at longer wavelengths. For this process we adopt the \citet{Dale14} FIR dust templates, with an assumption of no significant AGN emission, and an assumed dust radiation field of $\alpha_{\rm SF}=3$ for the diffuse ISM and $\alpha_{\rm SF}=1$ for the birth clouds. Since this re-emission process only makes use of the absorbed luminosity in the UV-NIR, the scaling is chosen to ensures energy balance. The $\alpha_{\rm SF}$ exponents represent the local interstellar radiation field the dust is exposed to, $0.3<U<10^5$, with $U=1$ being the local interstellar radiation field of the solar neighborhood. 
A power-law combination of local curves mimics the global dust emission, with a fraction ${\rm d}M_{\rm dust}$ of dust mass being heated by $U^{-\alpha_{\rm SF}}{\rm d}U$.
 The values adopted here for the screen and birth cloud components roughly correspond to effective dust temperatures of $20-25$~K and $50-60$~K, respectively. Note that emission from AGN can be included when using \prospect\ to fit the SEDs of observed galaxies; however, we do not use it in \shark\ as it requires significant additional modelling to scale the AGN SED templates with meaningful AGN properties. We leave this for future work.

Once the full generative spectrum has been created (by adding the attenuated stellar light and the dust emission together), we redshift to the observed frame using the full spectral resolution available. Finally, we pass the spectrum through a chosen number of available filters that span the FUV to FIR, giving our final reduced outputs. Storing the spectral information of all galaxies is impractical, so care must be taken that all filters of interest are specified at this stage. Only a subset of these filters are discussed in this work, and user defined filters can be added easily if required. We warn the reader that in this work we do not include nebular emission lines, which can make an impact on narrow bands. Hence, in this work we focus solely on broad band emission.


The highest level code \viperfish\ allows for a very simple interface between the HDF5 outputs created by \shark\ and \prospect. It effectively reduces a few hundred lines of {\sc R} code to a single call with the path to the relevant HDF5 file. This makes it trivial to post-process any \shark\ outputs at any time (it does not need to be run in parallel), and it is designed to scale naturally with the computing resources available, e.g. it can use multiple cores. 

\subsection{Optical depth and reddening calculation of \shark\ galaxies}

\begin{figure*}
\begin{center}
\includegraphics[trim=30mm 0mm 30mm 3mm, clip,width=0.99\textwidth]{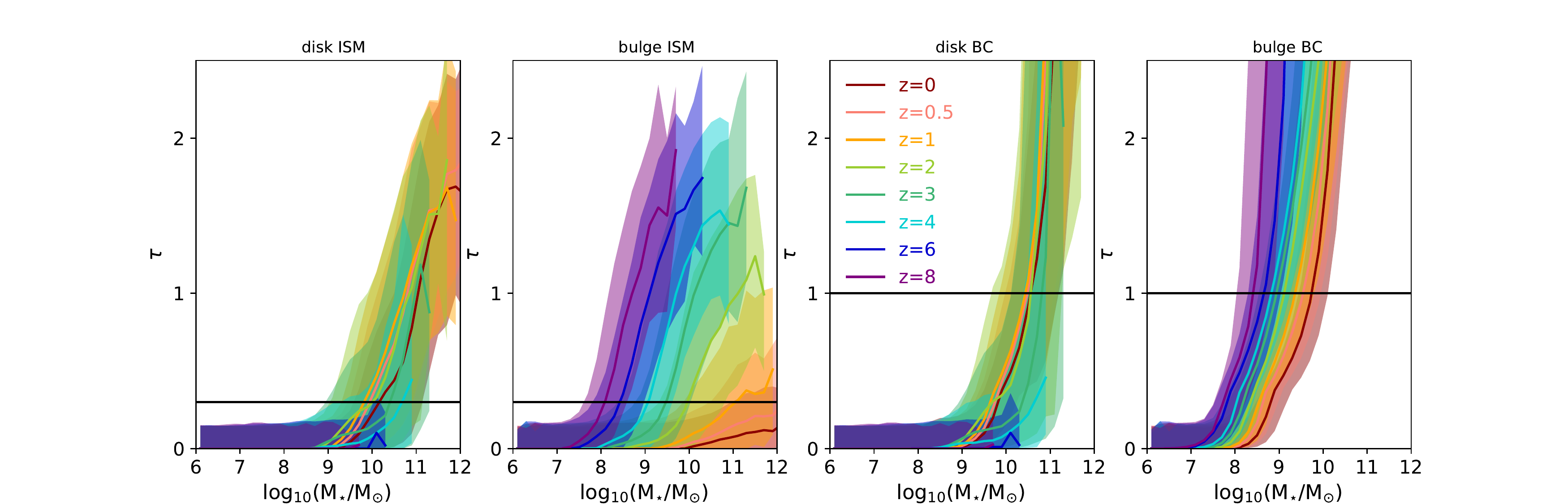}
\caption{Optical depth of dust in the diffuse ISM and birth clouds of the disks and bulges of galaxies, as labelled at the top of each panel, as a function of stellar mass from $z=0$ to $z=8$, as labelled, for the {\sc EAGLE}-$\tau$~RR14-steep attenuation model. Lines show the medians, while shaded regions show the $16^{\rm th}-84^{\rm th}$ percentile ranges. Horizontal lines show the default values adopted for the \citet{Charlot00} model.}
\label{FigTaus}
\end{center}
\end{figure*}

\begin{figure}
\begin{center}
\includegraphics[trim=11mm 3mm 15mm 11mm, clip,width=0.5\textwidth]{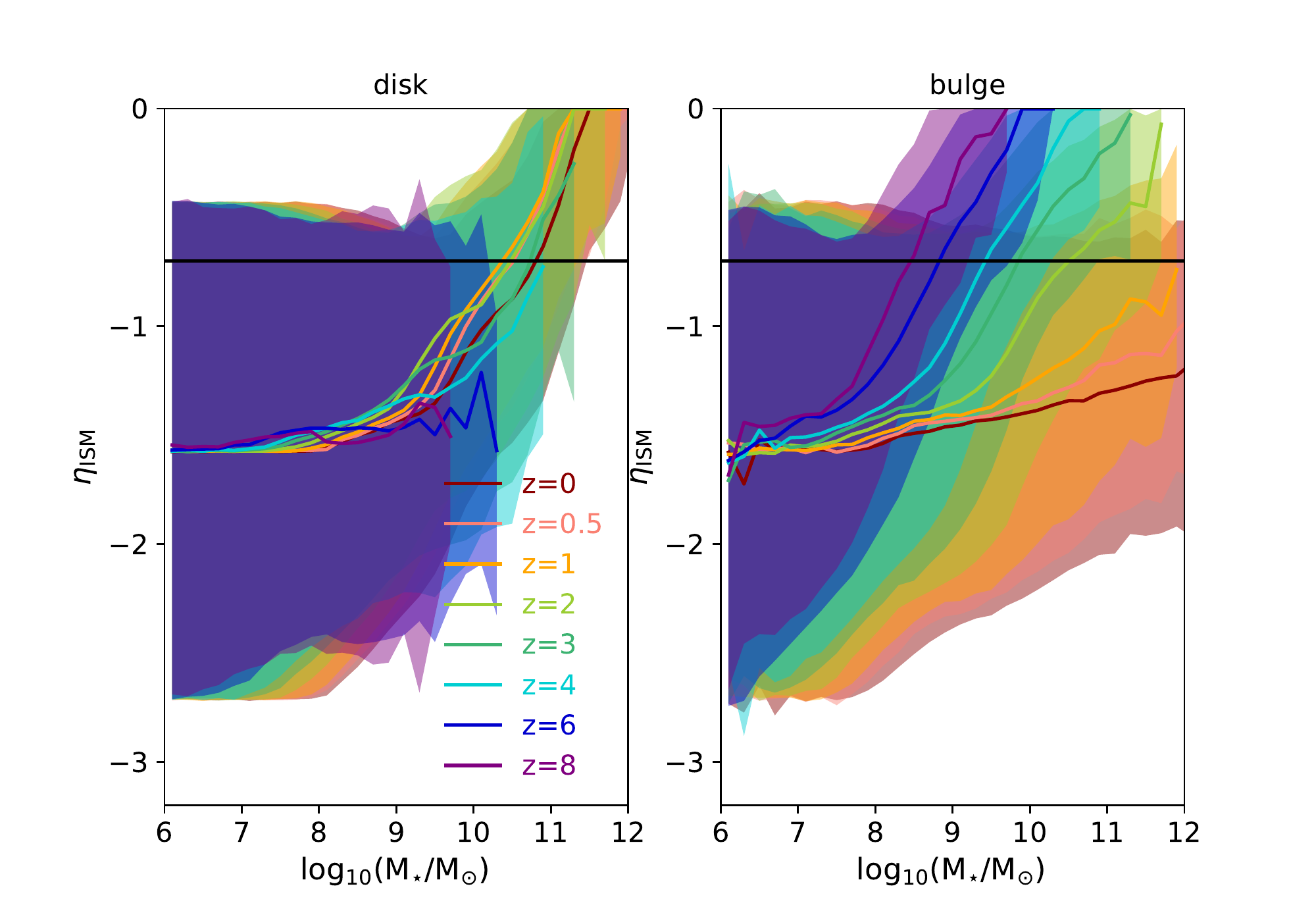}
\caption{Power-law index of the optical-depth dependence on wavelength in Eq.~\ref{taus_screen}, for the disks (left) and bulges (right) of \shark\ galaxies as a function of stellar mass from $z=0$ to $z=8$, as labelled, for the {\sc EAGLE}-$\tau$~RR14-steep attenuation model. Lines show the medians, while shaded regions show the $16^{\rm th}-84^{\rm th}$ percentile ranges. Horizontal lines show the default $\eta=-0.7$ in \citet{Charlot00}.}
\label{Slopes}
\end{center}
\end{figure}

\subsubsection{Attenuation due to the diffuse ISM}

\citet{Trayford17} used the RT code {\sc SKIRT} to compute the attenuation curve for each galaxy in the {\sc EAGLE} hydrodynamical simulation suite. From these curves, \citet{Trayford19} found that they can be parametrized using the \cite{Charlot00} model, with values for $\tau_{\rm ISM}$ and $\eta_{\rm ISM}$ varying with the dust column density in the line of sight (hence, considering the effects of inclination). \citet{Trayford19} in fact find that such parametrization is independent of redshift. Hence the redshift evolution obtained for the average optical depth and power-law index of Eq.~\ref{taus_screen} of galaxies is due to their dust surface density evolving. 

Trayford et al. computed the median and $1\sigma$ scatter relationship between $\tau_{\rm ISM}$, $\eta_{\rm ISM}$ and $\Sigma_{\rm dust}$, from which we sample. In \shark, we use each galaxy's dust surface density, $\Sigma_{\rm dust}$, to compute $\tau_{\rm ISM}$ and $\eta_{\rm ISM}$, and perturb the values by sampling from a gaussian distribution with width $\sigma$, where $\sigma$ is the $16^{\rm th}-84^{\rm th}$ percentile ranges predicted by \citet{Trayford19}. We compute $\Sigma_{\rm dust}$ for disks and bulges following Eqs.~\ref{eqsigmadustdisk}~and~\ref{eqsigmadustbulge}.

\subsubsection{Attenuation due to birth clouds} 

For the birth clouds we follow \citet{Lacey15}, who assume the birth cloud optical depth to scale with the gas metallicity and gas surface density of the cloud, but modify it to use the dust surface density of clouds rather than the metal surface density,

\begin{equation}
    \tau_{\rm BC} = \tau_{\rm BC,0}\,\left[\frac{f_{\rm dust}\, Z_{\rm gas}\, \Sigma_{\rm gas,cl}}{f_{\rm dust,MW}\, Z_{\odot}\, \Sigma_{\rm MW,cl}}\right],
\end{equation}

\noindent $f_{\rm dust}=M_{\rm dust}/M_{\rm Z}$ is the dust-to-metal mass ratio, $\tau_{\rm BC,0}=1$, $\Sigma_{\rm MW,cl}=85\,\rm M_{\odot}\,pc^{-2}$, $Z_{\odot}=0.0189$, and $f_{\rm dust,MW}=0.33$, so that in typical spiral galaxies $\tau_{\rm BC}\approx \tau_{\rm BC,0}$ as determined by \citet{Charlot00} and \citet{Kreckel13}. We compute the cloud surface density as 
$\Sigma_{\rm gas,cl}={\rm max}[\Sigma_{\rm MW,cl},\Sigma_{\rm gas}]$, with $\Sigma_{\rm gas}$ being the diffuse medium gas surface density, which is calculated as Eqs.~\ref{eqsigmadustdisk}~and~\ref{eqsigmadustbulge}, but using the gas masses of the disk and bulge, respectively. The reasoning behind this is that in the local group, galaxies ranging from metal-poor dwarfs to molecule-rich spirals seem to have giant molecular clouds (GMCs) with a constant gas surface density close to the value $\Sigma_{\rm MW,cl}$, which is surprisingly independent of galactic environment (see e.g. \citealt{Blitz07,Bolatto08}, and \citealt{Krumholz14} for a review). However, as the ambient ISM pressure increases, the GMC surface density must increase in order to maintain pressure balance with the surrounding ISM. Hence, it follows that $\Sigma_{\rm gas,cl}\approx \Sigma_{\rm gas}$ in those extreme environments \citep{Krumholz09}, which are expected to be more common at high redshift. We also impose the physical limit of $\tau_{\rm BC} \ge \tau_{\rm ISM}$. 

 For birth clouds we do not have a well informed choice for $\eta_{\rm BC}$, as we do for the diffuse ISM, and hence we adopt the default \citet{Charlot00} $\eta=-0.7$. Some models in the literature assume a more negative value of $-1.3$ (e.g. \citealt{daCunha08,Wild11}) due to the expected shell-like geometry of BCs. We find, however, that the use of a steeper $\eta_{\rm BC}$ does not affect our results in any significant manner.


\subsubsection{Summary of Attenuation models}

\begin{table}
        \setlength\tabcolsep{2pt}
        \centering\footnotesize
        \caption{Attenuation models tested. ``CF00'' refers to \citet{Charlot00}, and ``RR14'' refers to \citet{Remy-Ruyer14}. The dependence of the CF parameters on $\Sigma_{\rm dust}$ is taken from the parametrisation of {\sc EAGLE} galaxies by \citet{Trayford19}. Our default option is the model {\sc EAGLE}-$\tau$ RR14.}
        \begin{tabular}{@{\extracolsep{\fill}}l|ll|p{0.45\textwidth}}
                \hline
                \hline
            Name & Description\\
        \hline
    CF00 & Adopts default \citet{Charlot00} \\
    & parameters. \\
    {\sc EAGLE}-$\tau$ $f_{\rm dust}$ const & Adopts CF parameters depending on $\Sigma_{\rm dust}$, \\
    &using a constant $f_{\rm dust}$\\
    {\sc EAGLE}-$\tau$ RR14& Adopts CF parameters depending on $\Sigma_{\rm dust}$, \\
    (default) &using the RR14 best-fit $f_{\rm dust}-Z_{\rm gas}$relation.\\
    {\sc EAGLE}-$\tau$ RR14-steep& Adopts CF parameters depending on $\Sigma_{\rm dust}$,\\
    & using the RR14 $f_{\rm dust}-Z_{\rm gas}$ relation with \\
    &a steeper slope\\
    \hline
        \end{tabular}
        \label{tab:mods}
\end{table}

Table~\ref{tab:mods} shows all the attenuation models used here: (1) the simplest assumption, which corresponds to fixed \citet{Charlot00} parameters (which are therefore constant and do not depend on galaxy properties or inclination; referred to as CF00); (2) the {\sc EAGLE} attenuation parametrization of the \citet{Charlot00} parameters, assuming a constant fraction of the metals are locked in dust (referred to as
{\sc EAGLE}-$\tau$ $f_{\rm dust}$ const); (3) as model (2) but assuming the empirical relation of \citet{Remy-Ruyer14} between $M_{\rm dust}-M_{\rm Z}-Z_{\rm gas}$  (see thick dashed line in Fig.~\ref{DustDep}; referred to as {\sc EAGLE}-$\tau$ RR14); (4) as model (3) but using a steeper dependence of $M_{\rm dust}/M_{\rm Z}$ on $Z_{\rm gas}$ within the errors of the best fit relation in \citet{Remy-Ruyer14} (see thin, black dotted line in Fig.~\ref{DustDep}; referred to as {\sc EAGLE}-$\tau$ RR14-steep). Model (3) is our default model throughout the paper but we make it explicit in every figure caption which the model is shown.

\subsubsection{Stellar mass dependence and redshift evolution of the optical depth of \shark\ galaxies}

Fig.~\ref{FigTaus} shows the effective $V$-band optical depth, $\tau$, as a function of stellar mass at several redshifts for the disks and bulges of \shark\ galaxies. Here we adopt the attenuation model {\sc EAGLE}-$\tau$ RR14 (see Table~\ref{tab:mods}).

For the diffuse ISM, we obtain a steep increase of $\tau$ of galaxy's disks with stellar mass at $M_{\star}>10^{10}\,\rm M_{\odot}$ at $z=0$, below which $\tau \rightarrow 0$. This stellar mass threshold moves to lower stellar masses as redshift increases, up to $z\approx 1$ for disks and at all redshifts for bulges. In the latter, $\tau\lesssim 0.5$ for all galaxies at $z=0$ due to the gas fractions of bulges being very small. This changes at $z\gtrsim 1.5$ due to bulges hosting large gas reservoirs and undergoing starbursts. Although galaxies at high redshift are more metal poor, their gas surface density is increasing rapidly, causing the redshift evolution seen in \shark\ galaxies. For the BCs, we obtain a relatively sharp transition from small to large extinctions at $M_{\star}\approx 10^{10}\,\rm M_{\odot}$ in disks, which is dictated by the gas metallicity, and at the high mass galaxies, by the average gas surface density. This transition moves to lower stellar mass for bulges (which tend to be more compact and more metal rich than disks), and to progressively lower masses as the redshift increases, mostly driven by the evolution of the bulge gas surface density.
Adopting instead the attenuation models {\sc EAGLE}-$\tau$ RR14 or {\sc EAGLE}-$\tau$~$f_{\rm dust}$~const results in a shift of the $y$-axis values in both Figs.~\ref{FigTaus}~and~\ref{Slopes} to higher $\tau$ values, overall producing more attenuation (not shown here). 

\subsection{Example SEDs and SFHs}

\label{viperfishsec}
\begin{figure}
\begin{center}
\includegraphics[trim=0mm 10mm 5mm 23mm, clip,width=0.49\textwidth]{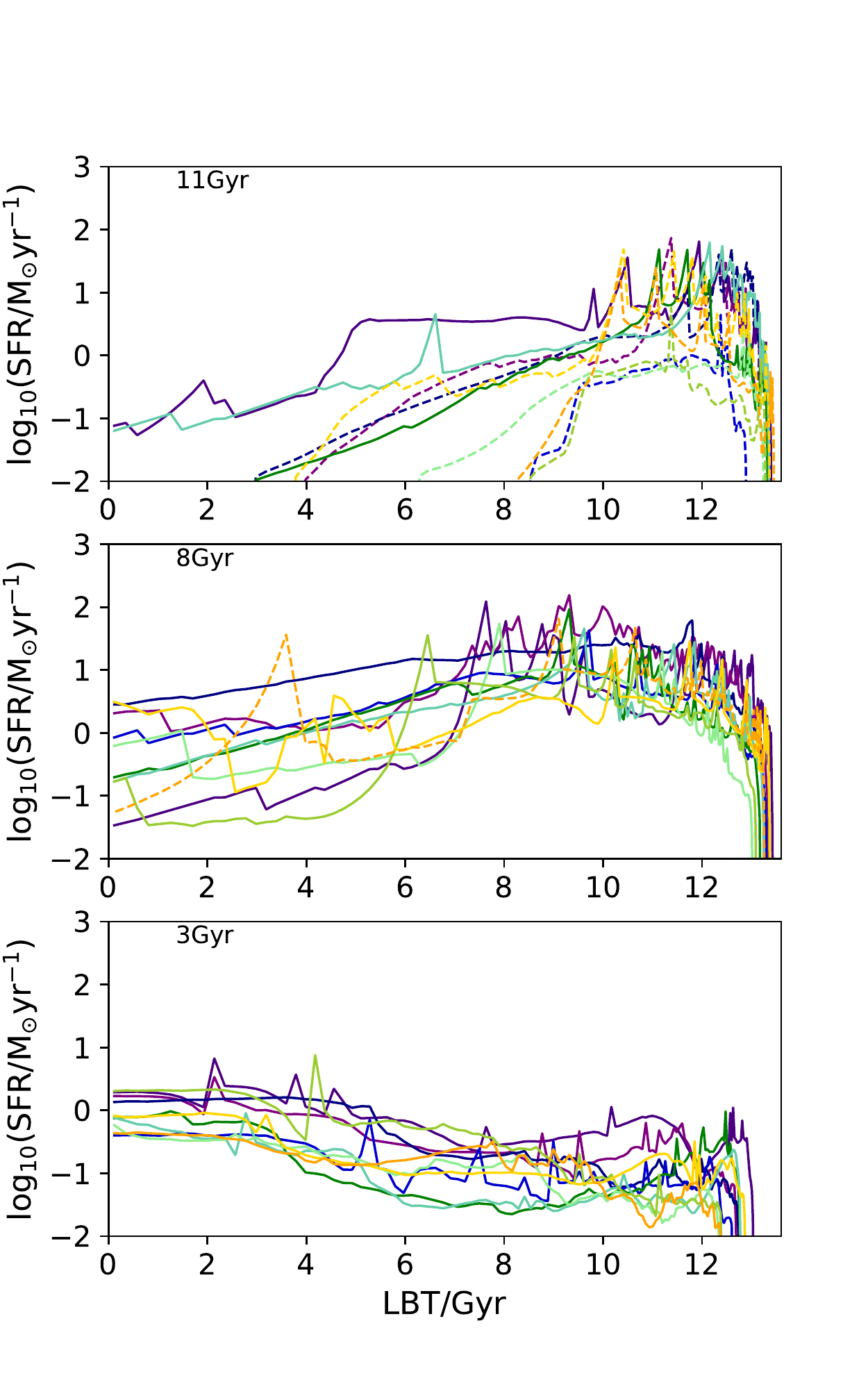}
\caption{Examples of the star formation rate as a function of lookback time (LBT) of \shark\ galaxies that by $z=0$ have stellar masses $>10^9\,\rm M_{\odot}$ and mean stellar-mass weighted ages $\pm 0.3$~Gyr from the value indicated in each panel. We show for each selection $10$ random examples, and show with solid lines those galaxies that by $z=0$ are centrals, and with dashed lines those that by $z=0$ are satellites.}
\label{SFHExamples}
\end{center}
\end{figure}

\begin{figure}
\begin{center}
\includegraphics[trim=0mm 10mm 7mm 25mm, clip,width=0.49\textwidth]{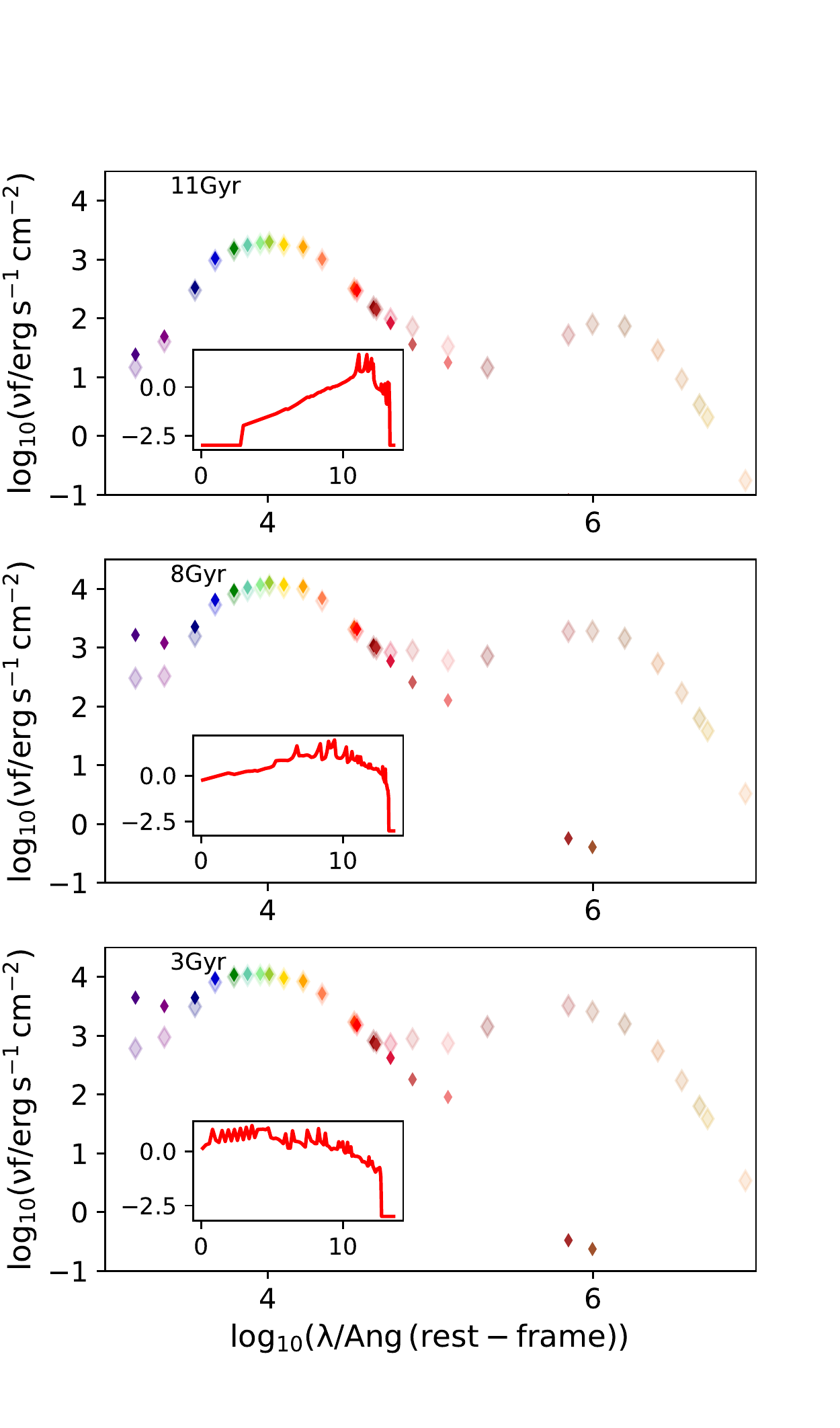}
\caption{Broadband photometry in 27 bands (in order of wavelength: GALEX FUV and NUV, SDSS ugriz, VISTA YJHK, WISE 1, IRAC 3.6$\mu$m, IRAC 4.5$\mu$m, WISE 2, IRAC 5.8$\mu$m, IRAC 8$\mu$m, WISE 3 and 4 and 
Herschel PACS 70$\mu$m, 100$\mu$m, 160$\mu$m, Herschel SPIRE 250$\mu$m, 350$\mu$m JCMT $450\mu$m, SPIRE 500$\mu$m and JCMT $850\mu$m; symbols) 
for $3$ \shark\ galaxies with stellar mass $>10^9\,\rm M_{\odot}$, randomly selected in bins 
of $\pm 0.3$~Gyr around the stellar-mass weighted age indicated at the top-left of each panel. Opaque and transparent diamonds show the intrinsic emission and the emission after we include the effects of dust, respectively. The insets show the SFR history (in units of $\rm log_{10}(SFR/M_{\odot}\, yr^{-1}$; a floor of $-3$ is applied for presentation purposes) of each of these galaxies as a function of lookback time (in Gyr).}
\label{SEDExamples}
\end{center}
\end{figure}

\begin{figure}
\begin{center}
\includegraphics[trim=2mm 10.5mm 6mm 13mm, clip,width=0.5\textwidth]{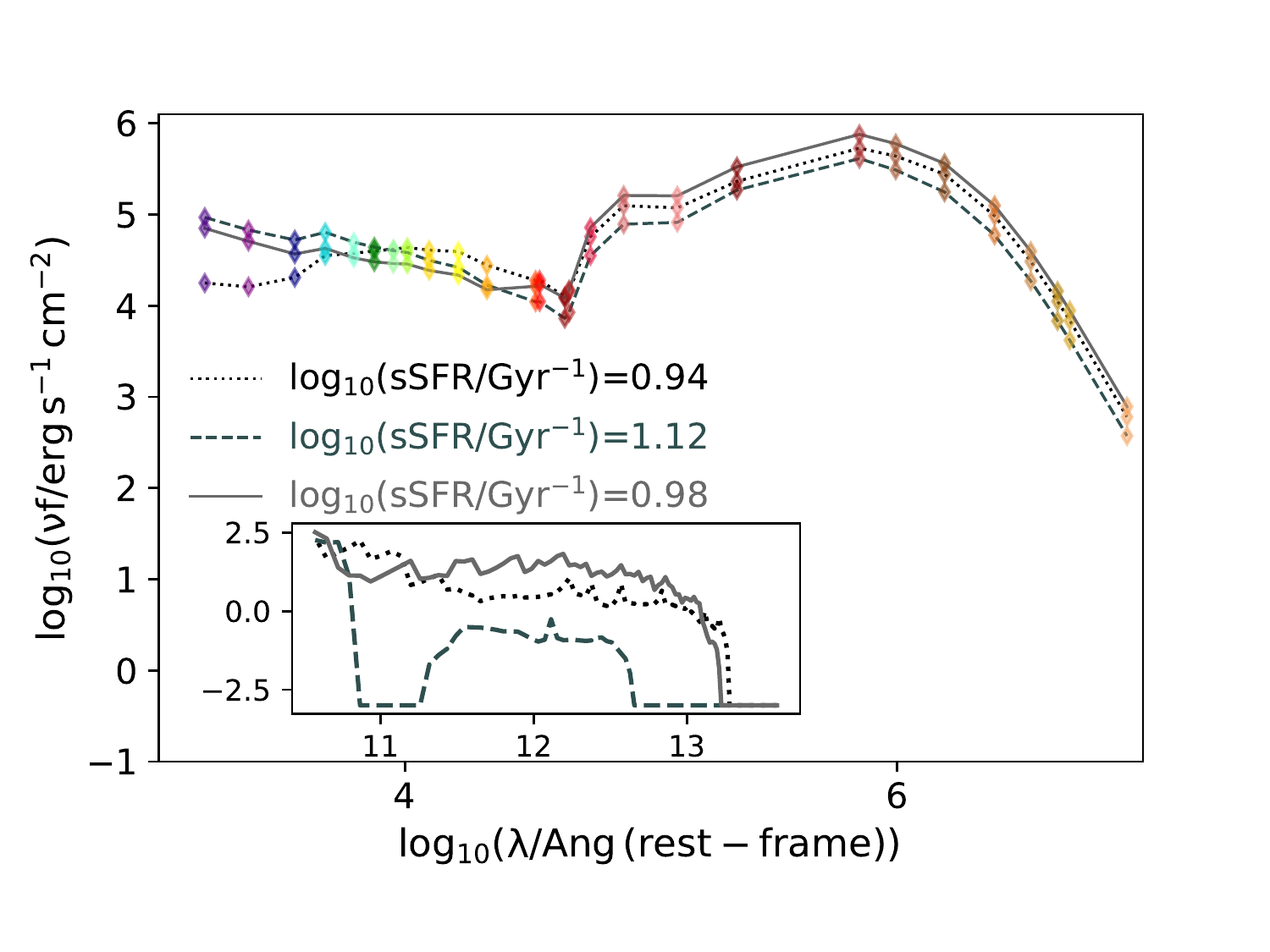}
\caption{Rest-frame broadband photometry (after including the effects of dust extinction and re-radiation) in 27 bands (as in Fig.~\ref{SEDExamples}) 
for $3$ \shark\ highly starburst galaxies with stellar masses $\approx 2-5\times 10^{10}\,\rm M_{\odot}$ and SFRs $250-500\,\rm M_{\odot}\, yr^{-1}$ at $z=2$ (sSFRs as labelled). Diamonds show the photometry. The insets show the SFR history, as in Fig.~\ref{SEDExamples}. These \shark\ galaxies correspond to SMGs, with their $850\mu$m emission being $7.5$~mJy (dotted line), $4.6$~mJy (dashed line) and $9.8$~mJy (solid line).}
\label{SEDExamplesz2}
\end{center}
\end{figure}

Fig.~\ref{SFHExamples} shows examples of SFHs of $10$ randomly selected \shark\ galaxies at $z=0$ that have stellar masses $>10^9\,\rm M_{\odot}$ and stellar-mass weighted ages at $\pm 0.3$~Gyr around the values labelled in each panel, which span from $11$~Gyr to $3$~Gyr. The SFHs of \shark\ galaxies look anything but the idealized exponentially decay or composite instantaneous-burst plus exponential decay, which are typically assumed in observations when performing SED fitting \citep{Mitchell13,daCunha08}. \citet{Pacifici12} used SFHs and ZFHs from SAMs as inputs for the SED fitting of observed galaxies. This makes an important difference in the recovered stellar mass and SFR of up to a factor of $0.6$~dex (see e.g. \citealt{Pacifici15}). This shows that using complex SFHs is important in the recovery of galaxy parameters. 

Many \shark\ galaxies experience early starbursts seen as short-lived peaks in the SFH (quite common at look-back times $\gtrsim 10$~Gyr). The latter are more common in galaxies that have older stellar populations by $z=0$ than younger ones. At look-back times $\lesssim 6$~Gyr, starbursts are much less common, mostly seen in galaxies that by $z=0$ are very young. Also note that old galaxies tend to show sharp cut-offs in their SFH associated to stripping of their hot gas as they become satellite galaxies. On the contrary, galaxies that are on average young by $z=0$, tend to have very extended SFHs, that in some cases continue to rise to $z=0$. Most central galaxies that by $z=0$ are old tend to have SFHs that drop towards $z=0$, but less sharply than for satellites (see for example solid lines vs. dashed lines in  Fig.~\ref{SFHExamples}).

\begin{figure}
\begin{center}
\includegraphics[trim=10mm 21mm 20mm 3mm, clip,width=0.5\textwidth]{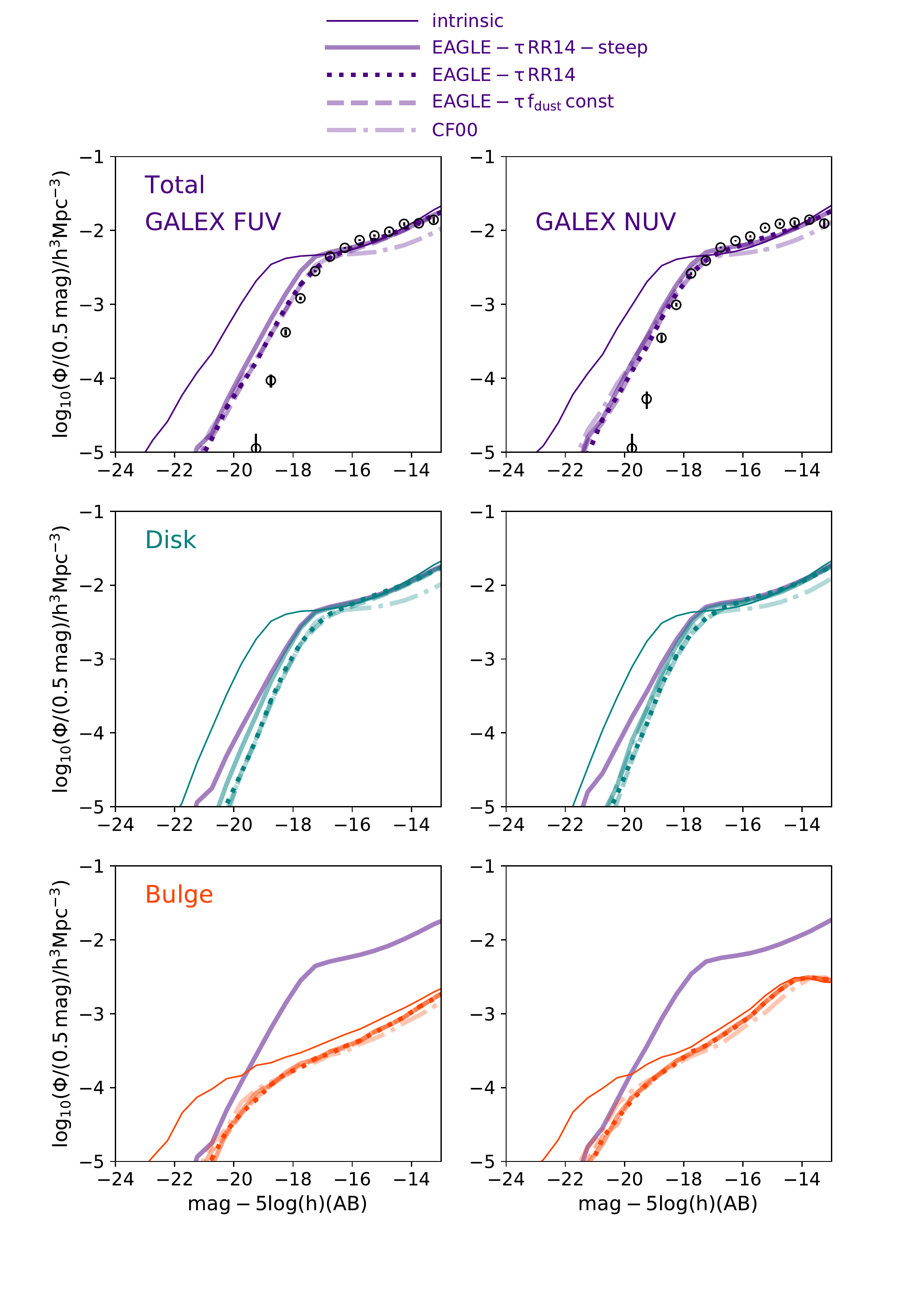}
\caption{Luminosity functions at $z=0$ for the GALEX FUV and NUV bands. Here, we include all galaxies in the \shark\ model of \citet{Lagos18b}, and show the intrinsic emitted light in thin, solid lines, and the four attenuation models of Table~\ref{tab:mods}, as labelled. The top panels show the total emission from galaxies, while the middle and bottom panels show the contribution from disks and bulges, respectively. In the middle and bottom panels we also show for guidance the UV LF of the {\sc EAGLE}-$\tau$~RR14-steep.
The symbols on the top panels show the observational measurements of \citet{Driver12}. Both \shark\ and observational luminosity functions are presented in bins of $(0.5)$~mag, and thus we do not normalize the $y$-axis by the adopted bin.}
\label{UVLFsz0}
\end{center}
\end{figure}

\begin{figure}
\begin{center}
\includegraphics[trim=3mm 21mm 10mm 28mm, clip,width=0.42\textwidth]{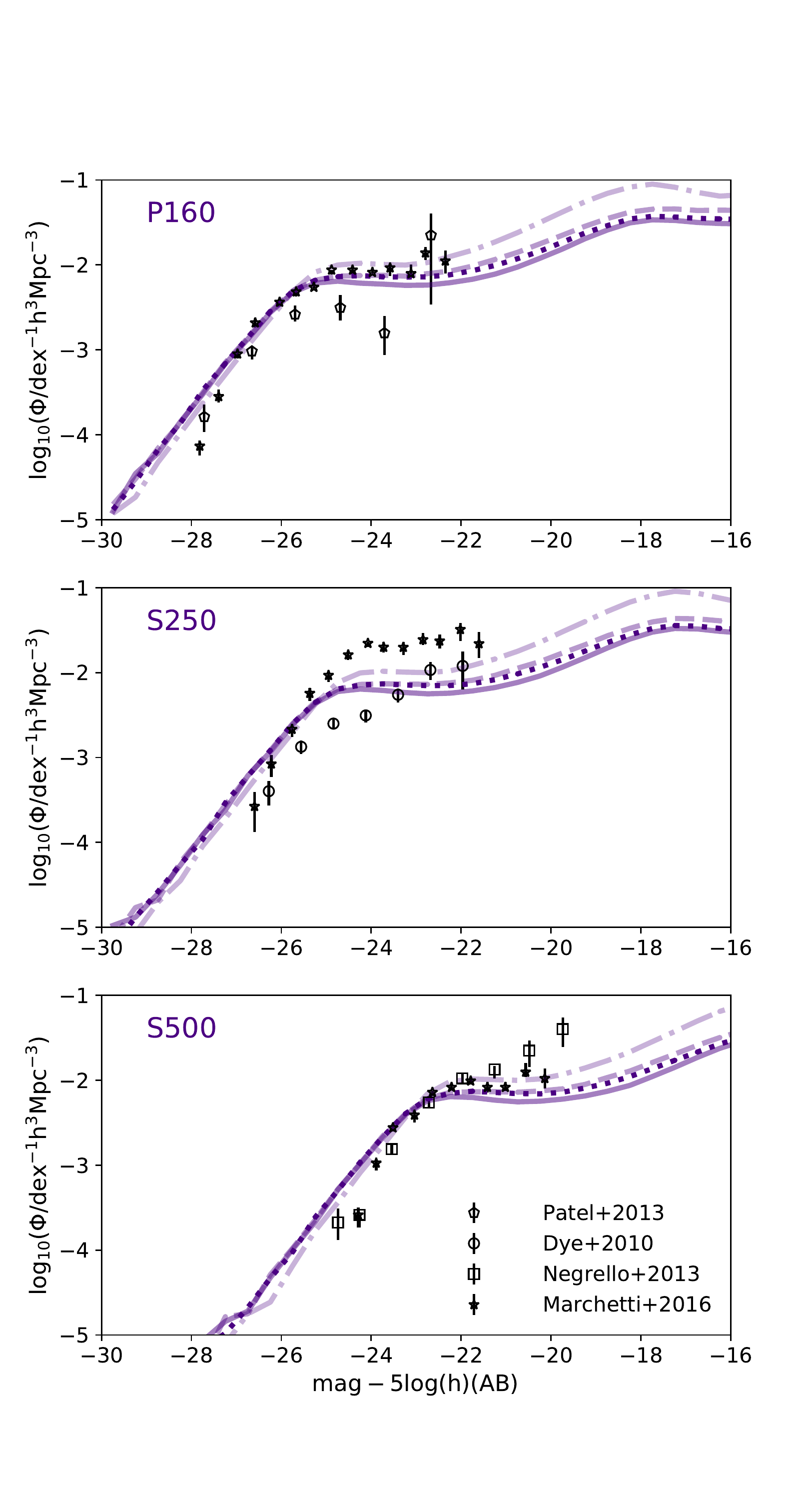}
\caption{Luminosity functions at $z=0$ for the
Herschel PACS band 160$\mu$m, and SPIRE bands 250$\mu$m and $500\mu$m. Here we show the total LF for all the galaxies in the \shark\ model of \citet{Lagos18b} using the four attenuation models of Table~\ref{tab:mods}, as labelled. 
The symbols show the observational measurements from \citet{Dye10,Patel13,Negrello13,Marchetti16}, as labelled. Unlike Fig.~\ref{UVLFsz0}, here we show the $y$ axis normalized by the bin size.}
\label{IRLFsz0}
\end{center}
\end{figure}

Fig.~\ref{SEDExamples} shows the broadband SED in $27$ bands for $3$ randomly selected \shark\ galaxies of different stellar ages. The SFHs of these galaxies are shown in the insets in each panel. Both the intrinsic emission and after dust attenuation and re-radiation are shown. As expected, young galaxies tend to have much more significant emission in the UV, which suffers from large extinction. Galaxies with ages $\gtrsim 11$~Gyr have very little intrinsic emission in the UV and little gas content, both of which result in a small extinction. 
We show in Fig.~\ref{SEDExamplesz2} the SEDs of three starburst galaxies at $z=2$ in the same $27$ bands of Fig.~\ref{SEDExamples}. These galaxies have widely different star formation histories, with one of them having significant star formation over the last $300\,\rm Myr$ but little before that. These galaxies differ significantly from the $z=0$ examples in that most of their emission happens at the FIR, and represent nice examples of sub-millimeter galaxies (SMGs) in \shark.

\section{Galaxy emission and the effects of dust extinction on the galaxy LF}\label{sec:LFs}



In this section we analyse the \shark\ predictions for the FUV-to-FIR emission of galaxies at $z=0$ and how this is affected by our new attenuation models. We specially focus on the properties of the different structural components of galaxies and the connection to their stellar populations and epochs. 



\subsection{The z=0 UV and FIR luminosity functions}\label{sec:uvz0}

Fig.~\ref{UVLFsz0} shows the $z=0$ GALEX FUV and NUV luminosity functions (LFs) predicted by \shark\ before and after dust attenuation is applied. We show four attenuation models corresponding to those in Table~\ref{tab:mods}. The top panels show the total LFs. We also show the observations of \citet{Driver12}.

Galaxies at $z=0$ emit several orders of magnitude more UV emission than is observed (thin, solid lines in Fig.~\ref{UVLFsz0}), meaning that extinction must play a very important role, particularly beyond the break of the LF, $L^*$. Adopting the CF00 extinction parameters leads to FUV and NUV LFs that are too shallow at the faint end ($>-17.5$ AB mags). The attenuation models based on the {\sc EAGLE} RT massively improve the predicted faint end of the LFs 
The attenuation models {\sc EAGLE}-$\tau$~$f_{\rm const}$ and {\sc EAGLE}-$\tau$~RR14 produce almost identical LFs,  due to most galaxies contributing to the UV LFs having $Z_{\rm gas}/Z_{\odot}>0.25$, which is the gas metallicity threshold above which galaxies converge to a constant dust-to-metal mass ratio (see Fig.~\ref{DustDep}). The extinction model {\sc EAGLE}-$\tau$~RR14-steep, on the other hand, predicts a slightly brighter break of the LF (by $\approx 0.2$~mag). This difference is due to galaxies in this variant deviating from the constant dust-to-metal ratio at $Z_{\rm gas}\approx 0.7 Z_{\odot}$ (see Fig.~\ref{DustDep}). Note that all the extinction models miss the sharp bright-end of the UV LFs, which indicate that \shark\ galaxies are slightly too star-forming and/or the attenuation for the brightest UV galaxies is too small. 
The obvious improvement obtained where going from the default CF00 to the {\sc EAGLE}-like extinction models justifies the need for the added complexity, and nicely confirms that our RT-motivated extinction models allow \shark\ to predict more realistic UV LFs. The latter becomes even clearer at higher redshifts ($\S$~\ref{sec:redshiftevoLF}).

The middle and bottom panels of Fig.~\ref{UVLFsz0}
 show the contribution from disks and bulges to the FUV and NUV LFs at $z=0$, respectively. Bulges are only important at the very bright end; these galaxies correspond to the few rare local starburst. Note that the attenuation models based on the {\sc EAGLE}-RT results produce virtually the same bulge UV LF, which is due to bulges having gas metallicities typically above $0.7\,Z_{\odot}$. This means that bulges have the same dust-to-metal ratio in the three {\sc EAGLE}-RT variants of Table~\ref{tab:mods}. This is not the case for disks, which is why the three {\sc EAGLE}-RT model variants produce different UV LFs. Because disks dominate over the whole magnitude range, we end up with visible differences in the total UV LFs.  
 
 The better match to the faint end of the UV LFs by the {\sc EAGLE}-$\tau$ attenuation models is the dependence of the gas surface density on stellar mass (which produces a differential optical depth): $z=0$ \shark\ galaxies of $M_{\star}\approx 10^8\,\rm M_{\odot}$ have $\Sigma_{\rm gas}\approx 10^{6.5}\,\rm M_{\odot}/kpc^2$, while $M_{\star}\approx 10^{10}\,\rm M_{\odot}$ galaxies have $\Sigma_{\rm gas}\approx 10^{7.3}\,\rm M_{\odot}/kpc^2$.
 
 The changes seen in the UV LF are expected to be also seen in FIR, as the light that is extincted by the dust is then re-radiated in the FIR. This is shown in Fig.~\ref{IRLFsz0} for the same $4$ attenuation models of Table~\ref{tab:mods}, but here we only show the total LF as we later analyse the contribution from disks and bulges. Significant differences are seen at the faint end of the FIR LFs of up to $\approx0.5$~dex in number density, but that regime unconstrained by observations. All models, however, predict a very similar bright-end, which agree very well with the observed LFs. We remind the reader that here we assume two effective dust temperatures for the diffuse ISM and BCs to re-emit the extincted light in the FIR when using the \citet{Dale14} templates. The values we adopt are typical of the local Universe and hence the agreement with the observations is not necessarily surprising.

\begin{figure*}
\begin{center}
\includegraphics[trim=15mm 14mm 30mm 25mm, clip,width=0.87\textwidth]{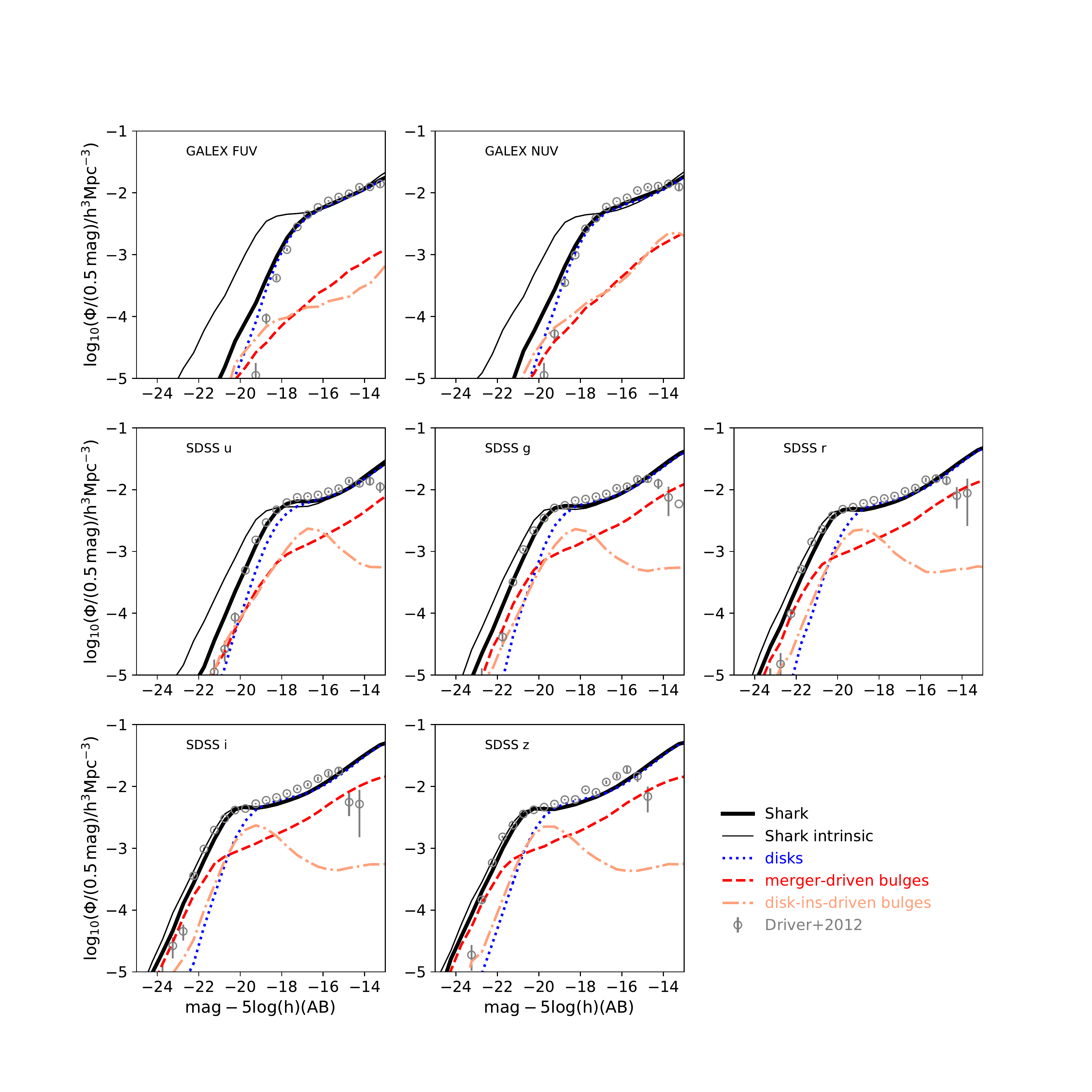}
\caption{Luminosity functions at $z=0$ for the GALEX FUV and NUV bands (top panels) and the SDSS u, r, g, i and z bands (middle and bottom panels), as labelled. Here, we include all galaxies in the \shark\ model and adopt the default extinction model {\sc EAGLE}$\tau$ RR14 (see Table~\ref{tab:mods}). We show as black thin and thick lines the emission before and after dust extinction. The dotted, dashed and dot-dashed lines show LFs of disks, bulges that formed predominantly via galaxy mergers and by disk instabilities, respectively. The symbols show the observational measurements of \citet{Driver12}. Both \shark\ and observational luminosity functions are presented in bins of $(0.5)$~mag, and thus we do not normalize the $y$-axis by the adopted bin.}
\label{opticalLFs}
\end{center}
\end{figure*}

\begin{figure}
\begin{center}
\includegraphics[trim=12mm 24mm 20mm 36mm, clip,width=0.51\textwidth]{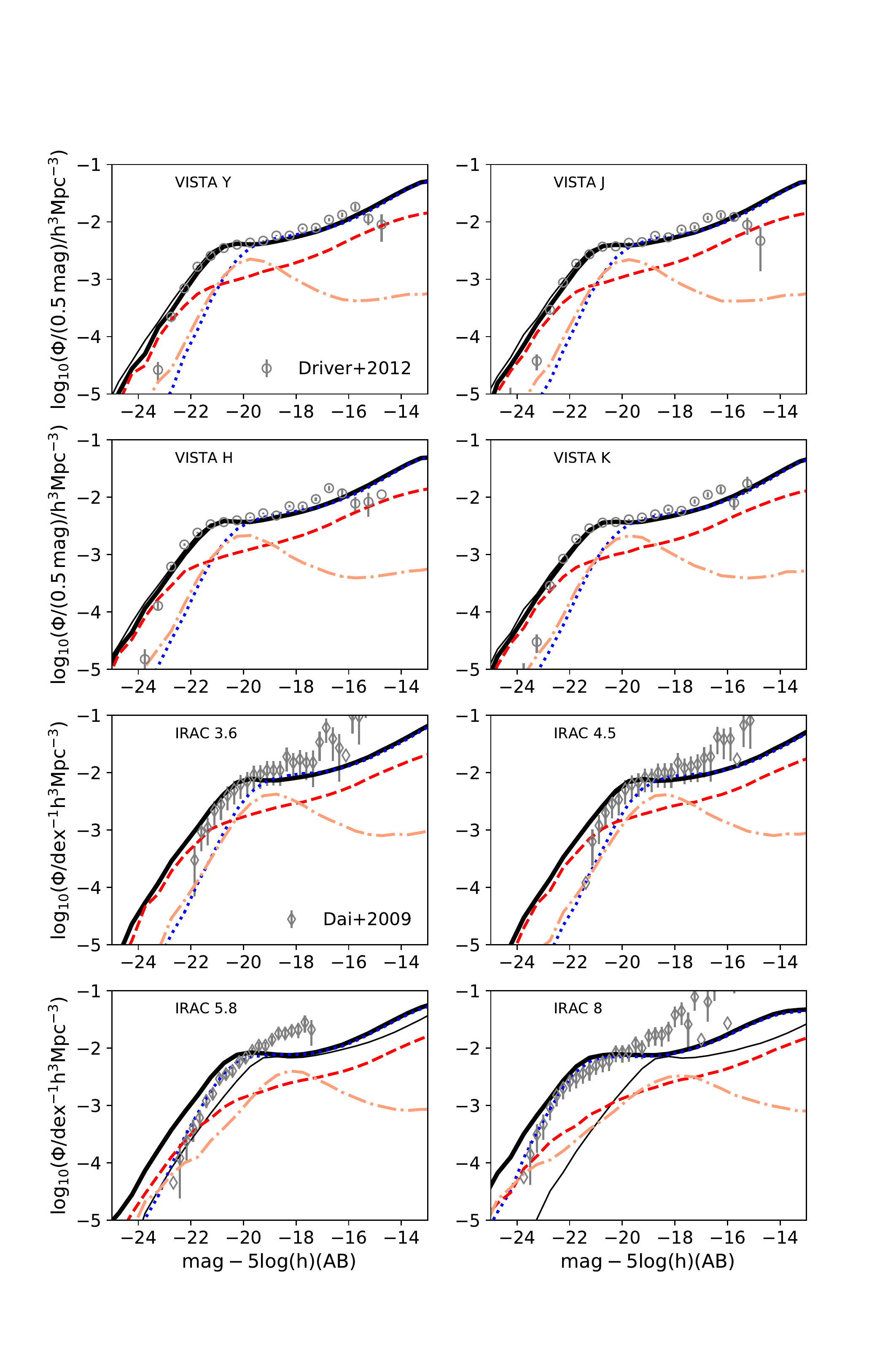}
\caption{Luminosity functions at $z=0$ for the UKIDDS Y, J, H,  K-bands, and IRAC 3.6$\mu$m, $4.5\mu$m, $5.8\mu$m and $8\mu$m, as labelled in each panel. As in Fig.~\ref{opticalLFs}, we adopt the default attenuation model {\sc EAGLE}$\tau$ RR14. Lines are in Fig.~\ref{opticalLFs}. Symbols show the observational measurements of \citet{Driver12} and \citet{Dai09}, as labelled.  Note that in the IRAC panels we show the number density normalized by the adopted $x$-axis bin.}
\label{NIRLFs}
\end{center}
\end{figure}

\begin{figure}
\begin{center}
\includegraphics[trim=12mm 22mm 20mm 35mm, clip,width=0.51\textwidth]{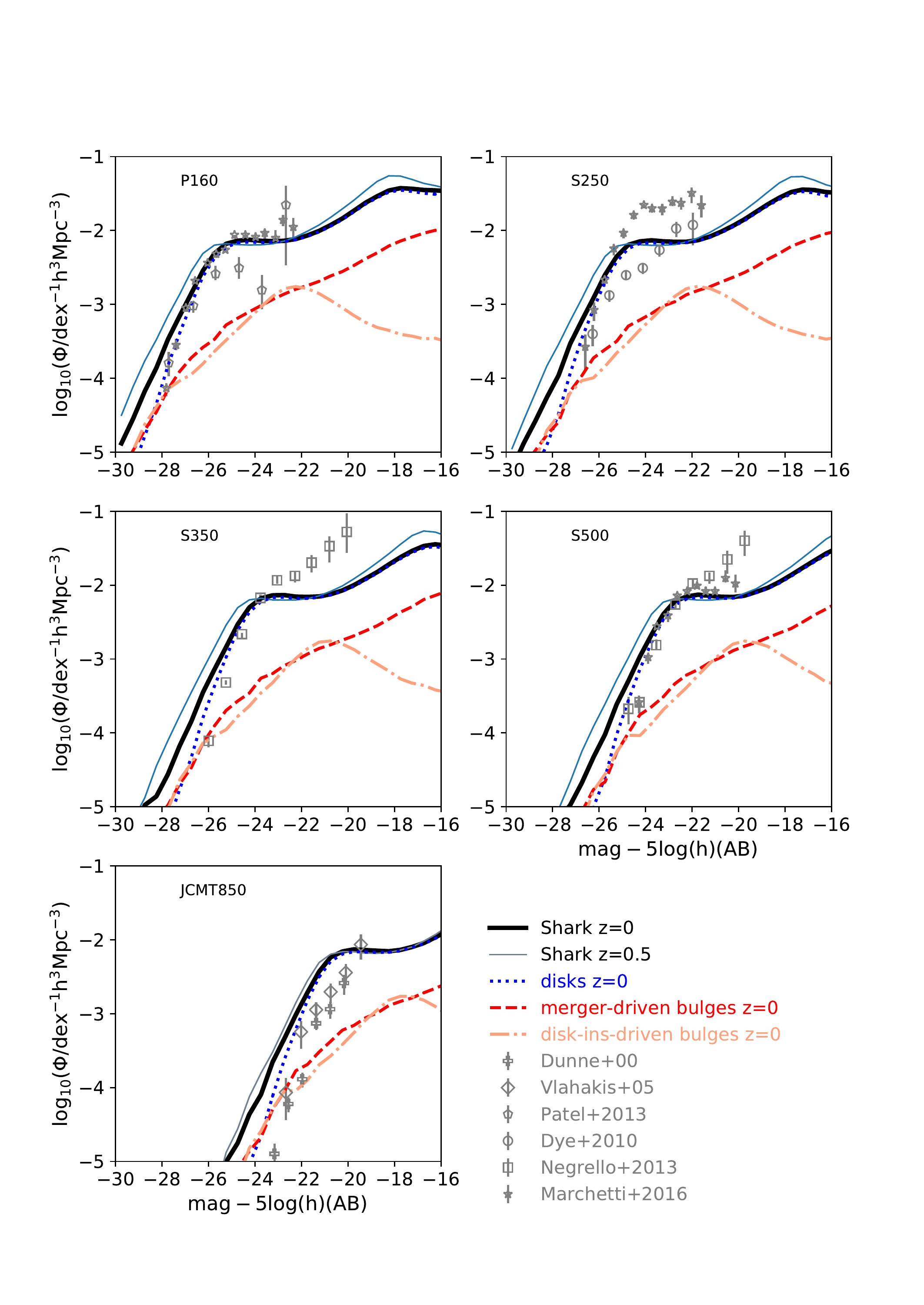}
\caption{Luminosity functions at $z=0$ for the  Herschel PACS band 160$\mu$m, SPIRE bands 250$\mu$m, 350$\mu$m, 500$\mu$m and the JCMT 850$\mu$m, as labelled in each panel. As in Fig.~\ref{opticalLFs}, we adopt the default attenuation model {\sc EAGLE}-$\tau$ RR14. We do not show intrinsic luminosities here, and instead the thin, solid line shows the $z=0.5$ \shark\ prediction, as a reference to the level of evolution expected on that redshift window, as some of the observational estimates are computed with all the galaxies at $z\le 0.5$. The symbols show the observational measurements of \citet{Dunne00}, \citet{Vlahakis05}, \citet{Dye10}, \citet{Patel13}, \citet{Negrello13} and \citet{Marchetti16}, as labelled. Unlike Fig.~\ref{opticalLFs}, here the $y$-axis is normalized by the adopted $x$-axis bin.}
\label{FIRLFs}
\end{center}
\end{figure}

\subsection{The $z=0$ FUV-to-FIR luminosity functions}

Fig.~\ref{opticalLFs} shows the UV and optical luminosity functions at $z=0$ compared to the measurements of \citet{Driver12} using the Galaxy and Mass Assembly(GAMA) survey. The thin lines show the intrinsic emission, while the thick lines show the emission after dust extinction and reprocessing. As we discussed in $\S$~\ref{sec:uvz0}, the effect of the latter is very important in the UV bands, shifting the luminosity function by up to $2$ magnitudes at the bright-end and in the FUV band. The effect becomes a lot weaker in the optical. For example, in the $r$-band the effect is only $\approx 0.3$~mag. 

The observations of \citet{Driver12} correspond to the observed luminosity functions and they should be compared to the thick lines. The agreement with the observations is remarkable across all the bands, considering that we do not use this information to tune the free parameters of the model. The latter is less obvious at the near-IR bands, as this luminosity correlates strongly with stellar mass, and as explained in $\S$~\ref{sharksec}, the $z=0$ SMF was used to tune the parameters. Thus, it is not necessarily surprising that the $z$-band LF agrees well with the observations. 

As discussed in $\S$~\ref{sec:uvz0}, \shark\ tends to produce slightly too many UV bright galaxies; $\approx 0.5-0.7$~dex more galaxies than \citet{Driver12} at a FUV $-19.3$ and NUV $-20$ magnitudes, due to the contribution of starbursts in \shark. This is seen in the dashed and dot-dashed lines in Fig.~\ref{opticalLFs}, which show the LFs of the bulges that formed predominantly by galaxy mergers and by disk instabilities in \shark, respectively. Both mechanisms of bulge formation contribute similarly to the number density of bright UV galaxies. Although this changes significantly as we move towards redder bands. In the $r$ to $z$ bands, bulges built by disk instabilities make a similar contribution as disks at the bright-end, which is much smaller that that of bulges built by galaxy mergers.

Stars in the disks of galaxies always dominate the faint-end of the luminosity functions, but their contribution beyond the break in the LF is a strong function of wavelength. The bluer the band, the higher the contribution from disks at the bright-end. In the extreme cases of the FUV and NUV luminosity functions, disks dominate the number density over all but the brightest luminosity bin, while in the $u$- and $g$-bands, they contribute about half of the luminosity above $L^*$. 
This contribution becomes negligible in the $z$-band, where the bright-end beyond $-21.5$~mag is primarily tracking the bulge content of galaxies. We later show  that this trend reverses for the mid- and FIR bands at low redshifts (fig.~\ref{FIRLFs}).

Fig.~\ref{NIRLFs} shows the $z=0$ luminosity functions for the $4$ UKIDSS bands, Y, J, H and K, an the IRAC $3.6\mu$m, $4.5\mu$m, $5.8\mu$m and $8\mu$m of \shark\ galaxies, compared to \citet{Driver12} and \citet{Dai09}. The agreement between the model and the observations in the UKIDSS and IRAC $3.6\mu$m, $4.5\mu$m bands is excellent, except in the brightest luminosity bin. Again, this is not surprising as \shark\ is tuned to fit the SMF at $z=0$. The overabundance of very bright galaxies is similar to the conclusion of \citet{Lagos18b} that the SMF has a high-mass end slope a bit too shallow compared to the observations, leading to slightly too many galaxies with stellar masses $\approx 10^{12}\,\rm M_{\odot}$, though still within the observational uncertainties. Note that here we see a continuation of the trend of the contribution from disks at the bright-end decreasing as the wavelength becomes longer. At the K-band, disks have a negligible contribution over the whole magnitude range above $L*$. The reasonable agreement at the IRAC $5.8\mu$m and $8\mu$m bands is more surprising and shows that our attenuation plus dust-remission models have a realistic effect on the UV light and re-emission at the mid IR. However, \shark\ does not reproduce perfectly the IRAC $5.8\mu$m and $8\mu$m LFs, with most tension seen at the faint end, and the bright end in the $5.8\mu$m band. These bands are particularly difficult as most of the emission comes from unidentified infrared emission (UIE), which is a ubiquitous component of the IR emission in galaxies and typically associated to polycyclic aromatic hydrocarbons \citep{Li12}. 

The LF of bulges built by disk-instabilities peaks below $L*$, but with the peak moving to brighter luminosities relative to $L^*$ as the wavelength shortens. This agrees with the overall picture of the stellar mass budget build up described in \citet{Lagos18b}, who showed that the stellar mass contribution from bulges built via disk instabilities peaks at stellar masses of $10^{10.3}-10^{10.8}\,\rm M_{\odot}$. Those galaxies contribute little to the UV luminosity functions, as $\approx 30-40$\% of them are passive (i.e. specific SFRs $>10$ times below the main sequence of star formation), while their contribution increases in the NIR bands as their stellar mass is large. The bottom panels of Fig.~\ref{NIRLFs} show the comparison with LFs measured in the IRAC bands at $z\approx 0$. The IRAC $3.6\mu$m and $4.5\mu$m, behave similarly to the UKIDDS bands, but the $5.8\mu$m band starts to show an increase in the contribution from disk emission, and the LF starts to be dominated by the re-emission of light by dust rather than the intrinsic stellar light. By the IRAC band $8\mu$m, disks are back to contributing most of the light, and to dominate even above $L^*$.

Fig~\ref{FIRLFs} shows the $z=0$ luminosity functions in the $160\mu$m, $250\mu$m, $350\mu$m, $500\mu$m bands  of the Herschel Space Observatory \citep{Pilbratt10}, and the James Clerk Maxwell Telescospe (JCMT) $850\mu$m band. 
We show observational measurements as symbols. 
Some of these LFs (e.g. those of \citealt{Marchetti16}) correspond to LFs measured in very wide redshift ranges ($z<0.5$); hence, we include the $z=0.25$ LF to show how much evolution is expected in that redshift window. 
Disks are the primary contributor over the whole magnitude range in the FIR bands, except in the brightest two bins, where starbursts either driven by galaxy mergers or disk instabilities, are significant.  
This is because at these wavelengths the re-emission of the UV light that was absorbed due to dust starts to become the most dominant source of light (see difference between the thin and thick lines in the bottom-right panel of Fig~\ref{NIRLFs}).

In the Herschel bands, we see that \shark's predictions agree well with the observations within the systematic uncertainties of the data. At the $850\mu$m, the model produces a bright end that is slightly too bright, but we will see in $\S$~\ref{cseds} that the total emission at this band agrees quite well with the observations, possibly indicating systematic effects are important.




To the knowledge of the authors, the agreement of \shark\ with the observed LFs in such a broad wavelength coverage is unprecedented and a success of the overall modelling included in \shark+\prospect. This implies that galaxies have roughly correct SFRs, gas content and gas metallicities (which were shown in \citealt{Lagos18b}), as well as sizes, which together provide realistic dust surface densities. We also remind the reader that the adopted empirical scalings (e.g. the dust-to-metal ratio vs. gas metallicity of \citealt{Remy-Ruyer14}) or theory-inspired relations (e.g. the attenuation parameters of \citet{Trayford19}  are not tuned to get the LFs correct. Instead, quite naturally they allow \shark\ to provide realistic multi-wavelength properties of galaxies.

\subsection{Redshift evolution of the UV and K-band LFs}\label{sec:redshiftevoLF}

\begin{figure}
\begin{center}
\includegraphics[trim=0mm 24mm 7mm 5mm, clip,width=0.45\textwidth]{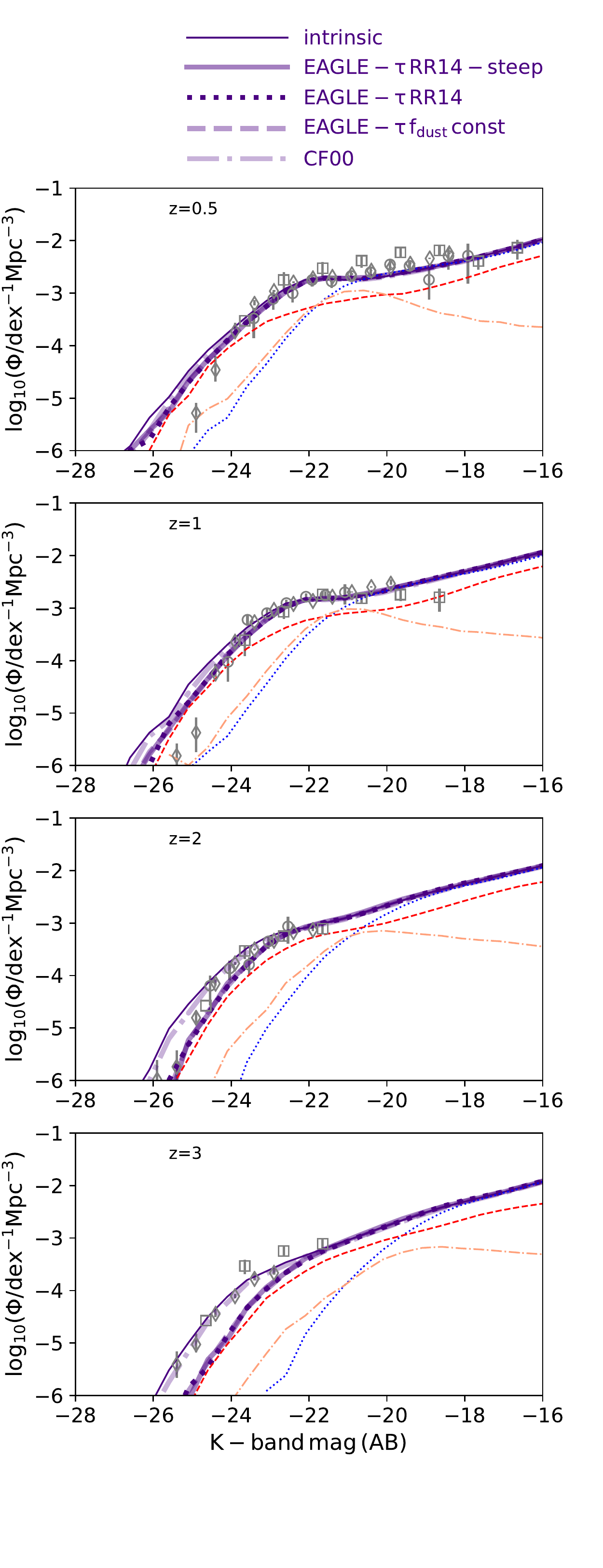}
\caption{K-band LF out to $z=3$, as labelled, for \shark\ after applying the extinction models of Table~\ref{tab:mods}. The thin, solid lines show the intrinsic emission. Observations from \citet{Pozzetti03,Saracco06,Cirasuolo10} are shown as circles, squared and diamonds, respectively. Because all attenuation models give similar predictions, we show the contribution from disks, bulges formed via galaxy mergers and via disk instabilities as thin dotted, dashed for the {\sc EAGLE}$-\tau$ RR14 model only.}
\label{KbandEvo}
\end{center}
\end{figure}

\begin{figure*}
\begin{center}
\includegraphics[trim=0mm 31mm 7mm 5mm, clip,width=0.329\textwidth]{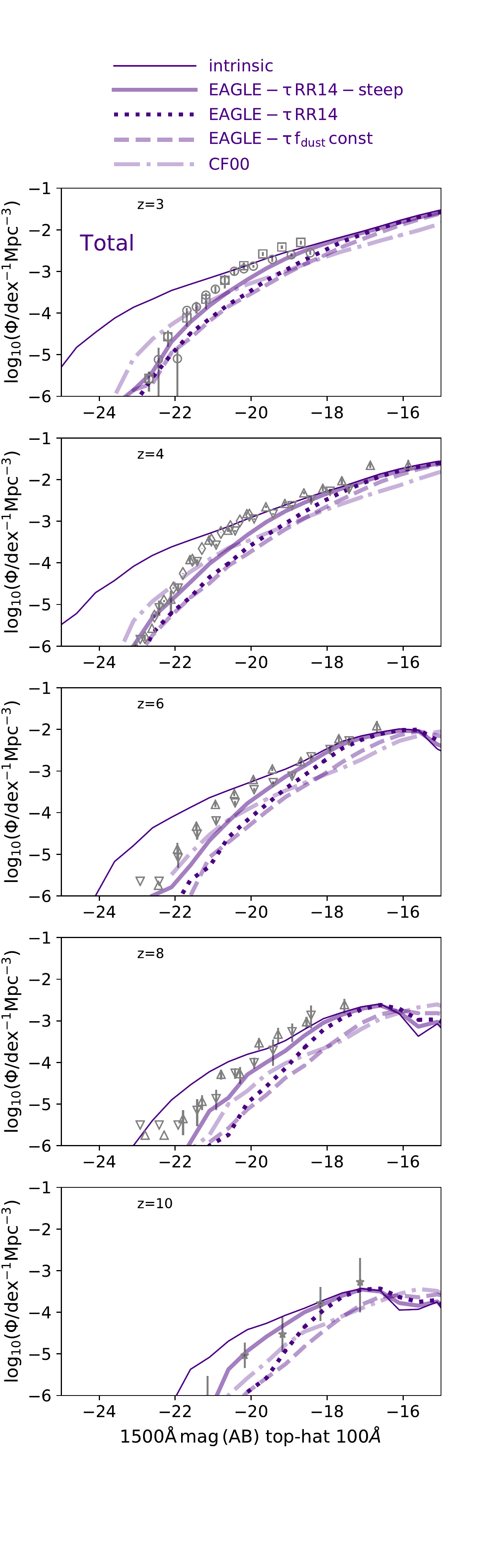}
\includegraphics[trim=0mm 31mm 7mm 5mm, clip,width=0.329\textwidth]{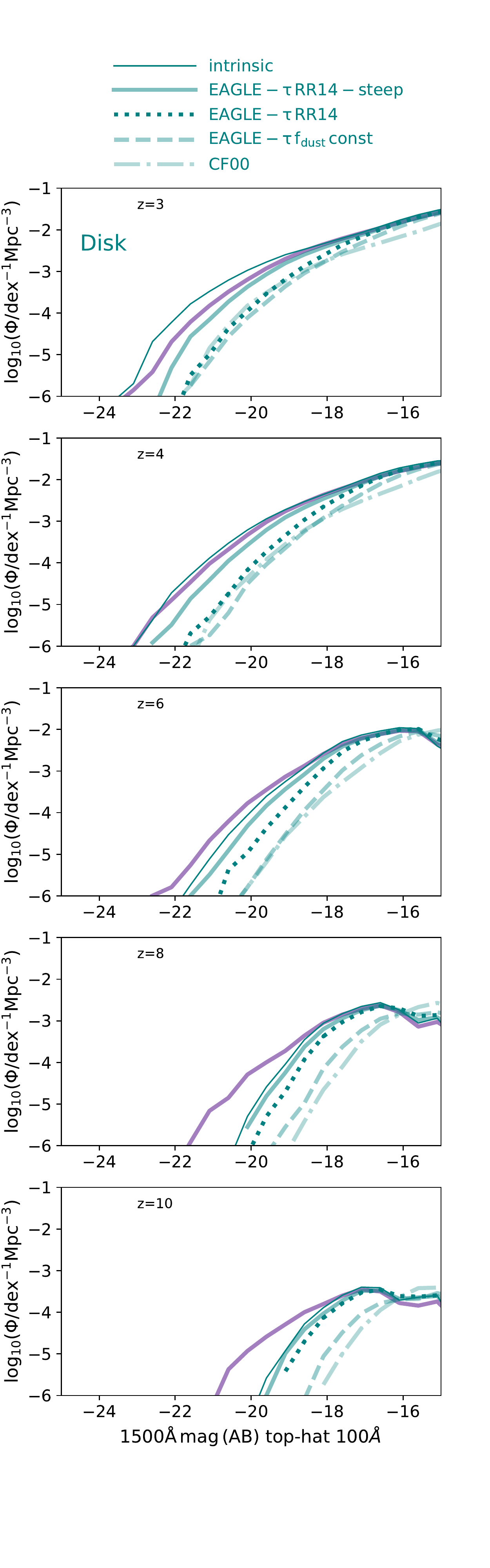}
\includegraphics[trim=0mm 31mm 7mm 5mm, clip,width=0.329\textwidth]{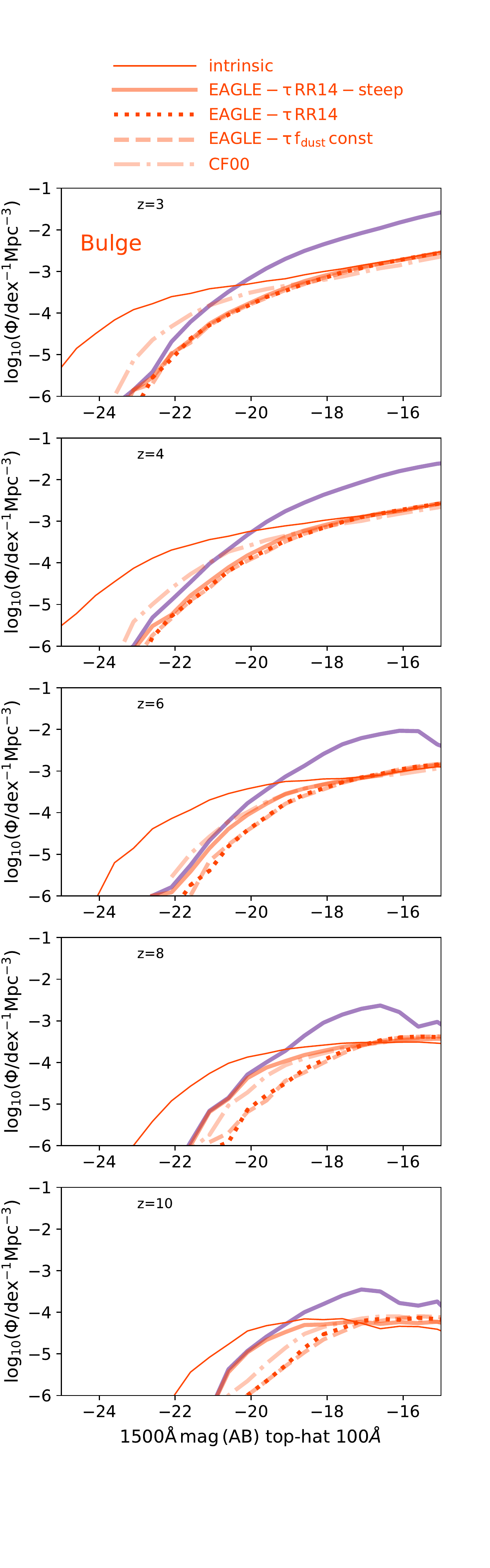}
\caption{Rest-frame UV LFs from $z=3$ to $z=10$, as labelled, showing the intrinsic emitted light in thin, solid lines, and the four attenuation models of Table~\ref{tab:mods}. The UV filter shown here is a top-hat filter of $100$\AA\ width around the $1500$\AA\ wavelength. The left panels show the total emission from galaxies, while the middle and right panels show the contribution from disks and bulges, respectively. We also show for guidance the UV LF of the {\sc EAGLE}-$\tau$~RR14-steep model in the middle and right hand panels. Observations from \citet{Sawicki06,Reddy09,Bouwens15,Finkelstein15,Oesch18} and Adams et al. (submitted) are shown as squares, circles, down-pointing triangles, up-pointing triangles, stars and thin diamonds, respectively, in the left panels. Note that it is only fair to compare the models in the left panels with the observations. The best performing attenuation model is the {\sc EAGLE}-$\tau$ RR14-steep. Note that the differences seen between the models in the left panel is mostly driven by the different by how different models predict the disk extinction, as the bulge is almost always highly attenuated.}
\label{UVEvo}
\end{center}
\end{figure*}

\begin{figure}
\begin{center}
\includegraphics[trim=4mm 2mm 9mm 10mm, clip,width=0.45\textwidth]{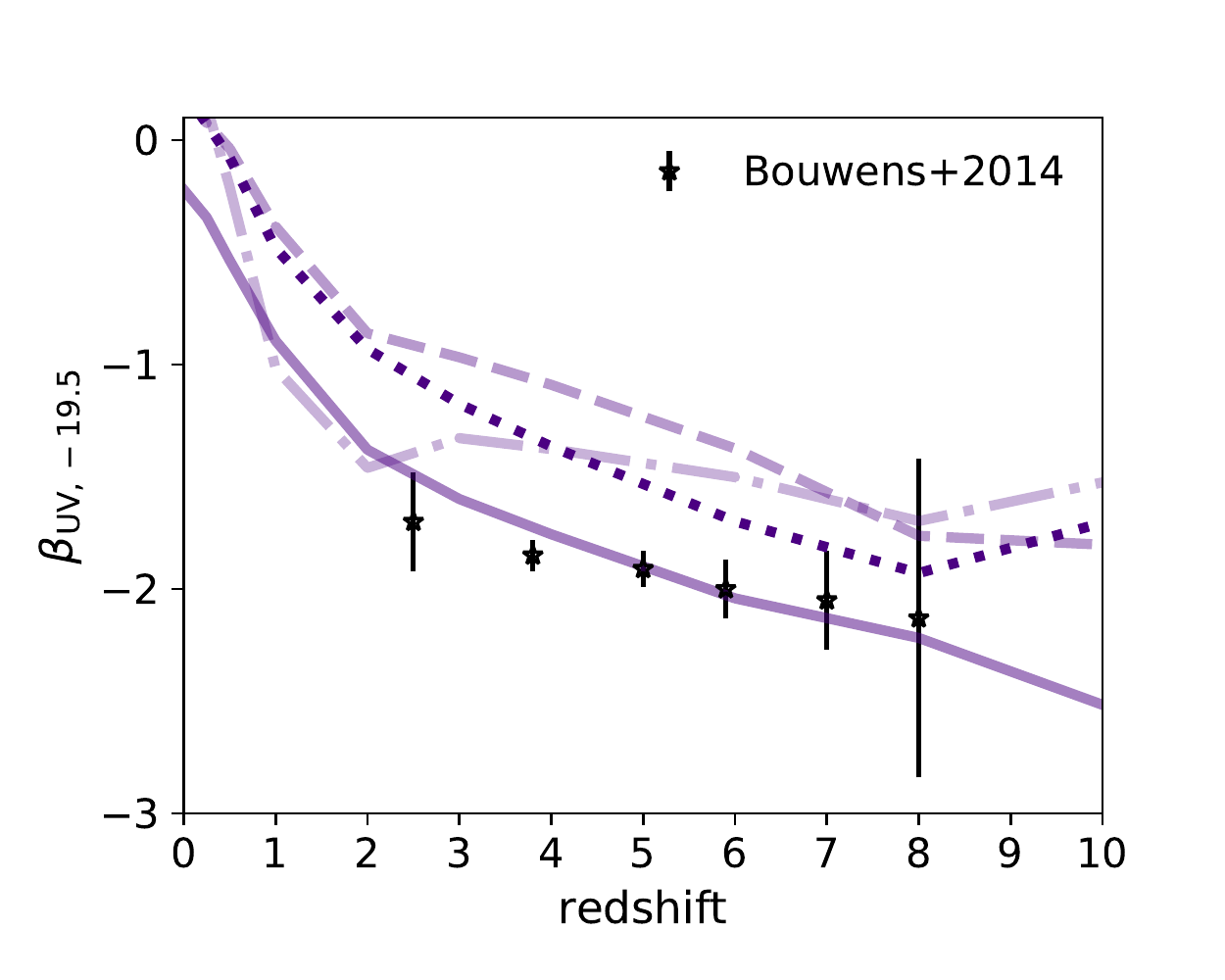}
\caption{The UV slope evolution of \shark\ galaxies with  a rest-frame $1500\AA$ magnitude of $[-19.7,-19.3]$ mags (AB),   computed as $\epsilon_{\nu}\propto \lambda^{2+\beta_{\rm UV}}$, for the $4$ attenuation models of Table~\ref{tab:mods}. Linestyles are in Fig.~\ref{UVEvo}. Symbols show the observations of \citet{Bouwens14}. The best performing attenuation model is the {\sc EAGLE}-$\tau$ RR14-steep.}
\label{UVslopesGals}
\end{center}
\end{figure}

We now focus on the evolution of the galaxy LF in two broadly studied bands: the rest-frame K- and FUV bands.
Fig.~\ref{KbandEvo} shows the $K$-band LF from $z=0.5$ to $z=3$ in \shark\ using the four extinction models of Table~\ref{tab:mods}. As expected, extinction is mostly unimportant in the $K$-band, except at $z=3$ where most of the stars are very young. The agreement with the observations, shown as symbols, is excellent. This is not necessarily surprising as the free parameters in \shark\ are chosen to provide a good fit to the $z=0,\, 1,\, 2$ stellar mass functions, which are strongly correlated with the rest-frame $K$-band luminosity. The tension seen at $z=3$ can in part be due to the BC03 SPs having a small contribution from Asymptotic Giant Branch (AGB) stars. Other SP models, such as those of \citet{Maraston05}, produce more $K$-band emission from AGB stars at $z\approx 3$ than BC03 (see \citealt{Gonzalez-Perez13} for a discussion). 

Because all the attenuation models produce very similar $K$-band LFs, we show the contribution from disks, and bulges formed via galaxy mergers and disk instabilities only for the {\sc EAGLE}$\tau$~RR14 attenuation model. Galaxy disks tend to dominate at the faint-end, with the luminosity below which they dominate becoming fainter as the redshift increases. Bulges driven by disk instabilities have a contribution to the $K$-band luminosity that increases strongly with time. At $z=3$, bulges built via disk instabilities make only a small contribution throughout the magnitude range studied here; as time passes by, they become more important, and by $z=0.5$ they play a significant role in shaping $L^*$. Bulges built via galaxy mergers on the other hand dominate the number density of galaxies over the whole magnitude range at $z=3$, but their dominance shifts to brighter luminosities at lower redshifts. Note, however, that they always play an important role, even at the faintest magnitudes, contributing $\approx 15-25$\% of the observed K-band luminosity in galaxies with $-20<M_{\rm K,rest-frame}(AB)<-16$. The integrated rest-frame $K$-band luminosity of galaxies is dominated by bulges even out to $z=8$. We come back to this in $\S$~\ref{cseds}.

The left panels of Fig.~\ref{UVEvo} show the total rest-frame UV LF evolution from $z=3$ to $z=10$ in \shark\ using the $4$ extinction models of Table~\ref{tab:mods}. We show both the intrinsic emission and the one after attenuation. The latter is the one that should be compared to observations. 
A general trend obtained for all models is that the attenuation in the brightest UV galaxies at $z=3$ and $z=4$ tends to be extremely large, reaching even $\approx 3-4$~mags in some cases, a lot higher than what the values in \shark\ at $z=0$ (see Figs.~\ref{UVLFsz0}~and~\ref{opticalLFs}). This shows that the extinction of the most star-forming galaxies tends to increase from $z=0$ out to $z=3$ and decrease towards higher redshift. This evolution is driven by these galaxies at $z=2-3$ being on average more dusty than those at $z=0$: they have dust surface densities peaking at higher values than at $z=0$ at fixed stellar mass (see Fig.~\ref{sigmadust}), and a tail of galaxies with extremely large dust surface densities, $\Sigma_{\rm dust}>10^{10}\rm M_{\odot}/kpc^2$.

Comparing the different attenuation models of Table~\ref{tab:mods}, it is clear that the model {\sc EAGLE}-$\tau$~RR14-steep provides the best agreement with the observations at all the redshifts of Fig.~\ref{UVEvo}. This is because this model produces the smallest $\tau$ in galaxies with $Z_{\rm gas}<0.5$, which most \shark\ galaxies are at $z>3$. The largest differences between models is seen for the bright galaxies, those with UV magnitudes $\lesssim -20$~mag. These galaxies have on average $0.25 < Z_{\rm gas}/Z_{\odot}<0.7$, which in the models {\sc EAGLE}-$\tau$~RR14 and  {\sc EAGLE}-$\tau$~$f_{\rm dust}$-const have the Milky-Way dust-to-metals ratio, while in the {\sc EAGLE}-$\tau$~RR14-steep model can have $>10$ times less dust per metals mass. Although a different dependence of the dust-to-metal ratio on gas metallicity could provide a better fit to the observations, we decide not to force the agreement and simply explore whether local Universe empirical relations allow \shark\ to provide a reasonable match. We caution the reader, however, that the effect of cosmic variance in the observations is very large, which for the area of the Hubble Deep Field ($2.6$~arcmin$^2$) is $\approx 77$\% at $z\approx 4$ according to the cosmic variance calculator of \citet{Driver10}. The latter is generally not included in the errorbars of the observations.

We remind the reader that we are assuming the dust-to-metal mass ratio to be invariant with time. \citet{Vijayan19} included explicit dust formation and destruction in the SAM {\sc L-galaxies} and predict the dust-to-metal ratio to evolve strongly, with $z=8$~and~$10$ values being about $1.5$~dex smaller than $z=0$ values at fixed stellar mass, which agrees with the observational inferences of \citet{DeVis19}. This not necessarily unexpected, as some sources of dust formation, such as AGB stars and formation in molecular clouds require at least few~$100$~Myr before they start to contribute. If we were to apply such an evolution, our fit to the UV LF would improve. However, other SAMs, e.g. \citep{Popping17}, after implementing similar models of dust formation and destruction find little to no evolution of the dust-to-metal ratio. These contradictory results therefore merit caution in using these relations. 

Other SAM results for the UV LF at high redshift (e.g. \citealt{Qiu19,Yung19}) provide better fits to the UV LFs than those in Fig.~\ref{UVEvo}. However, they tend to be tuned to the UV LFs at $z>3$, and it is unclear whether these models reproduce the panchromatic SEDs of galaxies and the lower redshift Universe observations simultaneously.
%
%

In the middle and right panels of Fig.~\ref{UVEvo} we split the UV LF into the contributions from galaxy disks and bulges, respectively. It is clear that the largest differences at $3\le z \le 6$ between different attenuation models in the total UV LF mostly come from how they predict the extinction for disks, with variations of up to $1.5$~mags at $z=3$ and $2$~mags at $z=6$ between the {\sc EAGLE}-$\tau$~RR14-steep and the other models. Note that at the faint end, magnitudes $>-17$, the {\sc EAGLE}-$\tau$~RR14-steep and {\sc EAGLE}-$\tau$~RR14 extinction models converge to the same answer, as these galaxies have $Z_{\rm gas}<0.25\,\rm Z_{\odot}$. By $z=8$~and~$10$ the {\sc EAGLE}-$\tau$~RR14-steep predicts almost no extinction in the case of disks, and hence there are only marginal differences between the intrinsic and attenuated UV LFs of disks in this model. The {\sc EAGLE}-$\tau$~$f_{\rm dust}$-const model produces a disk UV LF that is similar to the one obtained with the default CF00 parameters. 

We shift our focus now to bulges, which at these redshifts mostly correspond to central starbursts, and are the main channel of bulge formation. At $z=3, 4$, all the {\sc EAGLE}-$\tau$ extinction models produce more extinction than the default CF00 model, and in fact there are little differences between the three {\sc EAGLE}-$\tau$  models. This is because these starbursts have on average $Z_{\rm gas}>0.7\rm \, Z_{\odot}$. At $z\ge 6$ there are some significant differences, with the attenuation model {\sc EAGLE}-$\tau$~RR14-steep producing much smaller attenuation, due to these starbursts having $Z_{\rm gas}<0.7\,\rm Z_{\odot}$. Note, however, that even at $z=8$ and even at $z=10$, the extinction in starbursts galaxies is predicted to be significant, with typical values at the bright end of $\gtrsim 2$~mags. 

In Fig~\ref{UVslopesGals} we compare the predicted UV slopes of galaxies with an AB rest-frame UV magnitude of $-19.5\pm 0.2$, which we measure by fitting the spectrum in the range $0.1\mu\rm m<\lambda_{\rm rest}<0.3\mu \rm m$ with the function $\epsilon_{\nu}\propto \lambda^{2+\beta_{\rm UV}}$, which is equivalent to the fitting performed in observations with the flux in the wavelength space $f_{\lambda}\propto \lambda^{\beta_{\rm UV}}$. The two attenuation models based on RR14 produce similar evolution but with a zero-point offset of $\approx 0.3$. The other two attenuation models, CF00 and {\sc EAGLE}-$\tau$~$f_{\rm dust}$~const, produce weaker redshift evolution. We compare with the observations of \citet{Bouwens14} and find that the attenuation model {\sc EAGLE}-$\tau$~RR14-steep, which reproduces the UV LFs the best, also reproduces the observed UV slopes very well. This is very encouraging as it shows that an attenuation model based on local Universe dust-to-metal scaling relations is capable of reproducing the UV emission of galaxies even out to $z=10$.

\begin{figure*}
\begin{center}
\includegraphics[trim=23mm 19mm 28mm 32mm, clip,width=0.97\textwidth]{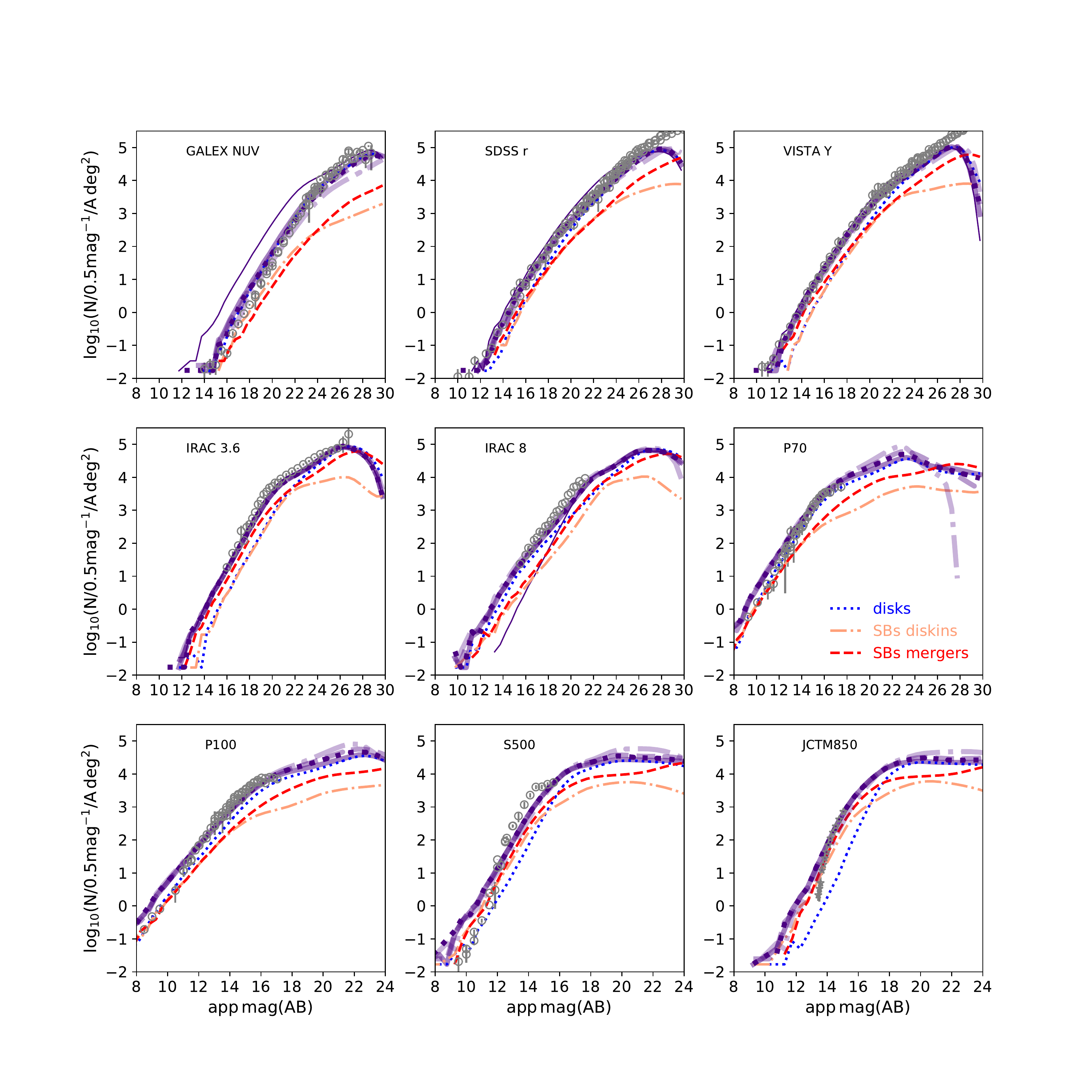}
\caption{Number counts for out \shark\ $107$~deg$^2$ deep lightcone and the $4$ attenuation models of Table~\ref{tab:mods} from the NUV to the $850\mu$m as labelled in each panel. Magnitudes are apparent AB. Indigo coloured lines are as in the left panel of Fig.~\ref{UVEvo}. We only show the intrinsic emission from the NUV to the IRAC $\mu$m band.  The contribution from disks, starbursts driven by galaxy mergers and via disk instabilities, respectively, are shown for the {\sc EAGLE}-$\tau$~RR14~steep only for clarity.  For ease of visualization we change the x-axis range in the bottom panels. The observations shown are from \citet{Driver16b}, except for the $850\mu$m in which we show the \citet{Geach17} data. The agreement with the observations is excellent, with all the models producing similar results, with differences becoming visible at faint magnitudes.}
\label{numbercounts}
\end{center}
\end{figure*}

\begin{figure}
\begin{center}
\includegraphics[trim=7mm 2mm 12mm 9mm, clip,width=0.45\textwidth]{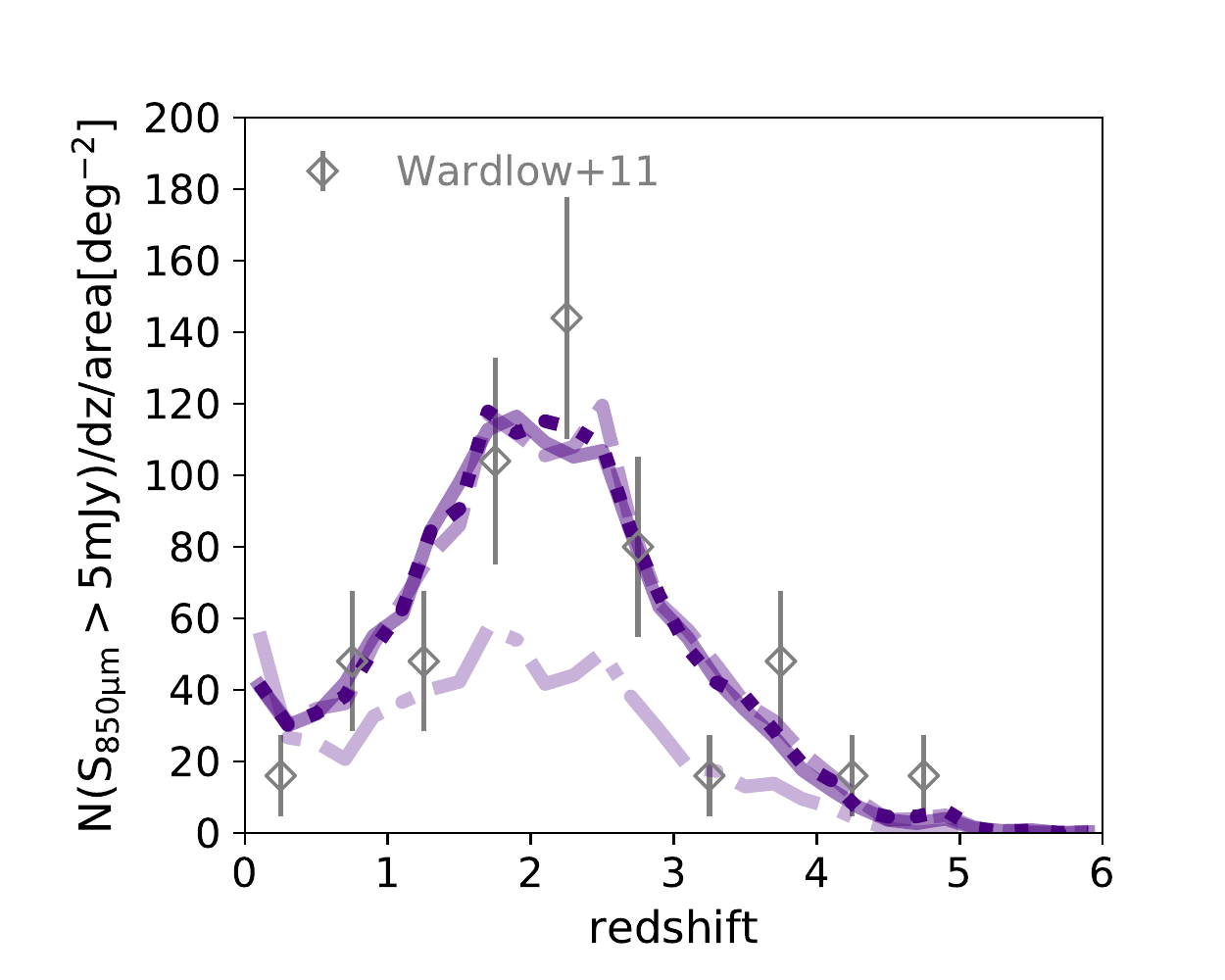}
\caption{Redshift distribution of \shark\ $850\mu$m galaxies with a flux $\ge 5$~mJy for the $4$ attenuation models of Table~\ref{tab:mods} (as labelled in the left panel of Fig.~\ref{UVEvo}). We also show as symbols the observations of \citet{Wardlow11}. Errorbars in the observations show the Poisson uncertanity. All the models based on the {\sc EAGLE} attenuation curves produce a distribution consistent with the observations.}
\label{zdist850}
\end{center}
\end{figure}

\begin{figure}
\begin{center}
\includegraphics[trim=4mm 24mm 9mm 5mm, clip,width=0.48\textwidth]{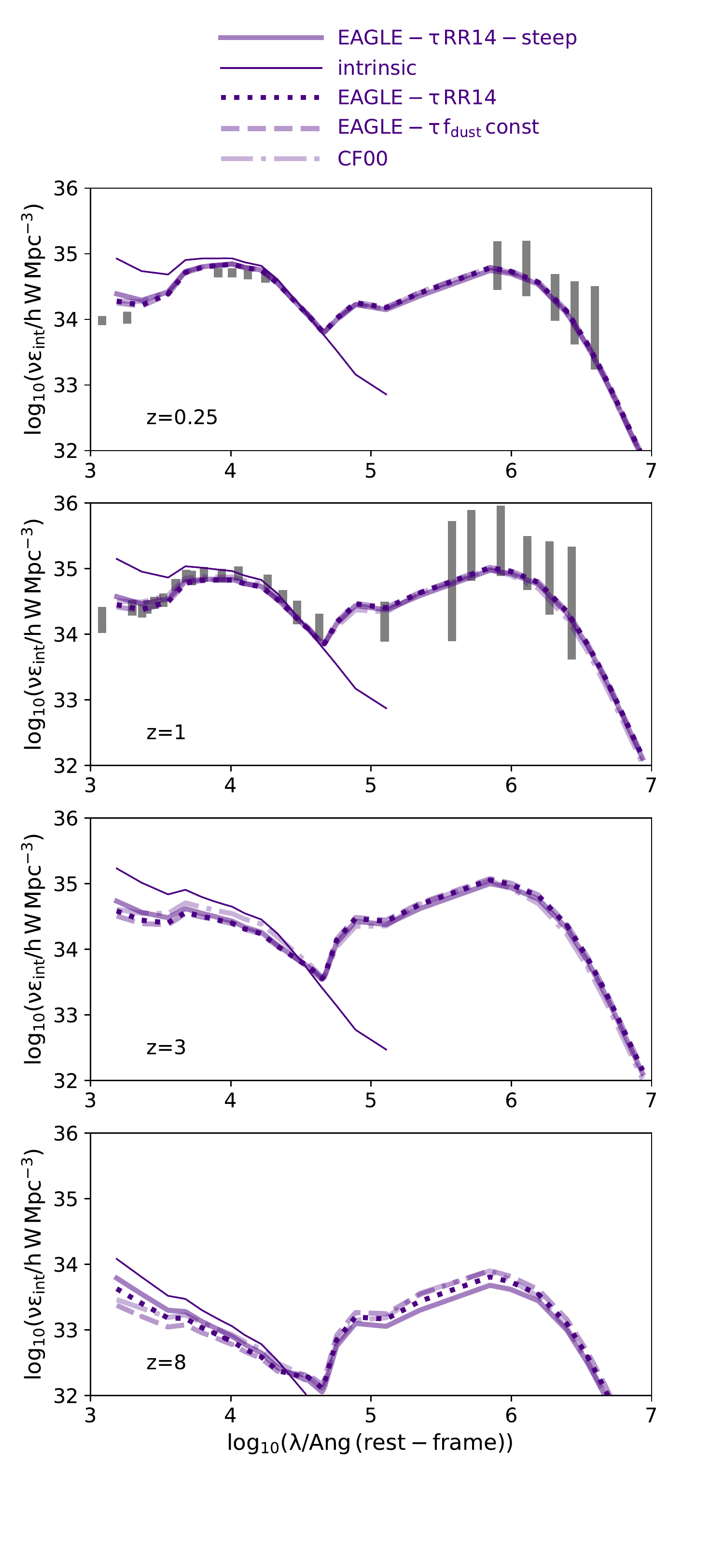}
\caption{Cosmic Spectral Energy Distribution at $z=0$, $z=1$, $z=3$ and $z=6$ for \shark\ using the $4$ attenuation models of Table~\ref{tab:mods}, as labelled.  Observational estimates from \citet{Andrews17} at $z=0$ and $z=1$ are shown as grey segments.}
\label{CSEDall}
\end{center}
\end{figure}

\begin{figure}
\begin{center}
\includegraphics[trim=4mm 2mm 9mm 10mm, clip,width=0.45\textwidth]{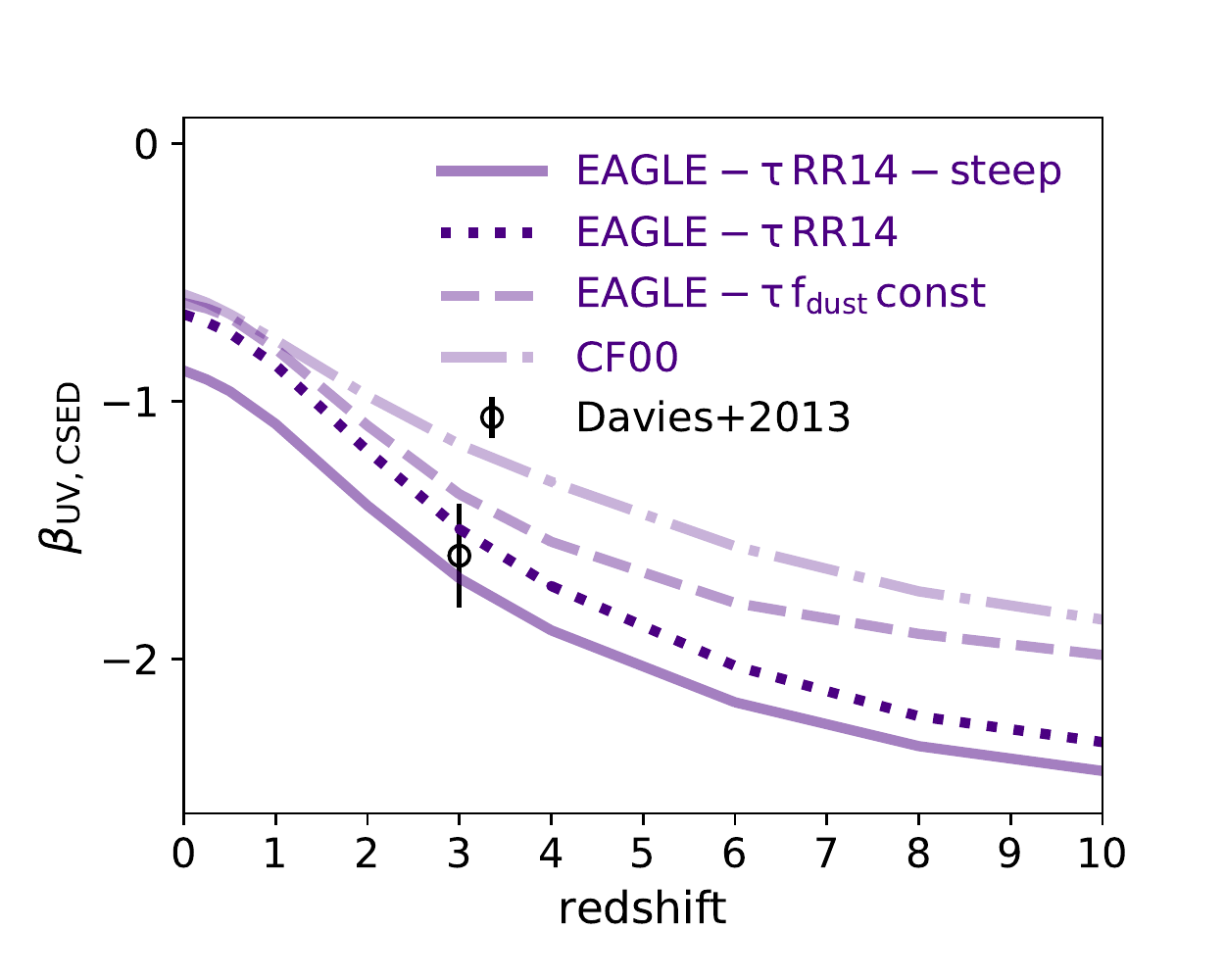}
\caption{The UV slope of the \shark\ CSED computed as $\epsilon_{\nu}\propto \lambda^{2+\beta_{\rm UV}}$ as a function of redshift for the $4$ attenuation models of Table~\ref{tab:mods}. We show the observational constraint of \citet{Davies13} from stacking of Lyman-break galaxies at $z=3$.}
\label{UVslopes}
\end{center}
\end{figure}

\begin{figure}
\begin{center}
\includegraphics[trim=4mm 15mm 9mm 28mm, clip,width=0.49\textwidth]{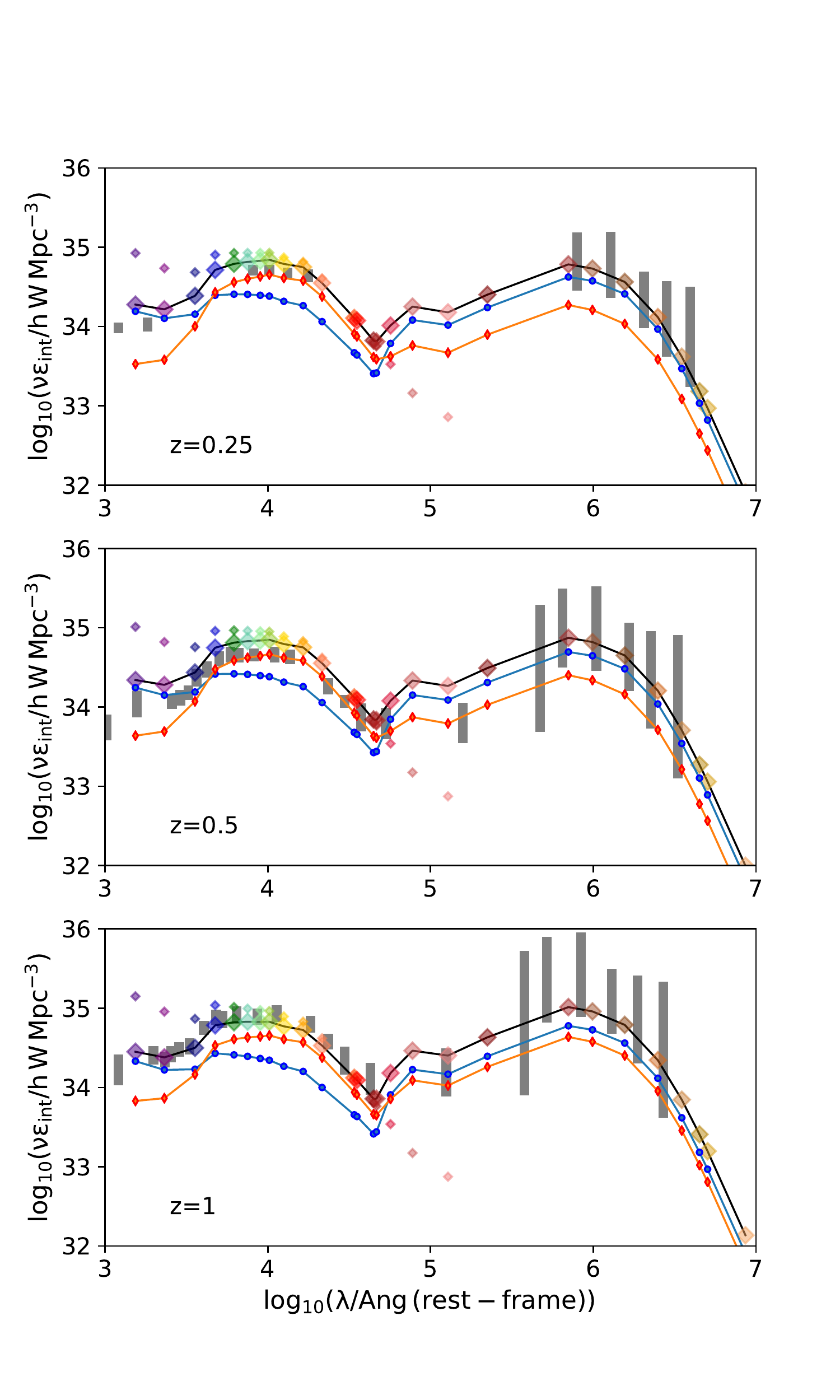}
\caption{Cosmic Spectral Energy Distribution at $z=0.25$, $z=0.5$ and $z=1$, as labelled, for \shark\ (small and large diamonds show the intrinsic and attenuated/remitted light, respectively) using the attenuation model {\sc EAGLE}-$\tau$~RR14 (see Table~\ref{tab:mods} for details). The contribution from emission of disk and bulge stars is shown as blue and red small symbols, respectively. Observational estimates from \citet{Andrews17} are shown as grey segments.}
\label{CSED}
\end{center}
\end{figure}

\begin{figure}
\begin{center}
\includegraphics[trim=4mm 32mm 9mm 46mm, clip,width=0.49\textwidth]{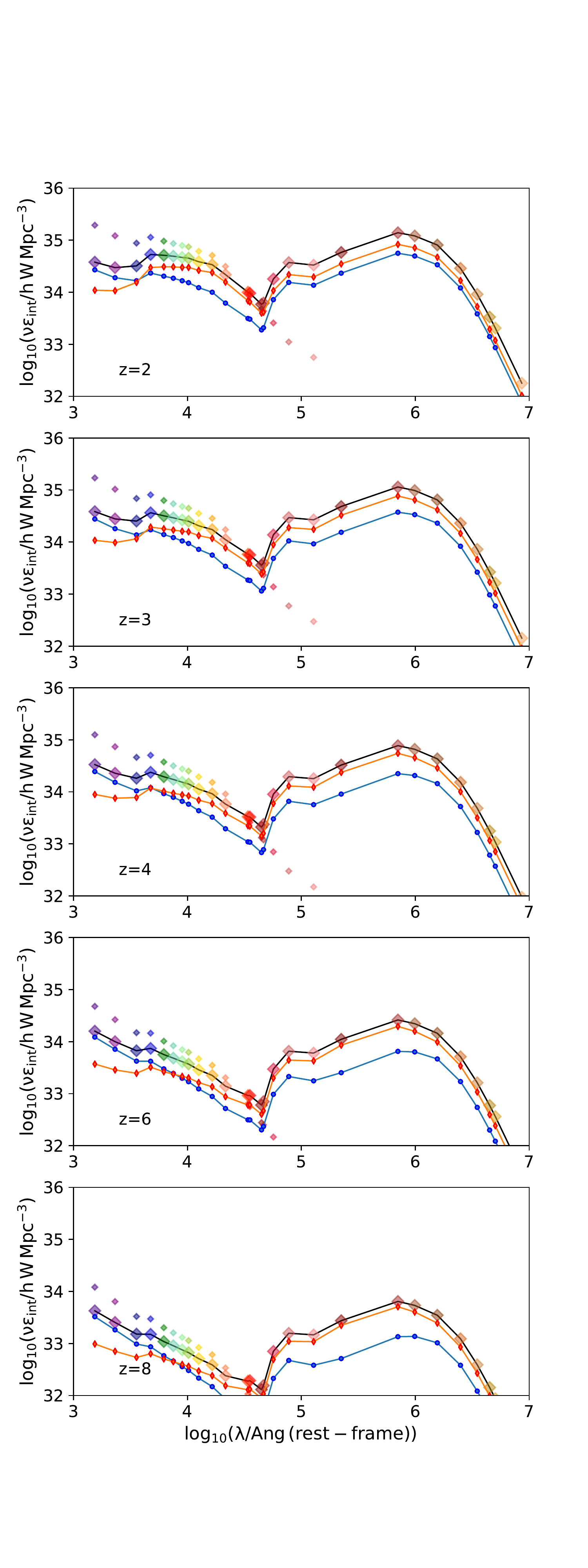}
\caption{As in Fig.~\ref{CSED} but for $z=2-8$.}
\label{CSEDhighz}
\end{center}
\end{figure}

\section{Number counts and the Cosmic SED across cosmic times}\label{cseds}

\subsection{Number counts}\label{sec:numbercounts}

Galaxy number counts are the most direct observable of galaxies: how many galaxies are observed in a given apparent magnitude in a given band. Because galaxies of different masses and at different cosmic epochs contribute to this observable, they have been difficult to reproduce in galaxy formation simulations (see \citealt{Somerville12,Lacey15} for a discussion). Another obvious difficulty is that constructing number counts necessarily requires to predict the galaxy population over the entire age of the universe and in a wavelength range as wide as possible. 

With the SED models presented here we can test \shark\ against the observed number counts. To do this, we build a lightcone of area $107$~deg$^2$ including all galaxies with a dummy magnitude, computed assuming a stellar mass-to-light ratio of $1$, $<32$ and at $0\le z\le 8$. We then use the method described in $\S$~\ref{sec:viperfish} to build SEDs. We refer to \citet{Chauhan19} for more details about our lightcone construction. Fig.~\ref{numbercounts} shows the predicted number counts from the NUV to the $850\mu$m of this lightcone for the $4$ attenuation models of Fig.~\ref{numbercounts}, compare with the observations of \citet{Driver16} and \citet{Geach17}.

The agreement with the observations is excellent across the entire wavelength range shown here and for all the attenuation models tested. Some tension is identified in the Herschel SPIRE bands, in which \shark\ tends to predict too few (many) galaxies with AB magnitudes $14-16$ ($<10$) by a factor of $\approx 2$ compared to \citet{Driver16b}. Interestingly, these differences are similar to those reported in \citet{Lacey15} for the GALFORM semi-analytic model. Recently, \citet{WangL19} showed that the Herschel number counts we show here likely suffer from systematic errors due to blending and confusion, and hence the tension with \shark\ {could be} due to those systematics. 

The truly unexpected result of Fig.~\ref{numbercounts} is that we are able to match the observed number counts in the UV-optical and FIR bands {\it simultaneously} without the need to invoke a varying IMF. \citet{Baugh05,Lacey15} showed that in GALFORM this was only achieved by invoking a top-heavy IMF during starbursts. In the case of a universal IMF the numbers of bright $850\mu$m galaxies in their work was consistently under-produced, and not only that, but they tended to lie at low redshift, in clear tension with the observations (which find a peak at $z\approx 2$). \shark\ assumes a universal \citet{Chabrier03} IMF and hence this shows that in a fully cosmological galaxy formation model, this is possible. In order to confirm this claim, we show in Fig.~\ref{zdist850} the predicted redshift distribution of bright $850\mu$m galaxies, fluxes $>5$~mJy, for the $4$ attenuation models of Fig.~\ref{tab:mods}, compared to the observations of \citet{Wardlow11}. The agreement is outstanding with all the models that use the {\sc EAGLE} attenuation curves, while for the model adopting the default CF00 parameters, the redshift distribution is less peaked at $z\approx 2$ than observations suggest. In any case, \shark\ captures well the redshift peak of the brightest $850\mu$m sources, and the tail towards high redshifts. We remind the reader that in all cases we assume an invariant relation between dust mass-gas metallicity and gas content that is informed by local Universe observations.

The reasons why \shark\ is able to reproduce the observed number counts from the UV to the FIR with a universal IMF and other models have not is difficult to pinpoint due to the many aspects that enter in the calculation: dust masses, gas metallicities, galaxy sizes, attenuation curves and dust temperature. Hence we here discuss some possibilities but warn the reader that these are not conclusive. An important quantity is the dust mass, which is tied to the gas metallicities and gas content. Both \shark\ and GALFORM reproduce well the gas content of galaxies, however, GALFORM predicts gas metallicities that are consistently too low compared to observations at $M_{\star}\lesssim 10^{10.5}\,\rm M_{\odot}$ by up to $1$~dex (see Fig.~11 in \citealt{Guo15}). Galaxy sizes may also be too large in GALFORM compared to observations (see Fig.~21 in \citealt{Lacey15}). Both these effects contribute to lowering the dust surface density. \shark\ on the other hand predicts sizes that agree with observations (by construction), and gas metallicities that are closer to those observed (but not perfect; see Figs.~10 and ~15 in \citealt{Lagos18b}). We are also assuming two constant dust temperatures for the BCs and diffuse dust, while in GALFORM this is computed self-consistently, which produces a dust temperature that weakly increases with redshift \citep{Cowley17}. The latter makes the $850\mu$m emission weaker at fixed total FIR luminosity.  
A definitive conclusion though is that the answer to whether a varying IMF is needed to reproduce simultaneously the UV-optical and FIR emission of galaxies or not is {\it model dependent}.

Fig.~\ref{numbercounts} also shows the contribution from star formation in disks and in bulges, the latter separated by triggering mechanism: galaxy mergers and disk instabilities. We show this for the {\sc EAGLE}-$\tau$~RR14~steep model only for the sake of clarity. As expected, the NUV is dominated by star formation in disks over the whole magnitude range, while the $r$-, $Y$ and IRAC $3.6\mu$m bands are dominated by bulges at bright magnitudes, transitioning to disks dominating at fainter magnitudes. The exact transition is wavelength dependent, moving from $\approx 18$ AB magnitudes in the $r$-band to $22$ in IRAC $3.6\mu$m. In the FIR the opposite trend takes place: going from the IRAC $8\mu$m to the $850\mu$m bands, we see the transition from bulge-dominated to disk-dominated emission moving to fainter magnitudes, with a transition of $12$ AB magnitudes in the IRAC $8\mu$m band to $18$~mags at $850\mu$m. In \shark, bright $850\mu$m galaxies (also referred to SMGs) are a mix of starbursts driven by galaxy mergers and by disk instabilities in almost equal numbers, with a slight dominance of galaxy mergers. 

\subsection{The Cosmic SED}

The integrated spectrum of galaxies at a given redshift is termed the cosmic SED (CSED), and holds important information of the star formation activity of galaxies, the amount of light that is absorbed and reprocessed by dust, and the type of galaxies that contribute to the light at different wavelengths. 

In this section we compare our predictions with the observations of \citet{Andrews17}, which are based on the GAMA survey \citep{Driver09}, as well as the re-analysis of the G10/COSMOS photometry and spectroscopy \citep{Davies15}. These measurements are for $0\le z\le 1$, and hence any higher redshift result can be considered a prediction of \shark.

We compute the predicted CSEDs of \shark\ by simply adding the light from all the galaxies at any given redshift. Truncating the integration to AB magnitudes $<-14$ does not have an effect on the predicted CSED, which shows that the integral is well converged for the resolution of our simulation.

\subsubsection{The effect of extinction in shaping the CSED}

Fig.~\ref{CSEDall} shows the predicted CSEDs of \shark\ at $z=0.25, 1, 3, 8$ for the four attenuation models of Table~\ref{tab:mods}. 
All the models predict a similar FIR CSED that at $z=0.25$ and $z=1$ agree {reasonably} well with the observations of \citet{Andrews17}. {The models tend to produce too much UV by $\approx 30-50$\% at $z=0.25$ compared to observations, though at $z=1$ the agreement is excellent. Given all the modelling that goes into predicting the UV, such as the adopted IMF, SP templates, SFH, ZFH and dust attenuation, and the effects in observations that are more difficult to include in the errorbars, such as cosmic variance, the UV LF faint-end slope and uncertain extrapolations, we consider this level of disagreement to be acceptable.} 

Some important differences {among models}  are seen at high redshift; by $z=8$, there are differences of up to $0.4$~dex in the power output at fixed wavelength at $10^5<\lambda_{\rm rest}/$\AA$<10^6$. This is due to the large differences in 
extinction predicted by our attenuation models in galaxies with gas metallicities $<0.75\,\rm Z_{\odot}$. 
The NIR is consistent among all the attenuation models at all redshifts. This is not surprising as the light at these wavelengths tends to trace stellar mass closely, which is the same for all models.

In the UV-end of the CSED, all models predict an important steepening of the UV slope with increasing redshift. Observations of individual high redshift galaxies show similar steepening of the UV compared to local galaxies (e.g. \citealt{Dunlop12}). Although the overall trends are qualitatively the same for the four attenuation models studied here, in the detail there are some important differences. In order to quantify them, we measure the UV slope of the CSED at different redshifts and show them in Fig.~\ref{UVslopes}. The {\sc EAGLE}-$\tau$~RR14 extinction model produces the strongest evolution with a difference of $1.8$ in $\beta_{\rm UV}$ between $z=0$ and $z=10$. We show in Fig.~\ref{UVslopes} the observational constraint of \citet{Davies13} from stacking of Ly-break galaxies, which seem consistent with the predictions of all the {\sc EAGLE}-$\tau$ attenuation models. The default CF00 attenuation model produces the weakest evolution, and in fact the values of $\beta_{\rm UV}$ at $z>5$ in this model are too large compared to \citet{Davies13}. 

\citet{Cowley19} analysed the CSED predictions of the GALFORM semi-analytic model, and unlike \shark, they find little evolution of $\beta_{\rm UV}$, with values that throughout redshift are close to $-2$. This is in clear tension with the observations at low redshift, as seen in Figs.~\ref{CSEDall}~and~\ref{CSED}, but at high redshift they are consistent with those in \shark\ (albeit some of our attenuation models produce bluer spectra). \citet{Somerville12} presented CSEDs using the Santa-Cruz SAM, and although they did not quantify the UV slope, their results seem to qualitatively support a strong redshift evolution of $\beta_{\rm UV}$. 

\subsubsection{Breaking down the Light budget in the CSED across cosmic time}

Fig.~\ref{CSED} shows the predicted CSED of \shark\ with the default {\sc EAGLE}-$\tau$~RR14 attenuation model at $0.25\le z\le 1$. Small diamonds show the intrinsic emitted light, while bigger diamonds show the predicted light after we include the effects of attenuation and re-emission in the IR. We find that \shark\ predicts a CSED that overall agrees very well with the observations through the whole wavelength range tested here, within the observational uncertainties. The level of agreement displayed by \shark\ is unprecedented to the knowledge of the authors. \citet{Cowley19} showed for the GALFORM semi-analytic model that their model variant with a universal IMF struggled to simultaneously reproduce the FUV-to-optical and FIR parts of the CSED, and a top-heavy IMF was required. Because our \shark\ model assumes a universal IMF, it suggests that this may be model dependent. This agrees with the findings discussed in $\S$~\ref{sec:numbercounts}.
\citet{Baes19} presented the CSEDs of the {\sc EAGLE} hydrodynamical simulations and showed excellent agreement at $z\lesssim 0.5$, but towards $z\approx 1$ they found {\sc EAGLE} to produce too little FIR emission.
 Hence, we consider the agreement seen in Fig.~\ref{CSED} to be a key success of \shark.
Some areas of tension at the $0.1-0.15$~dex level, however, remain. At $z=0.25$, \shark\ produces too much FUV emission, and at $z=0.5$, \shark\ tends to produce $0.1$~dex too much emission in the optical-to-NIR bands.

Fig.~\ref{CSED} shows the contribution from disks and bulges of galaxies to the total CSED. Bulges tend to dominate in the optical-to-NIR wavelength range at $z\le 1$, while disks dominate in the FUV-NUV and FIR ranges. The importance of bulges in the FIR emission, however, evolves strongly with redshift. This is because at $z=0$ we transition from bulges with no or little star formation, to centrally concentrated starbursts at $z\gtrsim 1$, which tend to be very dusty (see Fig.~\ref{FigTaus}). 

Fig.~\ref{CSEDhighz} shows the evolution of the CSED of \shark\ using the {\sc EAGLE}-$\tau$~RR14 attenuation model at $2\le z\le 8$. 
At these redshifts the FIR makes a more significant contribution to the integrated light than the FUV-NIR, with the peak of the CSED being at $10^{5.5}\lesssim \lambda_{\rm rest}/$\AA$\lesssim 10^{6.2}$. The slope of the CSED in the FUV-to-optical wavelength range becomes increasingly steeper with increasing redshift, due to both the very high star-formation activity in galaxies and their low metal and dust content (see Fig.~\ref{FigTaus}).

At $z\gtrsim 2$, bulges make up most of the FIR emission, due to their starburst and dusty nature, and their contribution continues to increase with increasing redshift. Disks, on the other hand, dominate at the FUV-NUV over the whole redshift range, and by $z=8$ they also dominate in the rest-frame $u$ and $g$ bands. Note that at the NIR, bulges dominate throughout the whole redshift range analysed here $0\le z \le 8$. 

\section{Conclusions}\label{conclusions}

We presented an exhaustive analysis of the spectral energy distribution (SED) predictions of the \shark\ semi-analytic model \citep{Lagos18b} at $0\le z\le 10$. We first introduced the modelling of galaxy's SEDs, which make use of the  \prospect\ software tool, which takes as input the SFH and ZFH of galaxies, and uses the \citet{Bruzual03} SPs to produce the intrinsic emitted light. We then use the parametric attenuation curves of \citet{Charlot00} to compute the amount of extinction, and re-emit that in the IR following the templates of \citet{Dale14} and energy conservations arguments. For the latter, we adopt an effective dust temperature for the diffuse ISM and birth clouds of $\approx 20-25$~K and $\approx 50-60$~K, which are fixed for the whole redshift range analysed in this paper.

To compute the appropriate \citet{Charlot00} extinction parameters of individual \shark\ galaxies, we make use of the predicted attenuation curves of the RT analysis of {\sc EAGLE} by \citet{Trayford19} and how these vary with the dust surface density of galaxies. We compute the dust content of \shark\ galaxies by applying the local Universe scaling relation between the dust mass, gas content and gas metallicity of \citet{Remy-Ruyer14}, and assume this relation to hold out to $z=10$. This method allows us to apply a physical model for the attenuation of UV-to-optical light and re-emission in the IR that scales with galaxy properties. After generating the FUV-to-FIR emission of \shark\ galaxies, we compare to observations  without re-tuning the model.

We summarize our findings below:

\begin{itemize}
\item Our model is capable of reproducing the wide diversity of observed galaxies, from galaxies that are almost metal free and have negligible attenuation, which tend to be abundant at high redshift and at low stellar masses, to SMGs, which are most prominent at around $1\le z\le 3$, but exist in the model out to $z=6$.
\item We tested different models for the conversion of gas mass and gas metallicity to dust mass within the observational uncertainties and find that these tend to produce different FUV LFs with the largest differences appearing at $z\ge 4$. Differences in the optical-to-NIR are negligible throughout $0\le z\le 10$, and in the FIR they are only important at faint magnitudes, below the current observational limits. 
\item \shark\ is capable of reproducing well the observed $z\approx 0$ LFs of galaxies from the FUV (GALEX) to the FIR ($850\mu$m). We compare our model with observed LFs in $27$ bands and found reasonable agreement in all of them. In a future paper (Bravo et al. in prep) we show that optical colours are also very well reproduced even at intermediate redshifts.
\item We analysed the rest-frame K-band and UV LFs out to $z=3$ and $z=10$, respectively, and found \shark\ to reproduce them reasonably well. We find that the rest-frame K-band LF above the knee is always dominated by bulges in galaxies while the rest-frame UV LF sees a strong evolution, from being dominated by star-forming galaxy disks throughout most of the magnitude range at $z\lesssim 4$ to a bigger contribution from low metallicity galaxy mergers-induced starbursts at the bright end at $z\gtrsim 4$. UV-bright galaxies display a strong evolution of their UV slope from $\approx -0.2$ at $z=0$ to $\approx -2.5$ at $z=10$, with some variations between the different adopted dust-to-gas mass scalings. We find the attenuation of UV-to-optical light to be maximal at $z\approx 1-2$. 
\item By building a deep, wide area lightcone of $107$~deg$^2$ with \shark\ galaxies, we compare the predicted number counts from the NUV to the $850\mu$m with observations and find unprecedented agreement. To confirm our SMG population is realistic, we also study the redshift distribution of bright, $>5$~mJy, SMGs and found that it peaks at $z\approx 2$ with a tail that extends out to $z\approx 6$, in very good agreement with observations. This is achieved {\it without} the need of invoking a top heavy IMF in starbursts and/or a redshift-dependent dust-gas mass-gas metallicity scaling showing that a fully cosmological galaxy formation model is capable of reproducing simultaneously the emission in the UV-optical to the FIR with a universal IMF.
\item We integrate the galaxy LFs at different redshifts to produce a CSED from $z=0$ out $z=10$ and find \shark\ to reproduce well the observed CSEDs at $z\le 1$, while there are no available observations at higher redshifts. \shark\ predicts the FIR emission to be dominated by star-forming disks at $z\lesssim 1.5$, and by starbursts at higher redshifts, even out to $z=10$. These starbursts are triggered by both disk instabilities and galaxy mergers, and we find that they contribute similarly to the IR emission. The rest-frame UV and NIR are dominated by star-forming disks and bulges at all redshifts, respectively. 
\end{itemize}

The success of our model makes it an ideal tool for future galaxy surveys from $z=0$ to $z=10$. Possible applications include understanding the galaxy populations different color-based selections isolate, how observationally-based environment metrics trace the underlying halo population, the bias of flux-selected galaxies in different bands, systematic effects in photometric redshift determinations, among many others. The interested reader is encouraged to contact the authors of this manuscript for access to the simulated lightcones. 

One of the most surprising results in this manuscript is the fact that we can simultaneously reproduce the UV-to-NIR and FIR properties of galaxies, including number counts and redshift distributions, without the need of varying the IMF of galaxies, which is unprecedented. The reason why previous models struggled with this and \shark\ does not is difficult to pinpoint as these models are complex and commonly a combination of processes are responsible for the differences seen among simulations. However, we discussed several possibilities, which we plan to explore in depth in the future, including (i) differences in the predicted gas metallicities and sizes among models (both of which affect the dust surface density) (ii) differences in the SFR function, particularly at $1\le z\le 3$, (iii) differences in the dust temperature evolution, and (iv) different attenuation curves. Nonetheless, we can certainly assert that the answer to what physical processes are required to simultaneously reproduce the FUV-to-FIR emission of galaxies is model dependent. 

\section*{Acknowledgements}

We thank Cedric Lacey, Carlton Baugh and Desika Narayanan for useful discussions about the results in this paper, and the anonymous referee for their constructive report. 
CL has received funding from the ARC Centre of
Excellence for All Sky Astrophysics in 3 Dimensions (ASTRO 3D), through project number CE170100013.
CL also thanks the MERAC Foundation for a Postdoctoral Research Award. JT and CL also thank the University of Western Australia for a Research Collaboration Award which facilitated the face-to-face interaction that led to this work.
This work was supported by resources provided by The Pawsey Supercomputing Centre with funding from the
Australian Government and the Government of Western Australia.
Cosmic Dawn Centre is funded by the Danish National Research Foundation.





\bibliographystyle{mn2e_trunc8}
\bibliography{SHArkIntro}

\begin{thebibliography}{106}
\expandafter\ifx\csname natexlab\endcsname\relax\def\natexlab#1{#1}\fi

\bibitem[{{Amarantidis} {et~al.}(2019){Amarantidis}, {Afonso}, {Messias},
  {Henriques}, {Griffin}, {Lacey}, {Lagos}, {Gonzalez-Perez}, {Dubois},
  {Volonteri}, {Matute}, {Pappalardo}, {Qin}, {Chary}, \&
  {Norris}}]{Amarantidis19}
{Amarantidis} S., {Afonso} J., {Messias} H., {Henriques} B., {Griffin} A.,
  {Lacey} C., {Lagos} C. d.~P., {Gonzalez-Perez} V. {et~al}, 2019, \mnras, 485,
  2694

\bibitem[{{Andrews} {et~al.}(2017){Andrews}, {Driver}, {Davies}, {Kafle},
  {Robotham}, {Vinsen}, {Wright}, {Bland-Hawthorn}, {Bourne}, {Bremer}, {da
  Cunha}, {Drinkwater}, {Holwerda}, {Hopkins}, {Kelvin}, {Loveday},
  {Phillipps}, \& {Wilkins}}]{Andrews17}
{Andrews} S.~K., {Driver} S.~P., {Davies} L.~J.~M., {Kafle} P.~R., {Robotham}
  A.~S.~G., {Vinsen} K., {Wright} A.~H., {Bland-Hawthorn} J. {et~al}, 2017,
  \mnras, 470, 1342

\bibitem[{{Baes} {et~al.}(2019){Baes}, {Tr{\v{c}}ka}, {Camps}, {Nersesian},
  {Trayford}, {Theuns}, \& {Dobbels}}]{Baes19}
{Baes} M., {Tr{\v{c}}ka} A., {Camps} P., {Nersesian} A., {Trayford} J.,
  {Theuns} T., {Dobbels} W., 2019, \mnras, 484, 4069

\bibitem[{{Baugh} {et~al.}(2005){Baugh}, {Lacey}, {Frenk}, {Granato}, {Silva},
  {Bressan}, {Benson}, \& {Cole}}]{Baugh05}
{Baugh} C.~M., {Lacey} C.~G., {Frenk} C.~S., {Granato} G.~L., {Silva} L.,
  {Bressan} A., {Benson} A.~J., {Cole} S., 2005, \mnras, 356, 1191

\bibitem[{{Blitz} {et~al.}(2007){Blitz}, {Fukui}, {Kawamura}, {Leroy},
  {Mizuno}, \& {Rosolowsky}}]{Blitz07}
{Blitz} L., {Fukui} Y., {Kawamura} A., {Leroy} A., {Mizuno} N., {Rosolowsky}
  E., 2007, Protostars and Planets V, 81

\bibitem[{{Blitz} \& {Rosolowsky}(2006)}]{Blitz06}
{Blitz} L., {Rosolowsky} E., 2006, \apj, 650, 933

\bibitem[{{Bolatto} {et~al.}(2008){Bolatto}, {Leroy}, {Rosolowsky}, {Walter},
  \& {Blitz}}]{Bolatto08}
{Bolatto} A.~D., {Leroy} A.~K., {Rosolowsky} E., {Walter} F., {Blitz} L., 2008,
  \apj, 686, 948

\bibitem[{{Bournaud} {et~al.}(2011){Bournaud}, {Chapon}, {Teyssier}, {Powell},
  {Elmegreen}, {Elmegreen}, {Duc}, {Contini}, {Epinat}, \&
  {Shapiro}}]{Bournaud11}
{Bournaud} F., {Chapon} D., {Teyssier} R., {Powell} L.~C., {Elmegreen} B.~G.,
  {Elmegreen} D.~M., {Duc} P.-A., {Contini} T. {et~al}, 2011, \apj, 730, 4

\bibitem[{{Bouwens} {et~al.}(2014){Bouwens}, {Illingworth}, {Oesch},
  {Labb{\'e}}, {van Dokkum}, {Trenti}, {Franx}, {Smit}, {Gonzalez}, \&
  {Magee}}]{Bouwens14}
{Bouwens} R.~J., {Illingworth} G.~D., {Oesch} P.~A., {Labb{\'e}} I., {van
  Dokkum} P.~G., {Trenti} M., {Franx} M., {Smit} R. {et~al}, 2014, \apj, 793,
  115

\bibitem[{{Bouwens} {et~al.}(2015){Bouwens}, {Illingworth}, {Oesch}, {Trenti},
  {Labb{\'e}}, {Bradley}, {Carollo}, {van Dokkum}, {Gonzalez}, {Holwerda},
  {Franx}, {Spitler}, {Smit}, \& {Magee}}]{Bouwens15}
{Bouwens} R.~J., {Illingworth} G.~D., {Oesch} P.~A., {Trenti} M., {Labb{\'e}}
  I., {Bradley} L., {Carollo} M., {van Dokkum} P.~G. {et~al}, 2015, \apj, 803,
  34

\bibitem[{{Bruzual} \& {Charlot}(2003)}]{Bruzual03}
{Bruzual} G., {Charlot} S., 2003, \mnras, 344, 1000

\bibitem[{{Ca{\~n}as} {et~al.}(2019){Ca{\~n}as}, {Elahi}, {Welker}, {del P
  Lagos}, {Power}, {Dubois}, \& {Pichon}}]{Canas18}
{Ca{\~n}as} R., {Elahi} P.~J., {Welker} C., {del P Lagos} C., {Power} C.,
  {Dubois} Y., {Pichon} C., 2019, \mnras, 482, 2039

\bibitem[{{Camps} {et~al.}(2016){Camps}, {Trayford}, {Baes}, {Theuns},
  {Schaller}, \& {Schaye}}]{Camps16}
{Camps} P., {Trayford} J.~W., {Baes} M., {Theuns} T., {Schaller} M., {Schaye}
  J., 2016, \mnras, 462, 1057

\bibitem[{{Capak} {et~al.}(2015){Capak}, {Carilli}, {Jones}, {Casey},
  {Riechers}, {Sheth}, {Carollo}, {Ilbert}, {Karim}, {Lefevre}, {Lilly},
  {Scoville}, {Smolcic}, \& {Yan}}]{Capak15}
{Capak} P.~L., {Carilli} C., {Jones} G., {Casey} C.~M., {Riechers} D., {Sheth}
  K., {Carollo} C.~M., {Ilbert} O. {et~al}, 2015, \nat, 522, 455

\bibitem[{{Casey} {et~al.}(2012){Casey}, {Berta}, {B{\'e}thermin}, {Bock},
  {Bridge}, {Budynkiewicz}, {Burgarella}, {Chapin}, {Chapman}, {Clements},
  {Conley}, {Conselice}, {Cooray}, {Farrah}, {Hatziminaoglou}, {Ivison}, {le
  Floc'h}, {Lutz}, {Magdis}, {Magnelli}, {Oliver}, {Page}, {Pozzi},
  {Rigopoulou}, {Riguccini}, {Roseboom}, {Sanders}, {Scott}, {Seymour},
  {Valtchanov}, {Vieira}, {Viero}, \& {Wardlow}}]{Casey12}
{Casey} C.~M., {Berta} S., {B{\'e}thermin} M., {Bock} J., {Bridge} C.,
  {Budynkiewicz} J., {Burgarella} D., {Chapin} E. {et~al}, 2012, \apj, 761, 140

\bibitem[{{Chabrier}(2003)}]{Chabrier03}
{Chabrier} G., 2003, \pasp, 115, 763

\bibitem[{{Charlot} \& {Fall}(2000)}]{Charlot00}
{Charlot} S., {Fall} S.~M., 2000, \apj, 539, 718

\bibitem[{{Chauhan} {et~al.}(2019){Chauhan}, {Lagos}, {Obreschkow}, {Power},
  {Oman}, \& {Elahi}}]{Chauhan19}
{Chauhan} G., {Lagos} C.~D.~P., {Obreschkow} D., {Power} C., {Oman} K., {Elahi}
  P.~J., 2019, arXiv:1906.06130, arXiv:1906.06130

\bibitem[{{Cirasuolo} {et~al.}(2010){Cirasuolo}, {McLure}, {Dunlop}, {Almaini},
  {Foucaud}, \& {Simpson}}]{Cirasuolo10}
{Cirasuolo} M., {McLure} R.~J., {Dunlop} J.~S., {Almaini} O., {Foucaud} S.,
  {Simpson} C., 2010, \mnras, 401, 1166

\bibitem[{{Cole} {et~al.}(2000){Cole}, {Lacey}, {Baugh}, \& {Frenk}}]{Cole00}
{Cole} S., {Lacey} C.~G., {Baugh} C.~M., {Frenk} C.~S., 2000, \mnras, 319, 168

\bibitem[{{Conroy}(2013)}]{Conroy13}
{Conroy} C., 2013, \araa, 51, 393

\bibitem[{{Cowley} {et~al.}(2017){Cowley}, {B{\'e}thermin}, {Lagos}, {Lacey},
  {Baugh}, \& {Cole}}]{Cowley17}
{Cowley} W.~I., {B{\'e}thermin} M., {Lagos} C. d.~P., {Lacey} C.~G., {Baugh}
  C.~M., {Cole} S., 2017, \mnras, 467, 1231

\bibitem[{{Cowley} {et~al.}(2019){Cowley}, {Lacey}, {Baugh}, {Cole}, {Frenk},
  \& {Lagos}}]{Cowley19}
{Cowley} W.~I., {Lacey} C.~G., {Baugh} C.~M., {Cole} S., {Frenk} C.~S., {Lagos}
  C. d.~P., 2019, \mnras, 487, 3082

\bibitem[{{Croton} {et~al.}(2016){Croton}, {Stevens}, {Tonini}, {Garel},
  {Bernyk}, {Bibiano}, {Hodkinson}, {Mutch}, {Poole}, \& {Shattow}}]{Croton16}
{Croton} D.~J., {Stevens} A.~R.~H., {Tonini} C., {Garel} T., {Bernyk} M.,
  {Bibiano} A., {Hodkinson} L., {Mutch} S.~J. {et~al}, 2016, \apjs, 222, 22

\bibitem[{{da Cunha} {et~al.}(2008){da Cunha}, {Charlot}, \&
  {Elbaz}}]{daCunha08}
{da Cunha} E., {Charlot} S., {Elbaz} D., 2008, \mnras, 388, 1595

\bibitem[{{Dai} {et~al.}(2009){Dai}, {Assef}, {Kochanek}, {Brodwin}, {Brown},
  {Caldwell}, {Cool}, {Dey}, {Eisenhardt}, {Eisenstein}, {Gonzalez}, {Jannuzi},
  {Jones}, {Murray}, \& {Stern}}]{Dai09}
{Dai} X., {Assef} R.~J., {Kochanek} C.~S., {Brodwin} M., {Brown} M.~J.~I.,
  {Caldwell} N., {Cool} R.~J., {Dey} A. {et~al}, 2009, \apj, 697, 506

\bibitem[{{Dale} {et~al.}(2014){Dale}, {Helou}, {Magdis}, {Armus},
  {D{\'{\i}}az-Santos}, \& {Shi}}]{Dale14}
{Dale} D.~A., {Helou} G., {Magdis} G.~E., {Armus} L., {D{\'{\i}}az-Santos} T.,
  {Shi} Y., 2014, \apj, 784, 83

\bibitem[{{Davies} {et~al.}(2013){Davies}, {Bremer}, {Stanway}, \&
  {Lehnert}}]{Davies13}
{Davies} L.~J.~M., {Bremer} M.~N., {Stanway} E.~R., {Lehnert} M.~D., 2013,
  \mnras, 433, 2588

\bibitem[{{Davies} {et~al.}(2015){Davies}, {Driver}, {Robotham}, {Baldry},
  {Lange}, {Liske}, {Meyer}, {Popping}, {Wilkins}, \& {Wright}}]{Davies15}
{Davies} L.~J.~M., {Driver} S.~P., {Robotham} A.~S.~G., {Baldry} I.~K., {Lange}
  R., {Liske} J., {Meyer} M., {Popping} A. {et~al}, 2015, \mnras, 447, 1014

\bibitem[{{Davies} {et~al.}(2019){Davies}, {Lagos}, {Katsianis}, {Robotham},
  {Cortese}, {Driver}, {Bremer}, {Brown}, {Brough}, {Cluver}, {Grootes},
  {Holwerda}, {Owers}, \& {Phillipps}}]{Davies19}
{Davies} L.~J.~M., {Lagos} C. d.~P., {Katsianis} A., {Robotham} A.~S.~G.,
  {Cortese} L., {Driver} S.~P., {Bremer} M.~N., {Brown} M.~J.~I. {et~al}, 2019,
  \mnras, 483, 1881

\bibitem[{{Davies} {et~al.}(2018){Davies}, {Robotham}, {Driver}, {Lagos},
  {Cortese}, {Mannering}, {Foster}, {Lidman}, {Hashemizadeh}, {Koushan},
  {O'Toole}, {Baldry}, {Bilicki}, {Bland -Hawthorn}, {Bremer}, {Brown},
  {Bryant}, {Catinella}, {Croom}, {Grootes}, {Holwerda}, {Jarvis}, {Maddox},
  {Meyer}, {Moffett}, {Phillipps}, {Taylor}, {Windhorst}, \& {Wolf}}]{Davies18}
{Davies} L.~J.~M., {Robotham} A.~S.~G., {Driver} S.~P., {Lagos} C.~P.,
  {Cortese} L., {Mannering} E., {Foster} C., {Lidman} C. {et~al}, 2018, \mnras,
  480, 768

\bibitem[{{De Lucia} \& {Blaizot}(2007)}]{DeLucia07}
{De Lucia} G., {Blaizot} J., 2007, \mnras, 375, 2

\bibitem[{{De Vis} {et~al.}(2019){De Vis}, {Jones}, {Viaene}, {Casasola},
  {Clark}, {Baes}, {Bianchi}, {Cassara}, {Davies}, \& {De Looze}}]{DeVis19}
{De Vis} P., {Jones} A., {Viaene} S., {Casasola} V., {Clark} C.~J.~R., {Baes}
  M., {Bianchi} S., {Cassara} L.~P. {et~al}, 2019, \aap, 623, A5

\bibitem[{{Driver} {et~al.}(2018){Driver}, {Andrews}, {da Cunha}, {Davies},
  {Lagos}, {Robotham}, {Vinsen}, {Wright}, {Alpaslan}, {Bland-Hawthorn},
  {Bourne}, {Brough}, {Bremer}, {Cluver}, {Colless}, {Conselice}, {Dunne},
  {Eales}, {Gomez}, {Holwerda}, {Hopkins}, {Kafle}, {Kelvin}, {Loveday},
  {Liske}, {Maddox}, {Phillipps}, {Pimbblet}, {Rowlands}, {Sansom}, {Taylor},
  {Wang}, \& {Wilkins}}]{Driver17}
{Driver} S.~P., {Andrews} S.~K., {da Cunha} E., {Davies} L.~J., {Lagos} C.,
  {Robotham} A.~S.~G., {Vinsen} K., {Wright} A.~H. {et~al}, 2018, \mnras, 475,
  2891

\bibitem[{{Driver} {et~al.}(2016{\natexlab{a}}){Driver}, {Andrews}, {Davies},
  {Robotham}, {Wright}, {Windhorst}, {Cohen}, {Emig}, {Jansen}, \&
  {Dunne}}]{Driver16b}
{Driver} S.~P., {Andrews} S.~K., {Davies} L.~J., {Robotham} A. S.~G., {Wright}
  A.~H., {Windhorst} R.~A., {Cohen} S., {Emig} K. {et~al}, 2016{\natexlab{a}},
  \apj, 827, 108

\bibitem[{{Driver} {et~al.}(2016{\natexlab{b}}){Driver}, {Davies}, {Meyer},
  {Power}, {Robotham}, {Baldry}, {Liske}, \& {Norberg}}]{Driver16}
{Driver} S.~P., {Davies} L.~J., {Meyer} M., {Power} C., {Robotham} A.~S.~G.,
  {Baldry} I.~K., {Liske} J., {Norberg} P., 2016{\natexlab{b}}, The Universe of
  Digital Sky Surveys, 42, 205

\bibitem[{{Driver} {et~al.}(2009){Driver}, {Norberg}, {Baldry}, {Bamford},
  {Hopkins}, {Liske}, {Loveday}, {Peacock}, {Hill}, {Kelvin}, {Robotham},
  {Cross}, {Parkinson}, {Prescott}, {Conselice}, {Dunne}, {Brough}, {Jones},
  {Sharp}, {van Kampen}, {Oliver}, {Roseboom}, {Bland-Hawthorn}, {Croom},
  {Ellis}, {Cameron}, {Cole}, {Frenk}, {Couch}, {Graham}, {Proctor}, {De
  Propris}, {Doyle}, {Edmondson}, {Nichol}, {Thomas}, {Eales}, {Jarvis},
  {Kuijken}, {Lahav}, {Madore}, {Seibert}, {Meyer}, {Staveley-Smith},
  {Phillipps}, {Popescu}, {Sansom}, {Sutherland}, {Tuffs}, \&
  {Warren}}]{Driver09}
{Driver} S.~P., {Norberg} P., {Baldry} I.~K., {Bamford} S.~P., {Hopkins} A.~M.,
  {Liske} J., {Loveday} J., {Peacock} J.~A. {et~al}, 2009, Astronomy and
  Geophysics, 50, 050000

\bibitem[{{Driver} \& {Robotham}(2010)}]{Driver10}
{Driver} S.~P., {Robotham} A.~S.~G., 2010, \mnras, 407, 2131

\bibitem[{{Driver} {et~al.}(2012){Driver}, {Robotham}, {Kelvin}, {Alpaslan},
  {Baldry}, {Bamford}, {Brough}, {Brown}, {Hopkins}, {Liske}, {Loveday},
  {Norberg}, {Peacock}, {Andrae}, {Bland-Hawthorn}, {Bourne}, {Cameron},
  {Colless}, {Conselice}, {Croom}, {Dunne}, {Frenk}, {Graham}, {Gunawardhana},
  {Hill}, {Jones}, {Kuijken}, {Madore}, {Nichol}, {Parkinson}, {Pimbblet},
  {Phillipps}, {Popescu}, {Prescott}, {Seibert}, {Sharp}, {Sutherland},
  {Taylor}, {Thomas}, {Tuffs}, {van Kampen}, {Wijesinghe}, \&
  {Wilkins}}]{Driver12}
{Driver} S.~P., {Robotham} A.~S.~G., {Kelvin} L., {Alpaslan} M., {Baldry}
  I.~K., {Bamford} S.~P., {Brough} S., {Brown} M. {et~al}, 2012, \mnras, 427,
  3244

\bibitem[{{Dunlop} {et~al.}(2012){Dunlop}, {McLure}, {Robertson}, {Ellis},
  {Stark}, {Cirasuolo}, \& {de Ravel}}]{Dunlop12}
{Dunlop} J.~S., {McLure} R.~J., {Robertson} B.~E., {Ellis} R.~S., {Stark}
  D.~P., {Cirasuolo} M., {de Ravel} L., 2012, \mnras, 420, 901

\bibitem[{{Dunne} {et~al.}(2000){Dunne}, {Eales}, {Edmunds}, {Ivison},
  {Alexander}, \& {Clements}}]{Dunne00}
{Dunne} L., {Eales} S., {Edmunds} M., {Ivison} R., {Alexander} P., {Clements}
  D.~L., 2000, \mnras, 315, 115

\bibitem[{{Dye} {et~al.}(2010){Dye}, {Dunne}, {Eales}, {Smith}, {Amblard},
  {Auld}, {Baes}, {Baldry}, {Bamford}, {Blain}, {Bonfield}, {Bremer},
  {Burgarella}, {Buttiglione}, {Cameron}, {Cava}, {Clements}, {Cooray},
  {Croom}, {Dariush}, {de Zotti}, {Driver}, {Dunlop}, {Frayer}, {Fritz},
  {Gardner}, {Gomez}, {Gonzalez-Nuevo}, {Herranz}, {Hill}, {Hopkins}, {Ibar},
  {Ivison}, {Jarvis}, {Jones}, {Kelvin}, {Lagache}, {Leeuw}, {Liske},
  {Lopez-Caniego}, {Loveday}, {Maddox}, {Micha{\l}owski}, {Negrello},
  {Norberg}, {Page}, {Parkinson}, {Pascale}, {Peacock}, {Pohlen}, {Popescu},
  {Prescott}, {Rigopoulou}, {Robotham}, {Rigby}, {Rodighiero}, {Samui},
  {Scott}, {Serjeant}, {Sharp}, {Sibthorpe}, {Temi}, {Thompson}, {Tuffs},
  {Valtchanov}, {van der Werf}, {van Kampen}, \& {Verma}}]{Dye10}
{Dye} S., {Dunne} L., {Eales} S., {Smith} D.~J.~B., {Amblard} A., {Auld} R.,
  {Baes} M., {Baldry} I.~K. {et~al}, 2010, \aap, 518, L10

\bibitem[{{Efstathiou} {et~al.}(1982){Efstathiou}, {Lake}, \&
  {Negroponte}}]{Efstathiou82}
{Efstathiou} G., {Lake} G., {Negroponte} J., 1982, \mnras, 199, 1069

\bibitem[{{Elahi} {et~al.}(2019{\natexlab{a}}){Elahi}, {Ca{\~n}as}, {Poulton},
  {Tobar}, {Willis}, {Lagos}, {Power}, \& {Robotham}}]{Elahi19a}
{Elahi} P.~J., {Ca{\~n}as} R., {Poulton} R. J.~J., {Tobar} R.~J., {Willis}
  J.~S., {Lagos} C. d.~P., {Power} C., {Robotham} A. S.~G., 2019{\natexlab{a}},
  Publications of the Astronomical Society of Australia, 36, e021

\bibitem[{{Elahi} {et~al.}(2019{\natexlab{b}}){Elahi}, {Poulton}, {Tobar},
  {Ca{\~n}as}, {Lagos}, {Power}, \& {Robotham}}]{Elahi19b}
{Elahi} P.~J., {Poulton} R. J.~J., {Tobar} R.~J., {Ca{\~n}as} R., {Lagos} C.
  d.~P., {Power} C., {Robotham} A. S.~G., 2019{\natexlab{b}}, Publications of
  the Astronomical Society of Australia, 36, e028

\bibitem[{{Elahi} {et~al.}(2018{\natexlab{a}}){Elahi}, {Power}, {Lagos},
  {Poulton}, \& {Robotham}}]{Elahi18b}
{Elahi} P.~J., {Power} C., {Lagos} C. d.~P., {Poulton} R., {Robotham} A. S.~G.,
  2018{\natexlab{a}}, Monthly Notices of the Royal Astronomical Society, 477,
  616

\bibitem[{{Elahi} {et~al.}(2018{\natexlab{b}}){Elahi}, {Welker}, {Power}, {del
  P Lagos}, {Robotham}, {Ca{\~n}as}, \& {Poulton}}]{Elahi18}
{Elahi} P.~J., {Welker} C., {Power} C., {del P Lagos} C., {Robotham} A.~S.~G.,
  {Ca{\~n}as} R., {Poulton} R., 2018{\natexlab{b}}, \mnras

\bibitem[{{Fanidakis} {et~al.}(2012){Fanidakis}, {Baugh}, {Benson}, {Bower},
  {Cole}, {Done}, {Frenk}, {Hickox}, {Lacey}, \& {Del P.~Lagos}}]{Fanidakis10b}
{Fanidakis} N., {Baugh} C.~M., {Benson} A.~J., {Bower} R.~G., {Cole} S., {Done}
  C., {Frenk} C.~S., {Hickox} R.~C. {et~al}, 2012, \mnras, 419, 2797

\bibitem[{{Finkelstein} {et~al.}(2015){Finkelstein}, {Ryan}, {Papovich},
  {Dickinson}, {Song}, {Somerville}, {Ferguson}, {Salmon}, {Giavalisco},
  {Koekemoer}, {Ashby}, {Behroozi}, {Castellano}, {Dunlop}, {Faber}, {Fazio},
  {Fontana}, {Grogin}, {Hathi}, {Jaacks}, {Kocevski}, {Livermore}, {McLure},
  {Merlin}, {Mobasher}, {Newman}, {Rafelski}, {Tilvi}, \&
  {Willner}}]{Finkelstein15}
{Finkelstein} S.~L., {Ryan} Jr. R.~E., {Papovich} C., {Dickinson} M., {Song}
  M., {Somerville} R.~S., {Ferguson} H.~C., {Salmon} B. {et~al}, 2015, \apj,
  810, 71

\bibitem[{{Geach} {et~al.}(2017){Geach}, {Dunlop}, {Halpern}, {Smail}, {van der
  Werf}, {Alexander}, {Almaini}, {Aretxaga}, {Arumugam}, {Asboth}, {Banerji},
  {Beanlands}, {Best}, {Blain}, {Birkinshaw}, {Chapin}, {Chapman}, {Chen},
  {Chrysostomou}, {Clarke}, {Clements}, {Conselice}, {Coppin}, {Cowley},
  {Danielson}, {Eales}, {Edge}, {Farrah}, {Gibb}, {Harrison}, {Hine}, {Hughes},
  {Ivison}, {Jarvis}, {Jenness}, {Jones}, {Karim}, {Koprowski}, {Knudsen},
  {Lacey}, {Mackenzie}, {Marsden}, {McAlpine}, {McMahon}, {Meijerink},
  {Micha{\l}owski}, {Oliver}, {Page}, {Peacock}, {Rigopoulou}, {Robson},
  {Roseboom}, {Rotermund}, {Scott}, {Serjeant}, {Simpson}, {Simpson}, {Smith},
  {Spaans}, {Stanley}, {Stevens}, {Swinbank}, {Targett}, {Thomson}, {Valiante},
  {Wake}, {Webb}, {Willott}, {Zavala}, \& {Zemcov}}]{Geach17}
{Geach} J.~E., {Dunlop} J.~S., {Halpern} M., {Smail} I., {van der Werf} P.,
  {Alexander} D.~M., {Almaini} O., {Aretxaga} I. {et~al}, 2017, \mnras, 465,
  1789

\bibitem[{{Gonzalez-Perez} {et~al.}(2014){Gonzalez-Perez}, {Lacey}, {Baugh},
  {Lagos}, {Helly}, {Campbell}, \& {Mitchell}}]{Gonzalez-Perez13}
{Gonzalez-Perez} V., {Lacey} C.~G., {Baugh} C.~M., {Lagos} C.~D.~P., {Helly}
  J., {Campbell} D.~J.~R., {Mitchell} P.~D., 2014, \mnras

\bibitem[{{Granato} {et~al.}(2000){Granato}, {Lacey}, {Silva}, {Bressan},
  {Baugh}, {Cole}, \& {Frenk}}]{Granato00}
{Granato} G.~L., {Lacey} C.~G., {Silva} L., {Bressan} A., {Baugh} C.~M., {Cole}
  S., {Frenk} C.~S., 2000, \apj, 542, 710

\bibitem[{{Griffin} {et~al.}(2018){Griffin}, {Lacey}, {Gonzalez-Perez},
  {Lagos}, {Baugh}, \& {Fanidakis}}]{Griffin18}
{Griffin} A.~J., {Lacey} C.~G., {Gonzalez-Perez} V., {Lagos} C.~d.~P., {Baugh}
  C.~M., {Fanidakis} N., 2018, ArXiv:1806.08370

\bibitem[{{Guo} {et~al.}(2016){Guo}, {Gonzalez-Perez}, {Guo}, {Schaller},
  {Furlong}, {Bower}, {Cole}, {Crain}, {Frenk}, {Helly}, {Lacey}, {Lagos},
  {Mitchell}, {Schaye}, \& {Theuns}}]{Guo15}
{Guo} Q., {Gonzalez-Perez} V., {Guo} Q., {Schaller} M., {Furlong} M., {Bower}
  R.~G., {Cole} S., {Crain} R.~A. {et~al}, 2016, \mnras, 461, 3457

\bibitem[{{Henriques} {et~al.}(2015){Henriques}, {White}, {Thomas}, {Angulo},
  {Guo}, {Lemson}, {Springel}, \& {Overzier}}]{Henriques15}
{Henriques} B.~M.~B., {White} S.~D.~M., {Thomas} P.~A., {Angulo} R., {Guo} Q.,
  {Lemson} G., {Springel} V., {Overzier} R., 2015, \mnras, 451, 2663

\bibitem[{{Koekemoer} {et~al.}(2011){Koekemoer}, {Faber}, {Ferguson}, {Grogin},
  {Kocevski}, {Koo}, {Lai}, {Lotz}, {Lucas}, {McGrath}, {Ogaz}, {Rajan},
  {Riess}, {Rodney}, {Strolger}, {Casertano}, {Castellano}, {Dahlen},
  {Dickinson}, {Dolch}, {Fontana}, {Giavalisco}, {Grazian}, {Guo}, {Hathi},
  {Huang}, {van der Wel}, {Yan}, {Acquaviva}, {Alexander}, {Almaini}, {Ashby},
  {Barden}, {Bell}, {Bournaud}, {Brown}, {Caputi}, {Cassata}, {Challis},
  {Chary}, {Cheung}, {Cirasuolo}, {Conselice}, {Roshan Cooray}, {Croton},
  {Daddi}, {Dav{\'e}}, {de Mello}, {de Ravel}, {Dekel}, {Donley}, {Dunlop},
  {Dutton}, {Elbaz}, {Fazio}, {Filippenko}, {Finkelstein}, {Frazer}, {Gardner},
  {Garnavich}, {Gawiser}, {Gruetzbauch}, {Hartley}, {H{\"a}ussler},
  {Herrington}, {Hopkins}, {Huang}, {Jha}, {Johnson}, {Kartaltepe},
  {Khostovan}, {Kirshner}, {Lani}, {Lee}, {Li}, {Madau}, {McCarthy},
  {McIntosh}, {McLure}, {McPartland}, {Mobasher}, {Moreira}, {Mortlock},
  {Moustakas}, {Mozena}, {Nandra}, {Newman}, {Nielsen}, {Niemi}, {Noeske},
  {Papovich}, {Pentericci}, {Pope}, {Primack}, {Ravindranath}, {Reddy},
  {Renzini}, {Rix}, {Robaina}, {Rosario}, {Rosati}, {Salimbeni}, {Scarlata},
  {Siana}, {Simard}, {Smidt}, {Snyder}, {Somerville}, {Spinrad}, {Straughn},
  {Telford}, {Teplitz}, {Trump}, {Vargas}, {Villforth}, {Wagner}, {Wand ro},
  {Wechsler}, {Weiner}, {Wiklind}, {Wild}, {Wilson}, {Wuyts}, \&
  {Yun}}]{Koekemoer11}
{Koekemoer} A.~M., {Faber} S.~M., {Ferguson} H.~C., {Grogin} N.~A., {Kocevski}
  D.~D., {Koo} D.~C., {Lai} K., {Lotz} J.~M. {et~al}, 2011, \apjs, 197, 36

\bibitem[{{Kreckel} {et~al.}(2013){Kreckel}, {Groves}, {Schinnerer}, {Johnson},
  {Aniano}, {Calzetti}, {Croxall}, {Draine}, {Gordon}, \&
  {Crocker}}]{Kreckel13}
{Kreckel} K., {Groves} B., {Schinnerer} E., {Johnson} B.~D., {Aniano} G.,
  {Calzetti} D., {Croxall} K.~V., {Draine} B.~T. {et~al}, 2013, \apj, 771, 62

\bibitem[{{Kregel} {et~al.}(2002){Kregel}, {van der Kruit}, \& {de
  Grijs}}]{Kregel02}
{Kregel} M., {van der Kruit} P.~C., {de Grijs} R., 2002, \mnras, 334, 646

\bibitem[{{Krumholz}(2014)}]{Krumholz14}
{Krumholz} M.~R., 2014, \physrep, 539, 49

\bibitem[{{Krumholz} {et~al.}(2009){Krumholz}, {McKee}, \&
  {Tumlinson}}]{Krumholz09}
{Krumholz} M.~R., {McKee} C.~F., {Tumlinson} J., 2009, \apj, 699, 850

\bibitem[{{Lacey} {et~al.}(2016){Lacey}, {Baugh}, {Frenk}, {Benson}, {Bower},
  {Cole}, {Gonzalez-Perez}, {Helly}, {Lagos}, \& {Mitchell}}]{Lacey15}
{Lacey} C.~G., {Baugh} C.~M., {Frenk} C.~S., {Benson} A.~J., {Bower} R.~G.,
  {Cole} S., {Gonzalez-Perez} V., {Helly} J.~C. {et~al}, 2016, \mnras, 462,
  3854

\bibitem[{{Lagos} {et~al.}(2018){Lagos}, {Tobar}, {Robotham}, {Obreschkow},
  {Mitchell}, {Power}, \& {Elahi}}]{Lagos18b}
{Lagos} C. d.~P., {Tobar} R.~J., {Robotham} A. S.~G., {Obreschkow} D.,
  {Mitchell} P.~D., {Power} C., {Elahi} P.~J., 2018, \mnras, 481, 3573

\bibitem[{{Lange} {et~al.}(2015){Lange}, {Driver}, {Robotham}, {Kelvin},
  {Graham}, {Alpaslan}, {Andrews}, {Baldry}, {Bamford}, {Bland-Hawthorn},
  {Brough}, {Cluver}, {Conselice}, {Davies}, {Haeussler}, {Konstantopoulos},
  {Loveday}, {Moffett}, {Norberg}, {Phillipps}, {Taylor},
  {L{\'o}pez-S{\'a}nchez}, \& {Wilkins}}]{Lange15}
{Lange} R., {Driver} S.~P., {Robotham} A.~S.~G., {Kelvin} L.~S., {Graham}
  A.~W., {Alpaslan} M., {Andrews} S.~K., {Baldry} I.~K. {et~al}, 2015, \mnras,
  447, 2603

\bibitem[{{Li} \& {Draine}(2012)}]{Li12}
{Li} A., {Draine} B.~T., 2012, \apjl, 760, L35

\bibitem[{{Madau} \& {Dickinson}(2014)}]{Madau14}
{Madau} P., {Dickinson} M., 2014, \araa, 52, 415

\bibitem[{{Magnelli} {et~al.}(2013){Magnelli}, {Popesso}, {Berta}, {Pozzi},
  {Elbaz}, {Lutz}, {Dickinson}, {Altieri}, {Andreani}, {Aussel},
  {B{\'e}thermin}, {Bongiovanni}, {Cepa}, {Charmandaris}, {Chary}, {Cimatti},
  {Daddi}, {F{\"o}rster Schreiber}, {Genzel}, {Gruppioni}, {Harwit}, {Hwang},
  {Ivison}, {Magdis}, {Maiolino}, {Murphy}, {Nordon}, {Pannella}, {P{\'e}rez
  Garc{\'{\i}}a}, {Poglitsch}, {Rosario}, {Sanchez-Portal}, {Santini}, {Scott},
  {Sturm}, {Tacconi}, \& {Valtchanov}}]{Magnelli13}
{Magnelli} B., {Popesso} P., {Berta} S., {Pozzi} F., {Elbaz} D., {Lutz} D.,
  {Dickinson} M., {Altieri} B. {et~al}, 2013, ArXiv:1303.4436

\bibitem[{{Maraston}(2005)}]{Maraston05}
{Maraston} C., 2005, \mnras, 362, 799

\bibitem[{{Marchetti} {et~al.}(2016){Marchetti}, {Vaccari}, {Franceschini},
  {Arumugam}, {Aussel}, {B{\'e}thermin}, {Bock}, {Boselli}, {Buat},
  {Burgarella}, {Clements}, {Conley}, {Conversi}, {Cooray}, {Dowell}, {Farrah},
  {Feltre}, {Glenn}, {Griffin}, {Hatziminaoglou}, {Heinis}, {Ibar}, {Ivison},
  {Nguyen}, {O'Halloran}, {Oliver}, {Page}, {Papageorgiou}, {Pearson},
  {P{\'e}rez-Fournon}, {Pohlen}, {Rigopoulou}, {Roseboom}, {Rowan-Robinson},
  {Schulz}, {Scott}, {Seymour}, {Shupe}, {Smith}, {Symeonidis}, {Valtchanov},
  {Viero}, {Wang}, {Wardlow}, {Xu}, \& {Zemcov}}]{Marchetti16}
{Marchetti} L., {Vaccari} M., {Franceschini} A., {Arumugam} V., {Aussel} H.,
  {B{\'e}thermin} M., {Bock} J., {Boselli} A. {et~al}, 2016, \mnras, 456, 1999

\bibitem[{{Mitchell} {et~al.}(2013){Mitchell}, {Lacey}, {Baugh}, \&
  {Cole}}]{Mitchell13}
{Mitchell} P.~D., {Lacey} C.~G., {Baugh} C.~M., {Cole} S., 2013, Monthly
  Notices of the Royal Astronomical Society, 435, 87

\bibitem[{{Negrello} {et~al.}(2013){Negrello}, {Clemens}, {Gonzalez-Nuevo}, {De
  Zotti}, {Bonavera}, {Cosco}, {Guarese}, {Boaretto}, {Serjeant}, {Toffolatti},
  {Lapi}, {Bethermin}, {Castex}, {Clements}, {Delabrouille}, {Dole},
  {Franceschini}, {Mandolesi}, {Marchetti}, {Partridge}, \&
  {Sajina}}]{Negrello13}
{Negrello} M., {Clemens} M., {Gonzalez-Nuevo} J., {De Zotti} G., {Bonavera} L.,
  {Cosco} G., {Guarese} G., {Boaretto} L. {et~al}, 2013, \mnras, 429, 1309

\bibitem[{{Nelson} {et~al.}(2018){Nelson}, {Pillepich}, {Springel},
  {Weinberger}, {Hernquist}, {Pakmor}, {Genel}, {Torrey}, {Vogelsberger},
  {Kauffmann}, {Marinacci}, \& {Naiman}}]{Nelson18}
{Nelson} D., {Pillepich} A., {Springel} V., {Weinberger} R., {Hernquist} L.,
  {Pakmor} R., {Genel} S., {Torrey} P. {et~al}, 2018, \mnras, 475, 624

\bibitem[{{Noll} {et~al.}(2009){Noll}, {Burgarella}, {Giovannoli}, {Buat},
  {Marcillac}, \& {Mu{\~n}oz-Mateos}}]{Noll09}
{Noll} S., {Burgarella} D., {Giovannoli} E., {Buat} V., {Marcillac} D.,
  {Mu{\~n}oz-Mateos} J.~C., 2009, \aap, 507, 1793

\bibitem[{{Oesch} {et~al.}(2018){Oesch}, {Bouwens}, {Illingworth}, {Labb{\'e}},
  \& {Stefanon}}]{Oesch18}
{Oesch} P.~A., {Bouwens} R.~J., {Illingworth} G.~D., {Labb{\'e}} I., {Stefanon}
  M., 2018, \apj, 855, 105

\bibitem[{{Ostriker} \& {Peebles}(1973)}]{Ostriker73}
{Ostriker} J.~P., {Peebles} P.~J.~E., 1973, \apj, 186, 467

\bibitem[{{Pacifici} {et~al.}(2012){Pacifici}, {Charlot}, {Blaizot}, \&
  {Brinchmann}}]{Pacifici12}
{Pacifici} C., {Charlot} S., {Blaizot} J., {Brinchmann} J., 2012, \mnras, 421,
  2002

\bibitem[{{Pacifici} {et~al.}(2015){Pacifici}, {da Cunha}, {Charlot}, {Rix},
  {Fumagalli}, {Wel}, {Franx}, {Maseda}, {van Dokkum}, {Brammer}, {Momcheva},
  {Skelton}, {Whitaker}, {Leja}, {Lundgren}, {Kassin}, \& {Yi}}]{Pacifici15}
{Pacifici} C., {da Cunha} E., {Charlot} S., {Rix} H.-W., {Fumagalli} M., {Wel}
  A.~v.~d., {Franx} M., {Maseda} M.~V. {et~al}, 2015, \mnras, 447, 786

\bibitem[{{Patel} {et~al.}(2013){Patel}, {Clements}, {Vaccari}, {Mortlock},
  {Rowan-Robinson}, {P{\'e}rez-Fournon}, \& {Afonso-Luis}}]{Patel13}
{Patel} H., {Clements} D.~L., {Vaccari} M., {Mortlock} D.~J., {Rowan-Robinson}
  M., {P{\'e}rez-Fournon} I., {Afonso-Luis} A., 2013, \mnras, 428, 291

\bibitem[{{Pilbratt} {et~al.}(2010){Pilbratt}, {Riedinger}, {Passvogel},
  {Crone}, {Doyle}, {Gageur}, {Heras}, {Jewell}, {Metcalfe}, {Ott}, \&
  {Schmidt}}]{Pilbratt10}
{Pilbratt} G.~L., {Riedinger} J.~R., {Passvogel} T., {Crone} G., {Doyle} D.,
  {Gageur} U., {Heras} A.~M., {Jewell} C. {et~al}, 2010, \aap, 518, L1

\bibitem[{{Planck Collaboration} {et~al.}(2016){Planck Collaboration}, {Ade},
  {Aghanim}, {Arnaud}, {Ashdown}, {Aumont}, {Baccigalupi}, {Banday},
  {Barreiro}, {Bartlett}, \& et~al.}]{Planck15}
{Planck Collaboration}, {Ade} P.~A.~R., {Aghanim} N., {Arnaud} M., {Ashdown}
  M., {Aumont} J., {Baccigalupi} C., {Banday} A.~J. {et~al}, 2016, \aap, 594,
  A13

\bibitem[{{Popping} {et~al.}(2017){Popping}, {Somerville}, \&
  {Galametz}}]{Popping17}
{Popping} G., {Somerville} R.~S., {Galametz} M., 2017, \mnras, 471, 3152

\bibitem[{{Poulton} {et~al.}(2018){Poulton}, {Robotham}, {Power}, \&
  {Elahi}}]{Poulton18}
{Poulton} R. J.~J., {Robotham} A. S.~G., {Power} C., {Elahi} P.~J., 2018,
  Publications of the Astronomical Society of Australia, 35, 42

\bibitem[{{Pozzetti} {et~al.}(2003){Pozzetti}, {Cimatti}, {Zamorani}, {Daddi},
  {Menci}, {Fontana}, {Renzini}, {Mignoli}, {Poli}, {Saracco}, {Broadhurst},
  {Cristiani}, {D'Odorico}, {Giallongo}, \& {Gilmozzi}}]{Pozzetti03}
{Pozzetti} L., {Cimatti} A., {Zamorani} G., {Daddi} E., {Menci} N., {Fontana}
  A., {Renzini} A., {Mignoli} M. {et~al}, 2003, \aap, 402, 837

\bibitem[{{Qiu} {et~al.}(2019){Qiu}, {Mutch}, {da Cunha}, {Poole}, \&
  {Wyithe}}]{Qiu19}
{Qiu} Y., {Mutch} S.~J., {da Cunha} E., {Poole} G.~B., {Wyithe} J. S.~B., 2019,
  Monthly Notices of the Royal Astronomical Society, 2156

\bibitem[{{Reddy} \& {Steidel}(2009)}]{Reddy09}
{Reddy} N.~A., {Steidel} C.~C., 2009, \apj, 692, 778

\bibitem[{{R{\'e}my-Ruyer} {et~al.}(2014){R{\'e}my-Ruyer}, {Madden},
  {Galliano}, {Galametz}, {Takeuchi}, {Asano}, {Zhukovska}, {Lebouteiller},
  {Cormier}, \& {Jones}}]{Remy-Ruyer14}
{R{\'e}my-Ruyer} A., {Madden} S.~C., {Galliano} F., {Galametz} M., {Takeuchi}
  T.~T., {Asano} R.~S., {Zhukovska} S., {Lebouteiller} V. {et~al}, 2014, \aap,
  563, A31

\bibitem[{{Romeo} \& {Mogotsi}(2018)}]{Romeo18}
{Romeo} A.~B., {Mogotsi} K.~M., 2018, \mnras, 480, L23

\bibitem[{{Romeo} \& {Wiegert}(2011)}]{Romeo11}
{Romeo} A.~B., {Wiegert} J., 2011, \mnras, 416, 1191

\bibitem[{{Santini} {et~al.}(2014){Santini}, {Maiolino}, {Magnelli}, {Lutz},
  {Lamastra}, {Li Causi}, {Eales}, {Andreani}, {Berta}, {Buat}, {Cooray},
  {Cresci}, {Daddi}, {Farrah}, {Fontana}, {Franceschini}, {Genzel}, {Granato},
  {Grazian}, {Le Floc'h}, {Magdis}, {Magliocchetti}, {Mannucci}, {Menci},
  {Nordon}, {Oliver}, {Popesso}, {Pozzi}, {Riguccini}, {Rodighiero}, {Rosario},
  {Salvato}, {Scott}, {Silva}, {Tacconi}, {Viero}, {Wang}, {Wuyts}, \&
  {Xu}}]{Santini13}
{Santini} P., {Maiolino} R., {Magnelli} B., {Lutz} D., {Lamastra} A., {Li
  Causi} G., {Eales} S., {Andreani} P. {et~al}, 2014, \aap, 562, A30

\bibitem[{{Saracco} {et~al.}(2006){Saracco}, {Fiano}, {Chincarini}, {Vanzella},
  {Longhetti}, {Cristiani}, {Fontana}, {Giallongo}, \& {Nonino}}]{Saracco06}
{Saracco} P., {Fiano} A., {Chincarini} G., {Vanzella} E., {Longhetti} M.,
  {Cristiani} S., {Fontana} A., {Giallongo} E. {et~al}, 2006, \mnras, 367, 349

\bibitem[{{Sawicki} \& {Thompson}(2006)}]{Sawicki06}
{Sawicki} M., {Thompson} D., 2006, \apj, 642, 653

\bibitem[{{Scoville} {et~al.}(2007){Scoville}, {Aussel}, {Brusa}, {Capak},
  {Carollo}, {Elvis}, {Giavalisco}, {Guzzo}, {Hasinger}, {Impey}, {Kneib},
  {LeFevre}, {Lilly}, {Mobasher}, {Renzini}, {Rich}, {Sanders}, {Schinnerer},
  {Schminovich}, {Shopbell}, {Taniguchi}, \& {Tyson}}]{Scoville07}
{Scoville} N., {Aussel} H., {Brusa} M., {Capak} P., {Carollo} C.~M., {Elvis}
  M., {Giavalisco} M., {Guzzo} L. {et~al}, 2007, \apjs, 172, 1

\bibitem[{{Scoville} {et~al.}(2016){Scoville}, {Sheth}, {Aussel}, {Vanden
  Bout}, {Capak}, {Bongiorno}, {Casey}, {Murchikova}, {Koda},
  {{\'A}lvarez-M{\'a}rquez}, {Lee}, {Laigle}, {McCracken}, {Ilbert}, {Pope},
  {Sanders}, {Chu}, {Toft}, {Ivison}, \& {Manohar}}]{Scoville16}
{Scoville} N., {Sheth} K., {Aussel} H., {Vanden Bout} P., {Capak} P.,
  {Bongiorno} A., {Casey} C.~M., {Murchikova} L. {et~al}, 2016, \apj, 820, 83

\bibitem[{{Somerville} {et~al.}(2012){Somerville}, {Gilmore}, {Primack}, \&
  {Dom{\'{\i}}nguez}}]{Somerville12}
{Somerville} R.~S., {Gilmore} R.~C., {Primack} J.~R., {Dom{\'{\i}}nguez} A.,
  2012, \mnras, 423, 1992

\bibitem[{{Somerville} {et~al.}(2015){Somerville}, {Popping}, \&
  {Trager}}]{Somerville15}
{Somerville} R.~S., {Popping} G., {Trager} S.~C., 2015, \mnras, 453, 4337

\bibitem[{{Trayford} {et~al.}(2017){Trayford}, {Camps}, {Theuns}, {Baes},
  {Bower}, {Crain}, {Gunawardhana}, {Schaller}, {Schaye}, \&
  {Frenk}}]{Trayford17}
{Trayford} J.~W., {Camps} P., {Theuns} T., {Baes} M., {Bower} R.~G., {Crain}
  R.~A., {Gunawardhana} M. L.~P., {Schaller} M. {et~al}, 2017, \mnras, 470, 771

\bibitem[{{Trayford} {et~al.}(2019){Trayford}, {Lagos}, {Robotham}, \&
  {Obreschkow}}]{Trayford19}
{Trayford} J.~W., {Lagos} C. d.~P., {Robotham} A. S.~G., {Obreschkow} D., 2019,
  arXiv:1908.08956, arXiv:1908.08956

\bibitem[{{Trayford} {et~al.}(2015){Trayford}, {Theuns}, {Bower}, {Schaye},
  {Furlong}, {Schaller}, {Frenk}, {Crain}, {Vecchia}, \&
  {McCarthy}}]{Trayford15}
{Trayford} J.~W., {Theuns} T., {Bower} R.~G., {Schaye} J., {Furlong} M.,
  {Schaller} M., {Frenk} C.~S., {Crain} R.~A. {et~al}, 2015, \mnras, 452, 2879

\bibitem[{{Vazdekis} {et~al.}(2016){Vazdekis}, {Koleva}, {Ricciardelli},
  {R{\"o}ck}, \& {Falc{\'o}n-Barroso}}]{Vazdekis16}
{Vazdekis} A., {Koleva} M., {Ricciardelli} E., {R{\"o}ck} B.,
  {Falc{\'o}n-Barroso} J., 2016, \mnras, 463, 3409

\bibitem[{{Vijayan} {et~al.}(2019){Vijayan}, {Clay}, {Thomas}, {Yates},
  {Wilkins}, \& {Henriques}}]{Vijayan19}
{Vijayan} A.~P., {Clay} S.~J., {Thomas} P.~A., {Yates} R.~M., {Wilkins} S.~M.,
  {Henriques} B.~M., 2019, arXiv e-prints, arXiv:1904.02196

\bibitem[{{Vlahakis} {et~al.}(2005){Vlahakis}, {Dunne}, \&
  {Eales}}]{Vlahakis05}
{Vlahakis} C., {Dunne} L., {Eales} S., 2005, \mnras, 364, 1253

\bibitem[{{Vogelsberger} {et~al.}(2019){Vogelsberger}, {Nelson}, {Pillepich},
  {Shen}, {Marinacci}, {Springel}, {Pakmor}, {Tacchella}, {Weinberger},
  {Torrey}, \& {Hernquist}}]{Vogelsberger19}
{Vogelsberger} M., {Nelson} D., {Pillepich} A., {Shen} X., {Marinacci} F.,
  {Springel} V., {Pakmor} R., {Tacchella} S. {et~al}, 2019, arXiv e-prints,
  arXiv:1904.07238

\bibitem[{{Wang} {et~al.}(2019){Wang}, {Pearson}, {Cowley}, {Trayford},
  {B{\'e}thermin}, {Gruppioni}, {Hurley}, \& {Micha{\l}owski}}]{WangL19}
{Wang} L., {Pearson} W.~J., {Cowley} W., {Trayford} J.~W., {B{\'e}thermin} M.,
  {Gruppioni} C., {Hurley} P., {Micha{\l}owski} M.~J., 2019, \aap, 624, A98

\bibitem[{{Wardlow} {et~al.}(2011){Wardlow}, {Smail}, {Coppin}, {Alexand er},
  {Brandt}, {Danielson}, {Luo}, {Swinbank}, {Walter}, {Wei{\ss}}, {Xue},
  {Zibetti}, {Bertoldi}, {Biggs}, {Chapman}, {Dannerbauer}, {Dunlop},
  {Gawiser}, {Ivison}, {Knudsen}, {Kov{\'a}cs}, {Lacey}, {Menten}, {Padilla},
  {Rix}, \& {van der Werf}}]{Wardlow11}
{Wardlow} J.~L., {Smail} I., {Coppin} K.~E.~K., {Alexand er} D.~M., {Brandt}
  W.~N., {Danielson} A.~L.~R., {Luo} B., {Swinbank} A.~M. {et~al}, 2011,
  \mnras, 415, 1479

\bibitem[{{Wild} {et~al.}(2011){Wild}, {Charlot}, {Brinchmann}, {Heckman},
  {Vince}, {Pacifici}, \& {Chevallard}}]{Wild11}
{Wild} V., {Charlot} S., {Brinchmann} J., {Heckman} T., {Vince} O., {Pacifici}
  C., {Chevallard} J., 2011, \mnras, 417, 1760

\bibitem[{{Xie} {et~al.}(2017){Xie}, {De Lucia}, {Hirschmann}, {Fontanot}, \&
  {Zoldan}}]{Xie17}
{Xie} L., {De Lucia} G., {Hirschmann} M., {Fontanot} F., {Zoldan} A., 2017,
  \mnras, 469, 968

\bibitem[{{Yung} {et~al.}(2019){Yung}, {Somerville}, {Finkelstein}, {Popping},
  \& {Dav{\'e}}}]{Yung19}
{Yung} L.~Y.~A., {Somerville} R.~S., {Finkelstein} S.~L., {Popping} G.,
  {Dav{\'e}} R., 2019, \mnras, 483, 2983

\end{thebibliography}







\bsp	
\label{lastpage}
\end{document}